\documentclass[11pt]{article}
\pdfoutput=1

\usepackage{my-j}
\usepackage[T1]{fontenc} 

\usepackage[vcentermath]{youngtab}
\usepackage{array,multirow}
\usepackage{amssymb}
\usepackage{amsfonts}
\usepackage{amsbsy}
\usepackage{amsmath}
\usepackage{amsthm}
\usepackage{graphicx}
\usepackage{epstopdf}

\def\Z{\mathbb{Z}}
\def\F{\mathbb{F}}

\def\C{\mathbb{C}}

\def\R{\mathbb{R}}

\def\P{\mathbb{P}}

\def\n3a{t}

\def\tr{{\mathrm{tr}}}

\def\O{\mathcal{O}}

\def\ge{{\mathfrak{e}}}
\def\gso{{\mathfrak{so}}}
\def\gsu{{\mathfrak{su}}}
\def\gsp{{\mathfrak{sp}}}
\def\gf{{\mathfrak{f}}}
\def\gg{{\mathfrak{g}}}

\def\aa{{\mathfrak{a}}}
\def\bb{{\mathfrak{b}}}

\def\neutral{0}
\def\hon{H_{\rm u}}

\newcommand{\eq}[1]{(\ref{#1})}

\title{Enhanced gauge symmetry in 6D F-theory models
and tuned elliptic Calabi-Yau threefolds}

\author{
Samuel B.\ Johnson and Washington Taylor \\
Center for Theoretical Physics\\
Department of Physics\\
Massachusetts Institute of Technology\\
77 Massachusetts Avenue\\
Cambridge, MA 02139, USA\\[0.7cm]
{\tt samj} {\rm at} {\tt mit.edu}\ \  ; \ {\tt wati} {\rm at} {\tt mit.edu}
}

\preprint{MIT-CTP-4645}

\abstract{
We systematically analyze the local combinations of gauge groups and
matter that can arise in 6D F-theory models over a fixed base.  We
compare the low-energy constraints of anomaly cancellation to explicit
F-theory constructions using Weierstrass and Tate forms,
and identify some new local
structures in the ``swampland'' of 6D supergravity and SCFT models
that appear consistent from low-energy considerations but do not have
known
F-theory realizations.  In
particular, we classify and carry out a local analysis of all
enhancements of the irreducible gauge and matter contributions from
``non-Higgsable clusters,'' and on isolated curves and pairs of
intersecting rational curves of arbitrary self-intersection.  Such
enhancements correspond physically to unHiggsings, and mathematically
to tunings of the Weierstrass model of an elliptic CY threefold.  We
determine the shift in Hodge numbers of the elliptic threefold
associated with each enhancement.  We also consider local tunings on
curves that have higher genus or intersect multiple other curves,
codimension two tunings that give transitions in the F-theory matter
content, tunings of abelian factors in the gauge group, and
generalizations of the ``$E_8$'' rule to include tunings and curves of
self-intersection zero.  These tools can be combined into an algorithm
that in principle enables a finite and systematic classification of
all elliptic CY threefolds and corresponding 6D F-theory SUGRA models
over a given compact base (modulo some technical caveats in various
special circumstances), and are also relevant to the classification of
6D SCFT's.  To illustrate the utility of these results, we identify
some large example classes of known CY threefolds in the
Kreuzer-Skarke database as Weierstrass models over complex surface
bases with specific simple tunings, and we survey the range of tunings
possible over one specific base.    }

\begin{document}
\maketitle

\flushbottom

\section{Introduction}

Compactifications have long played a fundamental role within string
theory, beginning as a necessary ingredient for approaching 4D
phenomenology. Evolving perspectives on string theory have yielded
increasingly rich classes of compactifications. With the advent of
F-theory \cite{Vafa-F-theory, Morrison-Vafa}, a spatially varying
axio-dilaton profile in type IIB string theory has enabled the
construction of lower-dimensional theories with a wide range of gauge
symmetries and matter content in a non-perturbative and unifying
framework.  In recent years, following extensive work on increasingly
phenomenologically coherent F-theory GUT models \cite{phenom-1,
  phenom-2}, there has been a renaissance of work on both
phenomenological and more foundational aspects of F-theory.  For 4D
$\mathcal{N}=1$ F-theory constructions, fluxes produce a
superpotential that lifts many moduli; non-perturbative 7-brane
world-volume dynamics are difficult to control; and other
constructions appear to give large classes of vacua with no F-theory
dual. These difficulties render the exploration and global
understanding of the full set of 4D ${\cal N} = 1$ compactifications
quite subtle. In 6D, on the other hand, the class of models resulting
from F-theory compactifications is tightly controlled.  In particular,
there is a  very close connection between the geometry of complex
surfaces and the physics of 6D supergravity theories; for example,
low-energy consistency conditions such as the Green-Schwarz mechanism
of anomaly cancellation \cite{gs-west, Sagnotti}, which are quite
stringent, can be mapped directly to geometric constraints
\cite{Grassi-Morrison, KMT,
  KMT-II, Grassi-Morrison-2, Grimm-Kapfer}.

The moduli space of 6D supergravities arising from F-theory
compactifications is a connected space, with different branches
connected through superconformal fixed points by tensionless string
transitions \cite{Seiberg-Witten, Morrison-Vafa, KMT-II}. The
possibility of a complete and explicit description of this space seems
in principle feasible. F-theory compactifications to 6D are realized
by compactifying type IIB string theory on a complex surface base $B$
that supports an elliptic Calabi-Yau threefold. In recent work,
significant progress has been made on a systematic classification of
all possible such compact smooth base surfaces $B$ \cite{clusters,
toric, martini, wang-non-toric}. If one can in addition
understand the allowed tunings of different elliptic fibrations over
each of
these bases, corresponding to F-theory models with distinct spectra,
one would have a very clear handle on this moduli space. Some progress
in this direction was made in \cite{large-h21}. One of the primary
goals of this paper is to expand the set of tools for the systematic
understanding of such tunings, which correspond physically to
enhancing the gauge and matter content of the supergravity
model. Recently, similar methods have been used to explore
six-dimensional superconformal field theories (SCFT's) \cite{SCFT-I,
  SCFT-II}. These non-compact F-theory models are in some ways
simpler, yet in other ways less constrained, than their supergravity
counterparts; for these theories as well, the classification problem
can be divided into two independent sub-problems: the classification
of the base geometry, and the classification of tuned enhancements of
gauge symmetry and matter content, given a base geometry.  Parts of
this paper that are relevant to SCFT's have significant overlap with
\cite{SCFT-II}, which appeared while this work was in progress.

Mathematically, the classification of 6D F-theory models corresponds
to the classification of Weierstrass models for elliptic Calabi-Yau threefolds.
Every elliptic
Calabi-Yau threefold has a Weierstrass model realization
\cite{Nakayama}, so this problem is tantamount to
(though slightly
distinct from) the problem of classifying elliptic Calabi-Yau threefolds.
The minimal model program for
surfaces and the work of Grassi \cite{Grassi} show that all base
surfaces that support elliptic CY threefolds can be constructed as
blow-ups of a small finite set of minimal surfaces.  It was proven by
Gross \cite{Gross} that the number of birational equivalence classes
of elliptically fibered Calabi-Yau threefolds is finite.  A more
constructive argument was given in \cite{KMT-II}, which shows that
there are a finite number of distinct families of Weierstrass models
corresponding to distinct 6D F-theory spectra; these Weierstrass
models can be in principle classified by first constructing all bases
$B$ and then finding all allowed tunings over each  $B$.    
It is the latter  problem, of tuning Weierstrass models over a given base
$B$, that we address in the present work.

There are thus three distinct classification problems that we wish to
make progress towards by the systematic study of allowed tunings in 6D
F-theory models: First, the classification of elliptic Calabi-Yau
threefolds; second, the related classification of 6D supergravity
models; and third, the classification of 6D SCFT's.  It has been
conjectured that all quantum consistent 6D supergravity theories and
all 6D superconformal field theories have realizations in F-theory
\cite{universality, KMT, SCFT-I}. Thus, a complete classification of
allowed tunings may help to solve all three of these classification
problems.
While we do not completely solve these classification problems here, we
develop a systematic description of possible local tunings, which we
incorporate into an algorithmic framework for approaching  these
problems, with unresolved issues isolated into specific technical
questions that can be addressed in further work.
Throughout the paper, we also identify and focus particular attention
on theories in the ``swampland'' \cite{Vafa-swampland}, which are
apparently consistent from low-energy considerations, yet do not have
an F-theory description.

The outline of this paper is the following. We first review the basics
of F-theory constructions in \S\ref{sec:tools}, including the
traditional interpretation in terms of a IIB theory with a varying
axio-dilaton profile.  We review the structure of non-Higgsable
clusters, 6D supergravity, and 6D SCFTs. We also include a brief
review of some aspects of algebraic geometry and toric geometry,
setting notation for the following calculations.  In
\S\ref{sec:strategy}, we present a more in-depth summary of our
classification strategy, the components of which are developed in the
following sections.  Sections \S\ref{sec:C-I} through
\S\ref{sec:C-III} contain the core results of the paper: they consist
respectively of constraints for tunings of enhanced gauge groups on
isolated divisors (curves), constraints for tunings on multiple-divisor
clusters, and constraints determining when two or more of the
previously discussed tunings may be achieved simultaneously. These
constraints are all local in that they require the simultaneous
consideration of local configurations of intersecting divisors in the base.  For
many of these cases, particularly those where the curves involved form
a non-Higgsable cluster, we explicitly check that local models for all
of the possible gauge groups allowed by the constraints can be
realized explicitly by tuning monomials in a Weierstrass model.    In
\S\ref{sec:matter} and \S\ref{sec:abelian} we describe the tunings of
exotic matter and abelian gauge factors, which go beyond the simple
Kodaira classification of gauge groups and generic matter. In
\S\ref{sec:algorithm}, we assemble all these pieces into a
systematic algorithm for determining a finite set of possible
Weierstrass models over a given base.  In some cases, this list may
represent a superset of the true set of allowed F-theory models;
certain combinations of local structures must be checked explicitly to
verify the existence of a global Weierstrass model.  In
\S\ref{sec:examples}, we illustrate the utility of these rules by
investigating two classes of tuned models in the Kreuzer-Skarke
database, and exploring the range of tunings possible over the
Hirzebruch surface base $B = \F_{12}$. Lastly, \S\ref{sec:conclusions}
reviews our results and their limitations, and explores some
potentially interesting future directions.

As this paper was in the final stages of completion, we learned of 
independent concurrent work by Morrison and Rudelius
\cite{Morrison-Rudelius} that also considers the generalization of the
$E_8$ rule to tuned gauge groups.

\section{Physical and geometric background}\label{sec:tools}

\subsection{F-theory preliminaries}
\label{sec:Tate}

We begin with a brief review of some basic aspects of F-theory; more
extensive pedagogical introductions can be found in
\cite{Morrison-TASI, Denef-F-theory, WT-TASI}.
In its original form, F-theory \cite{Vafa-F-theory, Morrison-Vafa} is
an attempt to capture and codify geometrically the $SL(2,\Z)$
S-symmetry of type IIB string theory in a way that greatly broadens
the class of manifolds on which the IIB theory can be compactified
while preserving symmetry, by allowing variation of the axiodilaton
field $\tau:=C_0+ie^{-\phi}$ to compensate for curvature in the
compactification space. Under the IIB $SL(2,\Z)$ symmetry, the
axiodilaton transforms as
\begin{equation}
\tau \rightarrow \tau'=\frac{a\tau+b}{c\tau+d}
\end{equation}
for $ab-cd =\pm 1$ (all four constants in $\Z$). Because the true
moduli space of IIB is a quotient of the $\tau$ plane by this
symmetry, and this is nothing but the (complex) moduli space of the
torus, the axiodilaton at each point in space-time can be identified with an elliptic
curve $\mathbb{E}$. Allowing this coupling to depend on position yields an elliptic
fibration (with a well defined global zero-section) over the compact
space $B$ of the IIB compactification. 
In practice, when we form an F-theory model by
starting with the IIB theory compactified on a ``base'' $B$, with
varying axio-dilaton profile as above, as a shorthand we say
``F-theory compactified on $X$'' where $X$ is the total space of the
elliptic fibration $\mathbb{E}\hookrightarrow X
\stackrel{\pi}{\rightarrow} B$.

The study of F-theory is morally the study of elliptically fibered
(with section) Calabi-Yau manifolds. (Throughout this paper, we focus
on (complex) threefolds.) Essential physical data of the resulting
low-energy theory on $\R^{1,5}$ are encoded in the singularities of the
elliptic fibration, namely the loci where one or both cycles of the
torus degenerate. Because singularities in the elliptic fibration are
accompanied by monodromies of $\tau$, it is natural to interpret these
singularity loci as the support of 7-branes within the
theory. In terms of the two
generators $T_1=\left( \begin{array}{cc} 
1 & 1 \\
0 & 1 \end{array} \right)$ and $T_2=\left( \begin{array}{cc}
0 & 1 \\
-1 & 0 \end{array} \right)$ of $SL(2,\Z)$, a monodromy of $pT_1+qT_2$
corresponds to a so-called $(p,q)$ 7-brane. (Only the familiar $(1,0)$
D7 brane is visible in perturbative IIB string theory.) Just as it was
discovered early on that when a stack of parallel D-branes coincide,
the open strings stretched between them have endpoint degrees of
freedom that fill out the gauge sector of an $SU(N)$ gauge theory; so
too was it realized that these generalized branes could encode more
general gauge symmetries. Although many perspectives exist on this
result (\cite{Morrison-Vafa, Denef-F-theory}), the most useful
is a long-known mathematical result that completely characterizes
codimension one singularities of elliptic fibrations: the Kodaira
classification (see {\it e.g.} \cite{bhpv}). 

\begin{table}
\begin{center}
\begin{tabular}{|c |c |c |c |c |c |}
\hline
Type &
ord ($f$) &
ord ($g$) &
ord ($\Delta$) &
singularity & nonabelian symmetry algebra\\ \hline \hline
$I_0$&$\geq $ 0 & $\geq $ 0 & 0 & none & none \\
$I_n$ &0 & 0 & $n \geq 2$ & $A_{n-1}$ & $\gsu(n)$  or $\gsp(\lfloor
n/2\rfloor)$\\ 
$II$ & $\geq 1$ & 1 & 2 & none & none \\
$III$ &1 & $\geq 2$ &3 & $A_1$ & $\gsu(2)$ \\
$IV$ & $\geq 2$ & 2 & 4 & $A_2$ & $\gsu(3)$  or $\gsu(2)$\\
$I_0^*$&
$\geq 2$ & $\geq 3$ & $6$ &$D_{4}$ & $\gso(8)$ or $\gso(7)$ or $\gg_2$ \\
$I_n^*$&
2 & 3 & $n \geq 7$ & $D_{n -2}$ & $\gso(2n-4)$  or $\gso(2n -5)$ \\
$IV^*$& $\geq 3$ & 4 & 8 & $\ge_6$ & $\ge_6$  or $\gf_4$\\
$III^*$&3 & $\geq 5$ & 9 & $\ge_7$ & $\ge_7$ \\
$II^*$& $\geq 4$ & 5 & 10 & $\ge_8$ & $\ge_8$ \\
\hline
non-min &$\geq 4$ & $\geq6$ & $\geq12$ & \multicolumn{2}{c|}{ does not occur in 
F-theory } \\ 
\hline
\end{tabular}
\end{center}
\caption[x]{\footnotesize  Table of 
codimension one
singularity types for elliptic
fibrations and associated nonabelian symmetry algebras.
In cases where the algebra is not determined uniquely by the orders
of vanishing of $f, g$,
the precise gauge algebra is fixed by monodromy conditions that can be
identified from the form of the Weierstrass model.
}
\label{t:Kodaira}
\end{table}

To discuss the Kodaira classification, it is necessary to recall a
convenient description of an elliptic curve. In the weighted
projective space $\P^{[2,3,1]}$, an elliptic curve can be written in
Weierstrass form by the equation
\begin{equation}
y^2=x^3+fx+g
\end{equation}
where $(x,y,t)$ are generalized homogeneous
coordinates on $\P^{[2, 3, 1]}$ with the respective weights
of the equivalence relation defining the projective space; we work in
an affine chart where $t=1$. Indeed, from this description it is
obvious that $f$ must have weight $4$ and $g$ weight $6$. Because $f$
and $g$ together determine the complex structure of the torus,
allowing this structure to change as a function of position $(z, w)$ on the
base $B$ amounts to promoting $f$ and $g$ to functions $f(z, w)$ and
$g(z, w)$ on the base. In fact, considering the defining equations as
living in $\P^{[2,3,1]}\times B$, these coefficients must be sections
of line bundles 
$f\in \O(-4K)$, $g\in \O(-6K)$ for the defining equation to yield a
Calabi-Yau total space, where $K$ is the canonical class of the base.

To understand the singularities of the defining equation, we rely on
the discriminant $\Delta$ to locate where its zeroes coincide with
those of its first derivatives. This locus is defined to be the set of
points in $B$ where the following equation holds: 
\begin{equation}
0=\Delta=4f^3+27g^2.
\end{equation}
Being generically codimension one, this locus corresponds to an
effective divisor
on $B$. On each irreducible component $\Sigma$ of
this divisor, a simple gauge algebra factor can reside. The Kodaira
classification makes the connection explicit by associating to a given
singularity a corresponding gauge algebra, according to the orders of
vanishing of $f, g,$ and $\Delta$ on the associated divisor in the
base. The possibilities are listed in table \ref{t:Kodaira}. As
discussed later, there are a few cases with ambiguities that arise
from monodromies of the defining equation in the fiber over the
singularity; the procedure that allows one to discriminate between
these cases is known as the Tate algorithm \cite{Tate, Bershadsky-all,
  kmss-Tate}.  Note that the Kodaira singularity type only
determines the Lie algebra of the resulting nonabelian group $G$.  In
many situations we will not be careful about this distinction; so in
general, for example, we may discuss tuning an SU($N$) gauge group, though
the actual group may have a quotient $G =$ SU($N$)$/\Gamma$ by a
discrete finite subgroup.  In a few cases where this distinction is
relevant we comment explicitly on the issue.

In some circumstances, it is convenient to describe Weierstrass models
starting from a more general form of the equation for an elliptic
curve on $\P^{[2,3,1]}$, known as the {\em Tate form}
\begin{equation}
y^2 + a_1 yx + a_3y = x^3 + a_2x^2 + a_4x + a_6 \,.
\label{eq:Tate}
\end{equation}
Here $a_k \in{\cal O} (-kK)$.
Given such a form, it is straightforward to transform into Weierstrass
form by completing the square in $y$ to remove the terms linear in
$y$, and then shifting $x$ to remove the quadratic term in $x$.  In
the resulting Weierstrass form, $f, g$ can then be expressed in terms
of the $a_k$ \cite{Bershadsky-all}.

The advantage of Tate form is that certain Kodaira singularity types
can be tuned more readily by choosing the sections $a_k$ to vanish to
a given order on a divisor of interest than by constructing the
corresponding Weierstrass model.  For example, if we wish to tune a
gauge algebra $\gsu(6)$ on a divisor $\Sigma$ defined in local coordinates by
$\Sigma =\{z = 0\}$,  in Weierstrass form $f$ and $g$ are described locally
by functions that can be expressed as power series in $z, f = f_0 +
f_1 z + f_2z^2$, etc..  The condition that $\Delta$ vanish to order
$5$ in $z$ while $f_0, g_0 \neq 0$ imposes a series of nontrivial
algebraic conditions on the $f_k, g_k$ coefficient functions.  While
these algebraic equations can be solved explicitly when $\Sigma$ is smooth
\cite{mt-singularities}, the resulting algebraic structures are rather
complex.  In Tate form, on the other hand, the classical algebras
$\gsp (n)$, $\gsu (n)$, and $\gso (n)$ can all be tuned simply by
choosing the leading coefficients in an expansion of the $a_k$ to
vanish to an appropriate order.  Table~\ref{t:Tate} 
 gives the orders
to which the $a_k$ must vanish to ensure the appropriate classical
algebra.  
Note
 that in each case we have only given the minimal required orders
of vanishing. 
Note also that while in most cases
tuning a Tate form guarantees the desired Kodaira singularity type of
the resulting Weierstrass model, there are some exceptions.  In some
cases the resulting Weierstrass model will have extra singularities;
we encounter some examples of this in \S\ref{sec:C-I}.
In other cases, there are Weierstrass models with a given gauge group
that do not follow from the Tate form \cite{mt-singularities}.  Thus,
the Weierstrass form is more complete, but in many cases the Tate
formulation gives a simpler way of constructing certain kinds of
tunings.  

It is important to emphasize that the coefficients in the Weierstrass
form map directly to neutral scalar fields in 6D F-theory models, so the Weierstrass form is useful in computing the spectrum of a theory and
verifying anomaly cancellation; this is much more difficult in Tate
form, where there is some redundancy in the parameterization for any
given Weierstrass model.

As an example of Tate form, we can tune an $\gsu(2)$ on the divisor
$\{ s
= 0\}$ in local coordinates by choosing the Tate model
\begin{equation}
y^2 +2 xy + s y = x^3 +  2 x^2 +  s x +  s^2 \,.
\end{equation}
Converting to Weierstrass form we have
\begin{equation}
y^2 = x^3 + (-3 + 2s) x + (2-2s + 5s^2/4) \,,
\end{equation}
and $\Delta = 99s^2 +{\cal O} (s^3)$, so indeed the discriminant
cancels to order $s^2$ and the Weierstrass model has a Kodaira type
$I_2$ singularity encoding an $\gsu(2)$ gauge algebra.

\begin{table}
\begin{center}
\begin{tabular}{|c |c |c |c |c |c | c |}
\hline
Group&
$a_1$ &
$a_2$ &
$a_3$ &
$a_4$ &
$a_6$ &
$\Delta$\\ \hline \hline
$\gsu (2) =\gsp (1)$ &
0 & 0 & 1 & 1 & 2 & 2\\
$\gsp (n)$ & 0 & 0 & $n$ & $n$ & $2n$ & $2n$\\
$\gsu (n)$ & 0 & 1 & 
$\lfloor n/2 \rfloor$ &$\lfloor (n + 1)/2 \rfloor$&
$n$ & $n$\\
$\gg_2$ & 1 & 1 & 2 & 2 &3 & 6\\
$\gso(7)$,
$\gso(8)^*$ & 1 & 1 & 2 & 2 & 4 & 6\\
$\gso (4n + 1)$,
$\gso (4n + 2)^*$ & 1 & 1 & $n$ & $n + 1$& $2n$ & $2n + 3$\\
$\gso (4n + 3),
\gso(4n + 4)^*$ & 1 & 1 & $n + 1$ & $n + 1$& $2n + 1$ & $2n + 4$\\
$\gf_4$ & 1 &  2 & 2 & 3 & 4 &  8\\
$\ge_6$ & 1 &  2 & 2 & 3 & 5 &  8\\
$\ge_7$ & 1 &  2 &  3 & 3 & 5 &   9\\
$\ge_8$ & 1 &  2 &  3 & 4 & 5 &   10\\
\hline
non-min. & 1 & 2 & 3 & 4 & 6 & 12\\ 
\hline
\end{tabular}
\end{center}
\caption[x]{\footnotesize Table of minimal vanishing orders needed for
  realizing algebras using Tate form.  Algebras marked with
  ${}^*$ require additional monodromy conditions.  In particular, for
  all the $\gso (N)$ algebras with $N$ even, there are monodromy
  conditions that remove the ambiguity between $\gso(2n -1)$ and
  $\gso(2n)$.
Table follows \cite{kmss-Tate}, with minor improvements as discussed
in \S\ref{sec:8}.   
}
\label{t:Tate}
\end{table}

In most cases, tuning a singularity in Tate form is equivalent to
tuning the same singularity in Weierstrass form in the most generic
way.  For example, $N \leq 5$, the Tate tuning of $\gsu(N)$ and the
Weierstrass tuning of $\gsu(N)$ are equivalent on a smooth divisor;
this can be seen explicitly by matching the terms in the analyses of
\cite{kmss-Tate, mt-singularities} using the dictionary provided on
page 22 of \cite{mt-singularities}.  There are more possibilities for
Weierstrass tunings beginning at $\gsu(6)$; note, however, that for
tuning on a curve $\Sigma$ of self-intersection $\Sigma \cdot \Sigma =
-2$, where $f_0, g_0$ are constant on $\Sigma$ since the normal bundle
is equal to the canonical bundle, the Tate and Weierstrass forms are
equivalent.  This fact will be relevant in the later analysis.
Examples of non-Tate tunings have recently been explored in
\cite{mt-singularities, ckpt, agrt, Klevers-WT} and in virtually all
known cases involve non-generic types of matter.

\subsection{Algebraic geometry preliminaries}\label{sec:AG}

To apply the Kodaira classification in various contexts, it is useful
to have available some well
known tools from algebraic geometry.  We first briefly review relevant
aspects of the geometry of the base surfaces in which we are
interested.  We then discuss general arguments that allow one to
deduce the existence of non-Higgsable clusters (NHCs): groups of
divisors over which even a generic fibration has a singularity that corresponds to a nontrivial gauge algebra. Then we introduce a few relevant aspects of
toric geometry that allow one to explicitly execute a given local
tuned gauge algebra enhancement (increasing the Kodaira singularity)
at the level of coordinates; generally, such computations can be used
to explicitly determine that a given tuned fibration is possible
either locally or globally in a geometry with a local or global toric
description.  We primarily focus on local constructions in this paper,
though in some situations global analysis on a toric base is also
relevant.

We are interested in complex surfaces $B$ that can act as the base of
an elliptically fibered Calabi-Yau threefold.  We thus focus on
rational surfaces that can be realized by blowing up $\P^2$ or $\F_m,
m \leq 12$ at a finite number of points.  We review a few basic facts
about such surfaces (for more details see {\it e.g.}
\cite{wang-non-toric}).  Divisors in a complex surface are integer
linear combinations of irreducible algebraic curves on $B$.  The set
of homology classes of curves in $B$ form a signature $(1, T)$ integer
lattice $ \Gamma= H_2 (B,\Z) =\Z^{1 + T}$ where $T =h^{1, 1} (B) -1$.
The intersection form on $\Gamma$ is unimodular, and for $T \neq 1$
can be written as diag ($+ 1, -1, -1, \ldots, -1$).  (For Hirzebruch
surfaces $\F_m$ with $m$ even, the intersection form is the matrix
$((0 1) (1 0))$.)  The canonical class $K$ satisfies $K \cdot K =
9-T$, and can be put into the form $(3, -1, -1, \ldots, -1)$ when the
intersection form is diagonal as above, and in the form $(2, 2)$ for
even Hirzebruch surfaces.  The set of effective curves, which can be
realized algebraically in $B$, form a cone in the homology lattice.
In F-theory, gauge groups can only be tuned on effective curves, so
these are the curves on which we focus attention.  As an example of a
set of allowed bases and their effective cones, the Hirzebruch
surfaces $\F_m$ have a cone of effective curves generated by the
curves $S, F$ where $S\cdot S= -m, S\cdot F = 1, F \cdot F = 0$, and
can support elliptic Calabi-Yau threefolds when $m = 0, \ldots, 8,
12$.

The Zariski decomposition \cite{Zariski} enables one to write $-kK$
of the base in an explicit form that allows one to read off minimal
(generic) degrees of vanishing of $f$, $g$ and $\Delta$ on a given
irreducible divisor. Given any effective divisor, in particular $-kK$,
we can expand it over the rational numbers as a combination of
irreducible effective divisors. We can write
\begin{equation}\label{eq:Zariski}
-kK=\sum_{i=1}^N \sigma_i\Sigma_i+X
\end{equation}
where $\{\Sigma_i\}$ is the set of irreducible effective divisors of negative
self-intersection, each of which must be rigid, and $X\cdot \Sigma_i, \Sigma_i
\cdot \Sigma_j \geq 0$. By the Riemann-Roch formula, curves of genus $0$ satisfy 
\begin{equation}
-2=2g-2=\Sigma\cdot(K+\Sigma)
\end{equation} 
(implying, {\it e.g.},
that a $-2$ curve $\Sigma$ satisfies $K\cdot \Sigma=0$). Taking the intersection
product of (\ref{eq:Zariski}) with a $-n$ curve $\Sigma=\Sigma_1$ yields (in the
case $N=1$) 
\begin{equation}
-k(n-2)\geq \sigma(-n)
\end{equation}
This immediately implies $\sigma\geq k(n-2)/n$, so that for $n\geq 3$, 
for example, we
have $\sigma \geq \frac{4}{3}$, $\geq 2$ for $k=4,6$, respectively. A
section of the line bundle ${\cal O} (-kK)$
thus vanishes to at least
order $\lceil \sigma \rceil$ on each $\Sigma$; therefore,
for $n = 3$ we are in case $IV$ of
the Kodaira classification, for which the algebra is $\gsu(3)$. (In
principle, one must perform an additional calculation using the Tate
algorithm  to distinguish this from $\gsu(2)$. We
will describe how this is done shortly.) This reasoning can be applied
to deduce the existence of all the non-Higgsable clusters
\cite{clusters}. These are
clusters of mutually intersecting divisors of self-intersections $\leq -2$ that
are forced, by this geometric mechanism, to support gauge algebras
even for a generic fibration. By demanding that no points in $B$ reach
a singularity type with ${\rm ord}\ f\geq 4,
{\rm ord}\ g\geq 6$, one can derive a complete set
of constraints for when these NHCs can be connected by $-1$
curves.\footnote{It is important to understand why $-1$ curves play
  a pivotal role here. 
Any two non-Higgsable Kodaira type singularities that are
  independently consistent can also be simultaneously realized on
  divisors that are separated by a $\geq 0$ curve. (For instance, when
  the base is a Hirzebruch surface $\F_n$, which is a $\P^1$ bundle
  over $\P^1$, the fiber is a $0$ curve.) On the other hand, two NHCs
  cannot be separated by a curve of self-intersection $<-1$, since
    then the resulting collection would itself be one larger
    NHC.  And not all combinations of NHC's can be separated by a $-1$
    curve. These facts isolate $-1$ curves as a particularly interesting
    intermediate situation whose cases must be studied with care.} The
  NHCs are listed in
  table~\ref{t:NHCs}. 

\begin{table}
\begin{center}
\begin{tabular}{| c | 
c | c |c |c |
}
\hline
Cluster & gauge algebra & $r$ & $V$ &  $H_{\rm charged}$ 
\\
\hline
(-12) &$\ge_8$ & 8 & 248 &0 \\
(-8) &$\ge_7$&  7 &  133 &0 \\
(-7) &$\ge_7$& 7 & 133 &28 \\
(-6) &$\ge_6$&   6 & 78 &0 \\
(-5) &$\gf_4$&   4 & 52 &0 \\
(-4) &$\gso(8) $&  4 & 28 &0 \\
(-3, -2, -2)  &  $\gg_2 \oplus \gsu(2)$&  3 & 17 &8\\
(-3, -2) &  $\gg_2 \oplus \gsu(2)$ &3 & 17 &8 \\
(-3)& $\gsu(3)$ &  2 & 8 &0 \\
(-2, -3, -2) &$\gsu(2) \oplus \gso(7) \oplus
\gsu(2)$&5 &  27 &16 \\
(-2, -2, \ldots, -2) & no gauge group & 0 & 0 &0 \\
\hline
\end{tabular}
\end{center}
\caption[x]{\footnotesize
List of ``non-Higgsable clusters'' of
  irreducible effective divisors with self-intersection $-2$ or below,
  and corresponding contributions to the gauge algebra and matter
  content of the 6D theory associated with F-theory compactifications
  on a generic elliptic fibration (with section) over a base
  containing each cluster.
The quantities $r$ and $V$ denote the rank and dimension of the
nonabelian gauge algebra, and $H_{\rm charged}$ denotes the number of
charged hypermultiplet matter fields associated with intersections
between the curves supporting the gauge group factors.
}
\label{t:NHCs}
\end{table}

For
genus $0$ curves that intersect their neighbors,
one can elaborate on the previous formula \ref{eq:Zariski}. Taking for example
 $N=3$ ({\it i.e.} including two neighbors),   and assuming
that the  curve $\Sigma$ intersects each of the two curves $\Sigma_{R,
L}$ with multiplicity one, we have $-kK =
\sigma_L\Sigma_L+\sigma\Sigma+\sigma_R\Sigma_R+X$. Intersecting
with the curve $\Sigma$, which we take to have self-intersection $-n$, yields
\begin{eqnarray}\label{eq:avg}
-k(n-2) & \geq & -n\sigma+\sigma_L+\sigma_R \nonumber \\ 
\sigma & \geq  & n^{-1}(k(n-2)+\sigma_L+\sigma_R) 
\end{eqnarray}
This inequality demonstrates that the orders of $f$ and $g$ on
neighboring divisors influence the minimum (generic) order of $f$ and
$g$ on $\Sigma$ itself; the higher these orders become on neighboring
divisors, the higher must be the order on $\Sigma$. We will see the
utility of this in many of the following calculations. 

We mention here that this kind of analysis can be rephrased in terms
of more explicit sheaves. Instead of speaking only of sections of
$\mathcal{O}(-kK)$ on the base, it is possible to describe the leading
nonvanishing term in $f, g$
around any given divisor in terms of sections of a line bundle over
that divisor. To this end, consider a divisor $\Sigma$ of
interest, which can locally be defined as the set $\{ z = 0 \}$ for
some coordinate $z$. Then any section $s\in \mathcal{O}(-kK)$ can be
expanded as a Taylor series in $z$: $s=\sum_{i=0}s_iz^i$ locally. As
derived in \cite{4D-NHC}, the leading nonvanishing coefficient $s_i$
in this expansion may be considered as a section of a sheaf defined
over the rational curve $\Sigma$; moreover this sheaf is explicitly
given as
\begin{eqnarray}\label{eq:sheaves}
s_i & \in & \mathcal{O}_{\Sigma = \P^1}(2k+(k-i)n-\sum_j\phi_j)
\end{eqnarray}
In this formula, $n$ is the self-intersection number of $\Sigma$ and
the sum adds the orders $\phi_j$ of $s$ on $\Sigma_j$ for all
neighbors $\Sigma_j$ of $\Sigma$ (with appropriate multiplicity if the
intersection has multiplicity greater than one). We will also have use
for this formulation in what follows. Just like the above Zariski
formula, it can be used to determine the minimal order of vanishing of
$f$ and $g$ on a divisor of interest, incorporating information about
the orders of $f$ and $g$ on neighboring divisors; this task is easily
accomplished by identifying the smallest $i$ such that $s_i\in
\mathcal{O}(m)$ for nonnegative $m$. This is the first nonvanishing
term in the expansion $s=\sum_is_iz^i$ and therefore the order of $s$
on $\Sigma$ is $i$. (One can check that this reproduces the above formula
\ref{eq:avg}.)

The preceding analysis is useful in determining the leading
nonvanishing terms in $f, g$ on each divisor and the corresponding
non-Higgsable gauge groups over the given base.  In order to analyze
tunings of the Weierstrass model over various divisors, while this
abstract approach is in principle possible to extend and implement, it
is helpful to have a more explicit presentation of the sections $s_i$
in terms of monomials. When
$\Sigma$ is a rational curve (equivalent to $\P^1$), and any other
curves with which it intersects are other rational curves connected by
single transverse intersections in a linear chain, we can give a
complete and explicit description of the local coordinates on $\Sigma$
and its neighbors using the framework of toric geometry.  In
particular, in this case, we may complete the local coordinate system
around $\Sigma =\{z = 0\}$ with a coordinate $w$ on $\Sigma$ (which
could be a local defining coordinate for one of $\Sigma$'s neighbors).
Then the statement $f_i\in\mathcal{O}(j)$ says that $f_i$ is an order
$j$ polynomial in $w$, and the expressions (\ref{eq:sheaves}) are
precisely reproduced by an analysis in local toric coordinates.
Furthermore, this expression holds for all values of $i$, not just the
first nonvanishing term, since the toric coordinates act as global
coordinates.  In the following analysis, therefore, we focus on
explicit local constructions of tunings in the toric context, and
freely use the language of toric geometry, which we now review
briefly.  Our use of toric geometry should always be understood as a
convient way to do calculations in local coordinates that are valid
for genus zero curves intersecting with multiplicity one.
This kind of local analysis thus allows us to compute tunings on sets
of curves that can be locally described torically, even if the full
base geometry is not a toric surface.  When the base is itself a
compact toric variety,  toric coordinates can be used to cover the
full base and we can completely control the Weierstrass model in terms
of monomials in the toric language.

Here we recall some notions and notations from toric geometry;
interested readers may consult excellent references such as
\cite{Fulton} for more background. Most of the relevant concepts are
described in this context and in more detail in \cite{toric}.  A toric
variety can be described by a fan, which for a two (complex)
dimensional variety is characterized by a collection of $r$ integral
vectors $\{v_i\}_{i=1}^r$ in the lattice $N=\Z^2$, each of which
represents a rational curve in a toric surface.  We restrict attention
to smooth toric varieties, where $v_i, v_{i + 1}$ span a unit cell in
the lattice, associated with a 2D cone describing a point in the toric
variety where a pair of local coordinates vanish.  A rational curve of
self-intersection $-n$ satisfies $nv_j = v_{j -1} + v_{j + 1}$.  A
compact toric variety also has a 2D cone connecting $v_r, v_1$.  The
principal formula we will borrow from toric geometry describes a basis
of sections of line bundles over a toric variety, with fixed vanishing
order on $D_j$,
\begin{equation}
\label{eq:toric-span}
{\cal S}(-kK)_{D_j,n_j}=
{\rm span}(\{m\in M \ | \ m\cdot v_i \geq -n \ \& \ m\cdot v_j = -k+n_j \})
\end{equation}
The $lhs$ denotes sections of $-kK$ that vanish to order exactly $n_j$
on the particular divisor $D_j$ associated to $v_j$, where $-K =D_j +
\sum_{i \neq j}^{}D_i $. (Taken together for all $n_j \geq 0$, this
reproduces the full collection of sections of ${\cal O} (-kK)$ without
poles.)  The additional constraints indexed by $i$ correspond to
conditions imposed from other toric rays.  The $rhs$ is the span of a basis of sections of $-nK$ with
the desired orders of vanishing. Finally, $M$ denotes the dual lattice
to $N =\Z^2$. In \S\ref{sec:C-I} and beyond, this formula is used
frequently.

Note that unlike in \cite{toric}, we are not necessarily performing a
global analysis of toric monomials.  For a local analysis on a
single divisor we only include the rays $i = j \pm 1$ adjacent to
$v_j$ in the toric fan, while by increasing the number of rays we can
include further adjacent divisors in a linear chain, or by including
all rays in the toric fan we can consider a global analysis on a toric
base $B$.  

\subsection{6D supergravity}

In the classification of 6D supergravity (SUGRA) vacua, one can bring
to bear the additional tool of anomaly cancellation, which turns out
to be quite powerful. The Green-Schwarz mechanism is possible in 6D if
and only if the anomaly polynomial factorizes, which can be rephrased
as a set of equations on various group theory quantities derived from
simple factors of the gauge group and their representations \cite{gs-west,
Sagnotti,
  Erler}. In fact, these equations are restrictive enough to strongly
constrain the set of possible 6D supergravity theories that can be
realized from F-theory or any other approach \cite{finite, KMT-II}.
These relations can furthermore determine uniquely the matter content
of the 6D theory in many cases.  Vectors in the anomaly polynomial,
which lie in the lattice of charged dyonic strings, map directly to
certain divisors in $H_2(B,\Z)$, which for F-theory constructions
enables computation of the low-energy spectrum of the theory and
associated constraints purely in terms of easily computed quantities
in the base; this greatly simplifies the implementation of anomaly
constraints in the F-theory context.
The close connection between F-theory geometry and 6D supergravity
theories is described in, among other places,
\cite{Grassi-Morrison, KMT-II, Grassi-Morrison-2}.  A detailed description of the
low-energy supergravity action for 6D
F-theory compactifications from the M-theory perspective 
is given in \cite{b-Grimm}.

For an F-theory compactification on a base $B$ with canonical class
$K$ and nonabelian gauge group factors $G_i$ associated with
codimension one singularities on divisors $\Sigma_i$, the anomaly
cancellation conditions are
\cite{gs-west, Sagnotti, Erler},
as summarized in \cite{KMT-II}
\begin{eqnarray}
H-V & = &  273-29T\label{eq:hv}\\
0 & = &     B^i_{\rm adj} - \sum_{\bf R}
x^i_{\bf R} B^i_{\bf R} \label{eq:f4-condition}\\
K \cdot  K & =   &9 - T  \label{eq:aa-condition}\\
-K \cdot \Sigma_i & =  & \frac{1}{6} \lambda_i  \left(  \sum_{\bf R}
x^i_{\bf R} A^i_{\bf R}-
A^i_{\rm adj} \right)  \label{eq:ab-condition}\\
\Sigma_i\cdot  \Sigma_i & =  &\frac{1}{3} \lambda_i^2 \left(  \sum_{\bf R} x_{\bf
  R}^i C^i_{\bf R}  -C^i_{\rm adj}\right)  \label{eq:bb-condition}\\
\Sigma_i \cdot \Sigma_j & = &  \lambda_i \lambda_j \sum_{\bf R S} x_{\bf R S}^{ij} A_{\bf R}^i
A_{\bf S}^j\label{eq:bij-condition}
\end{eqnarray}
where  $A_{\bf R},
B_{\bf R}, C_{\bf R}$ are group theory coefficients defined through
\begin{align}
\tr_{\bf R} F^2 & = A_{\bf R}  \tr F^2 \\
\tr_{\bf R} F^4 & = B_{\bf R} \tr F^4+C_{\bf R} (\tr F^2)^2 \label{eq:bc-definition}\,,
\end{align}
$\lambda_i$ are numerical constants associated with the different
types of gauge group factors ({\it e.g.}, $\lambda = 1$ for $SU(N)$, 2
for $SO(N)$ and $G_2$, $\ldots$), and where $x_{\bf R}^i$ and $x_{\bf
  R S}^{ij}$ denote the number of matter fields that transform in each
irreducible representation ${\bf R}$ of the gauge group factor $G_i$
and $({\bf R} , {\bf S})$ of $G_i \otimes G_j$ respectively. (The
unadorned ``tr'' above denotes a trace in the fundamental
representation.)  Note that for groups such as $SU(2)$ and $SU(3)$,
which lack a fourth order invariant, $B_{\bf R} = 0$ and there is no
condition \eq{eq:f4-condition}.  The group theory coefficients for
matter representations that appear in generic tunings of the various
nonabelian group factors and the values of $\lambda$ for different
groups are compiled in Appendix~\ref{sec:group-coefficients}. 

For groups such as SU($N$), $N > 2$, which have a quartic Casimir, the
coefficients $B_{\bf R}$ exist and are nonzero for most
representations.  For such groups, the anomaly conditions give three
independent constraints on the matter spectrum.  Thus, these
constraints can always be solved in terms of three basic
representations.  For each such group, generic F-theory tunings will
produce matter in a standard set of representations; for example, for
SU($N$), a generic tuning gives a combination of matter in the
fundamental, two-index antisymmetric, and adjoint representations.
For generic tunings, the number of adjoint representations is given by
the genus of the curve supporting the group.  Any other, more exotic,
representation will always be {\it anomaly equivalent}
\cite{KMT-II, Grassi-Morrison-2} to a linear
combination of the three basic representations.  We primarily focus
here on the generic representation content associated with minimal
(Tate) tunings; exotic matter is discussed in \S\ref{sec:matter}.
Note that many groups, particularly the exceptional groups and
SU(2), have no quartic Casimir, and thus (\ref{eq:f4-condition})
identically vanishes.  For these groups, there are only two
independent anomaly constraints, and the generic matter content
consists of
only the fundamental and adjoint representations.  This is discussed
further in \S\ref{sec:higher-genus}.

It is also worthwhile to comment briefly here on the purely
gravitational anomaly condition (\ref{eq:hv}).  For a global F-theory
model this constrains the number of moduli in the theory.  While we
are primarily focused on local constraints here, it must be kept in
mind that a global model must satisfy (\ref{eq:hv}), and in principle
this constraint can produce additional limits on what may be tunable
in a given Weierstrass model.  In fact, in most cases it seems that
the limits on tuning can be determined purely from combinations of
local constraints, so that the gravitational anomaly is generally
automatically satisfied by an F-theory model when all local
constraints are satisfied; it is not, however, proven at this time
that this must always be true.
We focus on deriving local constraints in this paper, but occasionally
reference the connection to the global gravitational anomaly
constraint.

In this paper we use the anomaly cancellation conditions to help
constrain the possibilities for F-theory tunings.  We are also
interested in exploring the ``swampland'' \cite{Vafa-swampland} of
models that appear consistent from known low-energy considerations but
are not realized in F-theory.  The 6D anomaly conditions as well as
other constraints such as the sign of the gauge kinetic term can be
used to strongly constrain 6D supergravity theories based on the
consistency of the low-energy theory.  All consistent F-theory models
should satisfy these constraints; otherwise F-theory would be an
intrinsically inconsistent theory of quantum gravity.  It has been
conjectured \cite{universality} that all consistent 6D ${\cal N} = 1$
supergravity theories have a description in string theory.  Given the
close correspondence between the low-energy theory and the geometry of
F-theory, and the fact that essentially all known
consistent 6D SUGRA spectra that come from string theory can be
realized in F-theory, it seems that F-theory may have the ability to
realize the full moduli space of consistent 6D supergravity theories.
Thus, we highlight particularly those cases where a given tuning seems
consistent from low-energy considerations but does not have a known
construction through an F-theory Weierstrass model.

\subsection{6D SCFTs}

In \cite{SCFT-I}, Heckman, Morrison, and Vafa proposed a method of
generating 6D SCFTs through F-theory. Here we perform only a cursory
review.  One of the crucial ingredients in the classification of
\cite{SCFT-I}, as in the classification of 6D supergravity theories,
is the set of non-Higgsable clusters, which form basic units for
composing 6D SCFTs.

To decouple gravity, F-theory is taken on a non-compact manifold
(cross $\R^{5,1}$) containing some set of seven-branes wrapped on
various closed cycles in the base. 
This defines a field theory, which should
flow to an SCFT under RG. 
Length scales are removed by
simultaneously contracting all the relevant
2-cycles (divisors) in the base geometry to zero
size. 
Whether this is possible in a given
geometry can be determined by investigating the adjacency matrix
\cite{SCFT-I} with entries defined by
\begin{equation}
A_{ij}:=-(D_i\cap D_j)
\end{equation}
If this matrix is positive definite, then all two-cycles can be
contracted simultaneously; otherwise, they cannot. It is interesting
to note that no closed circuit of two-cycles with nontrivial $\pi_1$
can satisfy this condition.  On a compact base, such cycles of
divisors always exist. 

The part of the classification that we carry out in this paper that
relates to tunings on local configurations of negative
self-intersection curves can be applied to the construction and
classification of 6D SCFT's.
In a recent and quite comprehensive work \cite{SCFT-II}, the authors
adopted a related (``atomic'') perspective on classifying 6D SCFTs
via the F theory construction. 
This work was posted during the
completion of this this paper, and overlaps with the relevant parts of
this work.  Where there is overlap, our results are in agreement with
those of \cite{SCFT-II}.  
Our investigation
differs in some aspects, mainly related to the fact that we do not
restrict to the study of SCFTs but are instead interested in using
these tunings in SUGRA as well, so that we are studying a much broader
range of possible tunings, including on curves of nonnegative
self-intersection, and computing Hodge number shifts, which are
irrelevant for 6D SCFTs.  Our results also extend those of
\cite{SCFT-II} in that while most of the computations in that paper
were based on field theory considerations, particularly anomaly
cancellation, we have also explicitly analyzed the local geometry in
all the cases relevant to 6D SCFTs.  This more detailed analysis confirms the close
correspondence between field theory and geometry in those situations
relevant to SCFTs, but also highlights some specific new
cases where  field theory and geometry seem to disagree.

The superconformal field theory perspective also suggests an additional class of
low-energy constraints that do not follow directly from anomalies.  In
particular, it was argued in \cite{SCFT-I} that the detailed constraints
identified in \cite{clusters} on the combinations of non-Higgsable
clusters that can be connected by a $-1$ curve can be understood from
an ``$E_8$'' rule stating that the global symmetry of the SCFT on any
contracted $-1$ curve not carrying a gauge group should be $E_8$, so
that the combination of gauge algebras of other curves intersected by
the $-1$ curve should be a subalgebra of $E_8$.  This logic suggests
that even tuned gauge groups on curves intersecting a $-1$ curve
without its own gauge algebra should obey the same constraint.  In
this paper (\S\ref{sec:generalized-e8}) we explore the extent to which
this extension of the $E_8$ rule holds for tuned F-theory models, and
speculate on an extension to curves of self-intersection 0.
While we find that the $E_8$ rule is satisfied for tuned models as
well as for NHC's, we also identify some cases of tunings
that satisfy this rule but do not admit realization in F-theory using
Tate-based tunings,
posing a puzzle for low-energy consistency conditions.

\subsection{Calabi-Yau threefolds}
\label{sec:Calabi-Yau}

One of the primary goals of this work is to use tunings as a means of
exploring and classifying the space of elliptic
Calabi-Yau threefolds.
For any given elliptically fibered CY threefold $X$ with a Weierstrass
description over a given base $B$, the Hodge numbers of $X$ can be
read off from the form of the singularities and the corresponding data
of the low-energy theory.  A succinct description
of the Hodge numbers of $X$ can be given using the geometry-F-theory
correspondence \cite{Morrison-Vafa, WT-Hodge, KMT}
\begin{eqnarray}
h^{1, 1}(X) & = & r + T +2 \label{eq:11}\\
h^{2, 1} (X) & = &  H_{\rm  \neutral}-1 = 272+ V -29T - H_{\rm  charged}
\label{eq:21}
\end{eqnarray}
Here, $T = h^{1, 1} (B) -1$ is the number of tensor multiplets in the
6D theory; $r$ is the rank of the 6D gauge group and $V$ is the number
of vector multiplets in the 6D theory, while $H_{\rm \neutral}$ and
$H_{\rm charged}$ refer to the number of 6D matter hypermultiplets
that are neutral/charged with respect to the
\emph{Cartan subalgebra} of the gauge group $G$.  The
relation (\ref{eq:11}) is essentially the Shioda-Tate-Wazir formula
\cite{stw}.  The equality (\ref{eq:21}) follows from the gravitational
anomaly cancellation condition in 6D supergravity, $H - V = 273-29T,$
which corresponds to a topological relation on the Calabi-Yau side
that has been verified for most matter representations with known
nongeometric counterparts \cite{Grassi-Morrison,
  Grassi-Morrison-2}. The nonabelian part of the gauge group $G$ can
be read off from the Kodaira types of the singularities in the
elliptic fibration according to Table~\ref{t:Kodaira} (up to the
discrete part, which does not affect the Hodge numbers and that we do
not compute in detail here).

One use of these conditions is to compute the shifts in Hodge numbers
for a given tuning of an enhanced gauge group on a given divisor or
set of divisors.  In many of the local situations we consider here, we
can directly compute the shift in the Hodge number $h^{2, 1}$ by
determining the number of complex degrees of freedom (neutral scalar
fields) that must be fixed in the Weierstrass model to realize the
desired tuning.  In other cases, where we do not have a local model,
we can use (\ref{eq:21}) to compute $h^{2, 1}= H_0 -1$ for a tuning
based simply on the spectrum of the theory.  Note that $h^{1, 1}$
follows simply from the gauge group and number of tensors, and does
not depend upon the detailed matter spectrum.  One  subtlety
is that in cases where a tuned group can be broken to a smaller group
without decreasing the rank, in particular for $G_2 \rightarrow
SU(3)$, $F_4 \rightarrow SO(8)$, and $SO(2N +1) \rightarrow SO(2N)$, the charged fields under the larger group that are
uncharged under the smaller group of equal rank (and which do not
carry charge under any other group) still contribute to
$H_0$ and $h^{2,
  1}(X)$ as neutral multiplets even from the larger group as they are
uncharged under the Cartan subalgebra
\cite{Morrison-Vafa}\footnote{Thanks to Y.\ Wang for discussions on
  this point}, so that the
Hodge numbers of the Calabi-Yau do not change in such a breaking.
  This phenomenon will be treated in more
detail elsewhere \cite{Berglund-note}.  Note that a somewhat related
situation in the ${\cal N} =
2$ 4D context is discussed in
\cite{Danielsson-Sundborg}.
In these situations, in the low-energy theory
the additional vector
fields in the larger nonabelian group cancel in the anomaly conditions
with the remaining
charged fields in a charged multiplet; {\it e.g.}, for $\gg_2
\rightarrow \gsu(3)$, a {\bf 7} $\rightarrow$ ${\bf 3} + 
\bar{\bf{3}} +
  {\bf 1}$, the $ {\bf 1}$ acts as a neutral scalar, and the ${\bf 3} +
\bar{\bf{3}}$ cancel the additional six vector bosons in $\gg_2$.  This is
relevant for many of the tunings discussed here.  For clarity, when
performing tunings we compute 
explicitly the shift in the number of completely uncharged hypermultiplets
$H_{\rm u}$; in all cases except tunings of $\gg_2,
\gf_4, \gso(2 N + 1)$ these correspond precisely to a shift in $h^{2,
  1}$, while in the case of $\gg_2$, {\it etc.}, the shift in $h^{2,
  1}$ should be that associated with the equal-rank tuning with
$\gsu(3)$ {\it etc.}
so
the shift in Hodge numbers can be determined by
considering the related model that is reached after a
rank-preserving breaking.  In the latter cases, where
$H_{\rm u}\neq H_0,$ we denote the shift in
$H_{\rm u}$ in brackets $[\Delta H_{\rm u}]$ to indicate this distinction.

In cases where we have a global toric model, there is a direct
relationship between $H_{\rm \neutral}$ and the number $W$
of Weierstrass
moduli given by the toric monomials in $M$ that describe $f, g$.  This
relationship is given by
\begin{equation}
H_{\rm \neutral} = W-w_{\rm aut} + N_{-2} \,,
\label{eq:h-neutral}
\end{equation}
where $N_{-2}$ is the number of $-2$ curves (on which the discriminant
does not identically vanish), and $w_{\rm aut}$ is
the number of automorphisms of the base, given by 2 for a generic base
with no toric curves of self-intersection $0$ or greater and adding $n
+ 1$ for every toric curve of self-intersection $n \geq 0$.  This
formula allows us to directly compute the  shift in $h^{2, 1}$ even in
local toric models by computing the local change in this quantity.
There is one further additional subtlety \cite{martini}, which is that
certain combinations of $-2$ curves form degenerate elliptic curves;
such configurations have an effective value of $N_{-2}$ that must be
decreased by 1.  We encounter this subtlety in
\S\ref{sec:-2}.

One of the goals of this paper is to continue to develop a systematic
set of tools for classifying elliptic Calabi-Yau threefolds through
F-theory.  This might seem like the reverse of the logical order: to
apply F theory, one needs to know about (elliptically fibered)
Calabi-Yaus.  But there are still many unanswered questions about
Calabi-Yau threefolds in general; for example, it is still unknown
whether there are a finite or infinite number of topological types of
non-elliptic Calabi-Yau threefolds.  Some evidence suggests
\cite{akms, WT-Hodge, Candelas-cs, Gray-hl, large-h21, Anderson-aggl}
that, particularly for large Hodge numbers, a large fraction of
Calabi-Yau threefolds and fourfolds that can be realized using known
construction methods are elliptically fibered.  Since the number of
elliptic Calabi-Yau threefolds is finite this suggests that the number
of Calabi-Yau threefolds may in general be finite, and that
understanding and classifying elliptic Calabi-Yau threefolds may give
insights into the general structure of Calabi-Yau manifolds.
As an example of how the methods developed here can be used in
classification of elliptic Calabi-Yau threefolds,
in \S\ref{sec:eg} we identify several large classes of known
Calabi-Yau threefolds in the Kreuzer-Skarke database as tunings  of
generic elliptic fibrations over allowed bases.

In the context of classification of Calabi-Yau threefolds, there is an
additional point that should be brought out.  Our classification is
essentially one of Weierstrass models, which contain various Kodaira
singularity types.  While any elliptic Calabi-Yau threefold has a
corresponding Weierstrass model, the Weierstrass models for any theory
with a nontrivial Kodaira singularity type, corresponding to a
nonabelian gauge group in the low-energy 6D F-theory model, have
singular total spaces.  The singularities in the total space must be
resolved to get a smooth Calabi-Yau threefold.  This resolution at the
level of codimension one singularities maps essentially to Kodaira's
original classification of singularities.  Resolutions at codimension
two, however, are much more subtle, and in many cases a singular
Weierstrass model can have multiple distinct resolutions at
codimension two, corresponding to different Calabi-Yau threefolds with
the same Hodge numbers but different triple intersection numbers.
There has been quite a bit of work in recent years on these
codimension two resolutions in the F-theory context \cite{mt-singularities, Esole-Yau,
  Lawrie-sn, Hayashi-ls, hlms, Esole-sy, Braun-sn}, but there is as
yet no complete and systematic description of what elliptic Calabi-Yau
threefolds can be related to a given Weierstrass model.    For the
purposes of classifying 6D F-theory models this distinction is
irrelevant, but it would be important in any systematic attempt to
completely classify all smooth elliptic Calabi-Yau threefolds.

\section{Outline of results}\label{sec:strategy}

The following three sections represent the core of this work.  In
them, we present and derive a set of fairly
simple rules that can be used to
determine which gauge symmetries and matter representations are
allowed, given the local geometric data of a set of one or more intersecting
divisors within a complex base surface appropriate for
F-theory. 
For each tuning over the local divisor geometry, we compare the
constraints given from low-energy consistency conditions to the
possibility of an explicit F-theory construction.
Before diving into details, we pause to delineate our
results and outline our methods and strategy.

The setting of our analysis is 6D F-theory, {\it i.e.} F-theory
compactified on a Calabi-Yau threefold that results from an elliptic
fibration with section over a two (complex)-dimensional base manifold $B$.
We focus on local combinations of effective divisors (curves) in $B$.
We focus particularly on smooth rational (genus 0) curves that
intersect pairwise in a single point. These cases are particularly amenable to study: we can analyze them locally using toric methods, they are the
only divisor combinations needed to tune elliptic Calabi-Yau
threefolds that arise as hypersurfaces in toric varieties as studied
in \cite{Kreuzer-Skarke}, and they are the only configurations needed
for analyzing 6D SCFTs.  For these combinations of curves, we carry
out a thorough analysis using both the field theory (anomaly) approach
and a local geometric approach for explicit construction of
Weierstrass and/or Tate models.  In these cases we can confirm that,
with a few notable exceptions that we highlight, the anomaly
constraints match perfectly with the set of configurations that is
allowed in a local Weierstrass model.  In addition to these cases
where we have both local geometry
and field theory control of the
configuration, we also consider more briefly more general
configurations needed to complete the classification of tunings over a
generic base, including higher genus curves
(\S\ref{sec:higher-genus}), exotic matter representations that can
arise for non-generic tunings on smooth curves or tunings on singular
curves (\S\ref{sec:matter}), and tuning of abelian gauge symmetries,
which requires global structure through the Mordell-Weil group
(\S\ref{sec:abelian}).

The results of our analysis could be applied in a variety of ways.
Most simply, they provide a toolkit for easily developing a broad
range of examples of 6D F-theory supergravity models and corresponding
elliptic Calabi-Yau threefolds and/or 6D SCFTs; given a base geometry
one can construct a set of tuned models with any particular desired
properties subject to constraints imposed by the base geometry.
More generally, these results can be used in a systematic
classification of 6D supergravity models or SCFTs.  A complete list of
toric bases that support 6D supergravity models was computed in
\cite{toric}.  The results presented here in principle give the local
information needed to construct all possible tunings on toric curves
over these bases, which could be used to compare with the
Kreuzer-Skarke database \cite{ks-data} to give an interpretation of
many of the constructions in that large dataset in terms of elliptic
fibrations and to identify those examples of Calabi-Yau threefold
that are  {\sl not} elliptically fibered.   
The broader set of constraints described
here for more general tunings in principle gives the basic components
needed for a systematic classification of all tunings, including on
non-toric curves over generic bases.  Combined with the systematic
classification of bases \cite{wang-non-toric}, this provides a
framework for the complete classification of all Weierstrass models
for elliptic Calabi-Yau threefolds.  A more detailed description of
how such an algorithm would proceed is given in \S\ref{sec:algorithm}.
Note that in this more general context, and even to some extent in the
more restricted toric context, our rules really only provide a
superset of the set of allowed tunings.  The local rules that we
provide, even when supported by local explicit Weierstrass
constructions, must be checked for a global tuning for compatibility
by explicit construction of a global Weierstrass model that satisfies
all the conditions needed for the tuning.  While we expect that at least in the
toric context, local rules are essentially adequate for determining
the set of allowed tunings, in a more global context this is less
clear.  For toric bases, there is an explicit description of the
Weierstrass model in terms of monomials \cite{toric}, so that, at
least for tuning over toric divisors in toric bases, the technology
for producing a global Weierstrass model is available.  For more
general bases, or tunings over non-toric curves, a concise and
effective approach for tuning Weierstrass models is not at present
known to the authors.

Within this setting, we summarize the results of the following sections.
These results can be summarized in terms of the following data:
given a base-independent local
collection of divisors $\{D_i\}$  with given genera
and self-
and mutual intersections, we determine a list
$\mathcal{L}[\{D_i\}]$ of the possible gauge symmetries over these
divisors, along with the matter representations and shifts in the
Hodge numbers $(\Delta h^{1,1}, \Delta h^{2,1})$ between the generic
and tuned models.
\begin{itemize}
\item Section \ref{sec:C-I}
analyzes tunings $\mathcal{L}[\Sigma]$
 for isolated
  divisors $\Sigma$ with $-12 \leq \Sigma \cdot \Sigma$. 
Curves of self-intersection below $-12$ cannot arise in valid F-theory
bases, and no tuning is possible over any curve with self intersection
below $-6$.
Local models are used to describe all tunings on genus 0 curves, and
tunings on
higher genus curves are constrained through anomalies.
\item Section \ref{sec:C-II} determines
$\mathcal{L}[C]$ for NHCs $C$ that consist of strings of multiple
  intersecting divisors. Explicitly, these are the
multi-curve NHCs $(-3,-2,-2)$,
  $(-2,-3,-2)$, $(-3,-2)$, and clusters of $-2$ curves of arbitrary
size. (There are in practice bounds on the size
and complexity of such $-2$  clusters that
  can appear F-theory SUGRA bases, some of which we discuss here). 
Local toric models are used for the NHCs with $-3$ curves, and 
a simple ``convexity'' feature is used to classify tunings over $-2$
clusters, the validity of which is checked in  Tate models in
\S\ref{sec:C-III}.
\item Section \ref{sec:C-III} analyzes multiple intersecting curves
  beyond the NHCs.  We show that there are only five combinations of
  gauge algebras (or families thereof) that can be tuned on
  intersecting pairs of divisors, and analyze the constraints on these
  combinations using local (largely Tate) methods.
We also consider constraints on tunings of multiple branes
intersecting a single brane, both when the single brane carries a
gauge group and
when it does not.  The latter case includes the ``$E_8$ rule'' \cite{SCFT-I}
governing
what gauge groups can be realized on divisors that intersect a $-1$
curve, which we generalize to include tunings, and a
similar but weaker rule for curves of  self-intersection zero.
\item Section \ref{sec:matter} gives some further rules that apply for
  tuning exotic matter representations with a finer tuning that leaves
  the gauge group (and $h^{1, 1}(X)$) invariant while modifying the
  matter content.  The underlying F-theory geometry and corresponding
  mathematical structure of non-Tate Weierstrass models is only
  partially understood at this point so this set of results may be
  incomplete.
\item Section \ref{sec:abelian} gives a guide to tuning abelian gauge
  factors over a given base.  While much is known and we can make some
  clear statements about tunings and constraints, this is also a
  rapidly evolving area of research and this set of results may also
  be improved by further progress in understanding such models.
\end{itemize}

\section{Classification I: isolated curves}\label{sec:C-I}

In this section we consider all possible tunings of enhanced groups on
individual divisors in the base.  In general, a divisor in the base is
a curve $\Sigma$ of genus $g$ and self-intersection $ \Sigma \cdot \Sigma = n$.  In
this section we concentrate on generic tunings of a given gauge group,
which means that the curve $\Sigma$ is generally smooth, and supports only
certain generic types of matter.  
For example, for $\gsu(N)$ a generic Weierstrass tuning on a genus 0 curve
will give matter only
in the fundamental ({\bf $N$})
and antisymmetric ({\bf $N (N -1)/2 $}) representations; when the genus
$g$ is nonzero, there are also $g$ adjoint ({\bf $N^2 -1$}) matter fields.
Further tunings that keep the gauge
group fixed but enhance the matter content are discussed in
\S\ref{sec:matter}.

For each type of curve and gauge group we consider both anomaly
constraints and the explicit tuning through the Weierstrass model of
the gauge group.  We focus primarily on rational (genus 0) curves.
For individual rational curves we find an almost perfect
matching between those tunings that are allowed by anomaly
cancellation and what can be realized explicitly in F-theory
Weierstrass models.  For curves of negative self-intersection that can
occur in non-Higgsable clusters, and for exceptional algebras tuned on
arbitrary rational curves, we compute the Hodge shifts
explicitly in Weierstrass models and confirm the match with anomaly
conditions.  
For curves of self-intersection $-2$ and above
supporting the
classical series $\gsu(N)$, $\gsp(N)$, and $\gso(N)$, we use the Tate method
to construct Weierstrass models explicitly, and anomaly cancellation
to predict the Hodge number shifts.

A summary of the allowed tunings on isolated genus 0 curves and
associated Hodge shifts are presented in Table \ref{t:isolated}.
These tunings listed are all those that may be allowed by anomaly
cancellation.  The details of the analysis for these cases are
presented in \S\ref{sec:3}-\S\ref{sec:general-curves}.  In
\S\ref{sec:higher-genus} we use anomaly cancellation to predict the
possible tunings and Hodge number shifts for tunings on higher genus
curves, though we do not compute these explicitly using local models
as the local toric methodology we use here is not applicable in those
cases.  The upshot of this analysis is that for genus 0 curves,
virtually everything in Table~\ref{t:isolated} that is allowed by
anomaly cancellation can be realized explicitly in Weierstrass models,
with the exception of some large SU($N$) tunings on curves of
self-intersection -2 or greater, as discussed explicitly in
\S\ref{sec:general-curves}.

\begin{table}[h]
\begin{center}
\begin{tabular}{|c|c|c|}
\hline
$\gg$ & matter & $(\Delta h^{1,1},\Delta \hon)$ \\
\hline
$\gsu (2)$ &$ (6n + 16) {\bf 2} $ & $(1, -12n-29 ) $\\
$\gsu(N)$ & $((8-N) n + 16){\bf N}$
\hspace*{0.7in} & $(N-1, 
-(\frac{15 N
  -N^2}{2}) n-(15 N + 1)) $
\\
&\hspace*{0.1in}
$+(n+2)\frac{{\bf N}({\bf
    N-1})}{{\bf 2}} $ &\hspace*{0.1in}
\\
$\gsp(N/2)$ & $((8-N)n + 16){\bf N}  $
\hspace*{0.9in}
 & $\left(N/2,
\left[-(\frac{15 N
  -N^2 -2}{2}) n-
\left( \frac{
31 N -N^2}{2} \right]\right)  
\right)$ \\
& \hspace*{0.1in}
$+ (n+1)\left({\bf \frac{{\bf N}({\bf N}-1)}{2} -1} \right) $ &\\
$\gso(N)$ & $(n + (N -4) ){\bf N}
$\hspace*{0.9in}  & $\left( \lfloor N/2\rfloor -r_*,-h_*-
(N+ 16)n -(\frac{N^2 -7 N + 128}{2})\right)$\\
& \hspace*{0.1in}
$+(n+4)2^{5-\lfloor (N+1)/2\rfloor} {\bf S}$&\\
$\gg_2$ & $(3n+10){\bf 7}
$ & $(2-r_*, [-7(3n+8)-h_*])$\\
$\gf_4$ & $(n+5){\bf 26}$ & $(4 -r_*,[-26(n+3) -h_*])$\\
$\ge_6$ & $(n+6){\bf 27}$ 
& $(6 -r_*,(78-h_*)-27(n + 6) )$\\
$\ge_7$ & 
$(4+n/2) {\bf 56}$ & $(7 -r_*, (133-h_*) -28  (n+ 8))$\\
\hline
\end{tabular}
\end{center}
\caption[x]{\footnotesize  Possible tunings of gauge groups with
  generic matter on a curve
  $\Sigma$ of self-intersection $n$, together with matter and shifts
  in Hodge numbers, computed from anomaly cancellation conditions. 
For algebras that can be
  obtained from Higgsing chains from $\ge_8$, these matter contents
were previously computed in
  \cite{Bershadsky-all}. 
$r_*, h_*$ are the rank and difference $V-H_{\rm charged}$ of any non-Higgsable gauge
  factor that may exist on the curve before tuning.
Note that
  tunings are impossible when the formula for the multiplicity of
  representations yields a negative number or a
  fraction. (Multiplicities of $\frac{1}{2}$ are allowed when the
  representation in question is self-conjugate.)
$N$ for $\gsp(N/2)$ is assumed to be even.  $\hon$ is total number of
  uncharged scalars. In cases that admit rank-preserving breaking the
  Hodge numbers are given by those of the model after the breaking;
  brackets indicate cases where $\hon \neq H_0$.}
\label{t:isolated}
\end{table}

To make the method of analysis completely transparent, we carry out an
explicit computation of the possible tunings on a $-3$ curve using
both anomaly and Weierstrass methods in \S\ref{sec:3}.  This example
demonstrates how such anomaly calculations and local toric
calculations are done in practice. It also serves to highlight the
non-trivial agreement between these completely distinct methods: both at the
gross level of which algebras can be tuned, but also at the detailed
level of Hodge number shifts. The corresponding calculations for curves of
self-intersection $-4$ and below, as well as the multiple-curve
non-Higgsable clusters, can be found in Appendix \ref{sec:appendix}.

Before proceeding with the example calculation, we should note that
the core of this section's results, Table \ref{t:isolated}, can be
found to a large extent in a corresponding table in
\cite{Bershadsky-all}. Our version differs from that in
\cite{Bershadsky-all} in two respects: we include algebras that are
not subalgebras of $\ge_8$, and we also include shifts in Hodge
numbers that result from implementing these tunings. This extra
information is essential in order to use these tunings as an
organizational tool to search through the set of elliptically fibered
threefolds.  Finally, our analysis of local Weierstrass models allows
us to determine that virtually all of these configurations that are
allowed by anomaly analysis are actually realizable locally in
F-theory, with some specific possible exceptions that we highlight.

Following the extended example of tunings on a $-3$ curve, we give a
general analysis of tunings on curves of self-intersection $-2$ and
above; these cases can be uniformly described in a single framework.
In these sections and in the Appendix, we discuss all possible tunings
except for $\gso(N)$, because these tunings are 
particularly delicate.  $\gso(N)$ tunings are separately
described in
\S\ref{sec:8}.

\subsection{Extended example: tunings on a $-3$ curve}
\label{sec:3}

Let us begin with an extended example that will illustrate many of the features of the following computations. On an isolated $-3$ curve, the minimal gauge algebra is $\gsu(3)$, which can be enlarged as
\begin{equation} \begin{array}{rccccccccccccc}
{\gg}=\ \  & \gsu(3) &\longrightarrow & \gg_2 & \longrightarrow &  \gso(7) & \longrightarrow &  \gso(8) & \longrightarrow &  \gf_4 & \longrightarrow & \ge_6 & \longrightarrow \ge_7  \\
(f,g)= \ \ & (2,2) & \longrightarrow &  \{ & &  (2,3) & & \} & \longrightarrow & \{ & (3,4) & \} & \longrightarrow & (3,5)  \end{array}
\end{equation}
The middle three and subsequent two gauge algebras are distinguished
by monodromy of the singularity, as per the Kodaira classification; we
will describe this in detail below. These tuned algebras and their
associated matter all fall in a Higgsing chain from $\ge_7$.   The complete set of tunings {\it a priori} allowed on
a $-3$ curve also includes $\gso(N)$ for $8 < N\leq 12$, but these
  will be discussed in the following section.

\subsubsection{Spectrum and Hodge shifts from anomaly cancellation}

First we will perform an anomaly calculation; then we will discuss a
local toric model (essentially $\F_3$ with the $+ 3$ curve removed) on
which we can implement these tunings. 
A tabulation of the relevant anomaly coefficients $A_R, B_R, C_R$ and
$\lambda$ values is given in Appendix~\ref{sec:group-coefficients}
Taking the ``C'' condition, we
find: 
\begin{eqnarray}
\Sigma \cdot \Sigma & = & \frac{\lambda^2}{3}\left(\sum_RC_R-C_{Adj}\right) \nonumber \\
-3 & = &\frac{1}{3} \left(\sum_RC_R-9\right) \nonumber \\
0 & = & \sum_RC_R
\end{eqnarray}
Since all coefficients $C_R > 0$ for $\gsu(3)$ (which follows from the
definition of $C_R$ and the absence of a quartic Casimir), this implies that no
matter transforms under this gauge group; there is only the vector
multiplet in the adjoint ${\bf 8}$. This in turn implies that the
presence of this gauge algebra contributes to the quantity $H_0$
(\ref{eq:21})
by an amount $h_*=V-H_{\rm charged}$ and $r_*=2$ (this
algebra's rank) to $h^{1,1}$. Since the gauge algebra $\gsu(3)$ (with
no matter) corresponds to the generic elliptic fibration over a $-3$
curve ({\it i.e.}, $-3$ is an NHC), we conclude that all shifts between the
generic case and a tuned case in the Hodge numbers $(\Delta
h^{1,1},\Delta h^{2, 1}\sim \Delta H_0)=(\Delta r, \Delta (V-H_{\rm charged})$ must be calculated
as $(\Delta h^{1,1},\Delta H_0)=(r_{\rm tuned}-2,V_{\rm tuned}-H_{\rm
charged,
  tuned}-8)$, as denoted in Table~\ref{t:isolated}.  

With this most generic case in mind, let us calculate the
corresponding quantities for $\gg_2$. Assuming only fundamental
matter\footnote{This is the generic matter type expected for $\gg_2$.
  More generally,
other C coefficients are $\geq 5/2$ and therefore the presence of even
one hypermultiplet in one of these non-fundamental representations
makes it impossible to satisfy the C condition on any negative
self-intersection curve.}, with a multiplicity $N_f$, anomaly calculation gives
\begin{eqnarray}
\Sigma \cdot \Sigma & = & \frac{\lambda^2}{3}\left(\sum_RC_R-C_{Adj}\right) \nonumber \\
-3 & = &\frac{4}{3} \left(\frac{N_f}{4}-\frac{5}{2}\right) \nonumber \\
N_f & = & 1
\end{eqnarray}
The contribution to $\hon$ is (recall that
the adjoint of $\gg_2$  has
dimension $14$ and the fundamental has dimension $7$): 
\begin{equation}
\Delta  \hon =14-7-8=+7-8= [-1]
\end{equation}
In other words, implementing this tuning decreases $\hon$ by one in comparison to the generic case.  Note that, as mentioned in \S\ref{sec:Calabi-Yau},  one
of the charged scalars in the ${\bf 7}$ of $\gg_2$ will really act as a
neutral scalar for purposes of computing $h^{2, 1}(X)$, since it can
be used to break the gauge group without reducing rank.  We continue
to treat this scalar as charged, without contributing to $\hon$,
here and in the rest of the paper, but this caveat should be kept in
mind for all $\gg_2$, $\gf_4$
and $\gso(2n + 1)$ tunings, and is indicated by the notation $[-1]$.

For $\gso(7)$, $C_{Adj}=3$, which implies \cite{finite} that the only
relevant representations on negative self-intersection curves are
${\bf 7_f}$ and ${\bf 8_s}$.  Since $C_f=0$ and $C_s=\frac{3}{8}$, we
have
\begin{eqnarray}
\Sigma \cdot \Sigma & = & \frac{\lambda}{3}\left(\sum_RC_R-C_{Adj}\right) \nonumber \\
-3 & = &\frac{4}{3} \left(N_s\frac{3}{8}-3\right) \nonumber \\
N_s & = & 2
\end{eqnarray}
One can
then use the ``A'' condition to demonstrate\footnote{In calculating
  $K\cdot \Sigma$, we use $(K+\Sigma)\cdot \Sigma= 2g-2=-2$ for a genus 0 curve
  (topologically $\mathbb{P}^1$).}  that $N_f=0$:
\begin{eqnarray}
K\cdot \Sigma & = & \frac{\lambda}{6}\left(A_{adj}-\sum_RA_R\right) \nonumber \\
1 & = & \frac{1}{3}\left(5-N_s-N_f \right) \nonumber \\
3 & = & \left(5-2-N_f \right) \nonumber\\
N_f & = &0
\end{eqnarray}
With knowledge of the representation content in hand, we can compute the change in $\hon$:
\begin{equation}
\Delta \hon = \Delta(V-H)=21-2\times 8 -8 =-3 \,.
\end{equation}
Note that the absence of fundamental matter in this case means that
there is no rank-preserving breaking $\gso(7) \rightarrow \gg_2$, so
that the shift in $\hon= H_0$ is not denoted in brackets.
A similar calculation for $\gso(8)$ yields $N_s =2$, $N_f=1$, hence
\begin{equation}
\Delta \hon =\Delta (V-H)=28 - 3\times 8 -8= +4-8=-4
\end{equation}
Proceeding to $\gf_4$, we find again that only fundamental matter is possible on a $(-3)$-curve, and
\begin{eqnarray}
\Sigma \cdot \Sigma & = & \frac{\lambda}{3}\left(\sum_RC_R-C_{Adj}\right) \nonumber \\
-3 & = &\frac{6^2}{3} \left(\frac{N_f}{12}-\frac{5}{12}\right) \nonumber \\
-3 & = & N_f -5 \nonumber \\
N_f & = & 2
\end{eqnarray}
Recalling that the dimensions of the fundamental and adjoint are $26$ and $52$, respectively, we find
\begin{equation}
\Delta \hon = 52 - 2 \times 26 -8 =  [-8]
\end{equation}
For $\ge_6$, we find
\begin{eqnarray}
\Sigma \cdot \Sigma & = & \frac{\lambda^2}{3}\left(\sum_RC_R-C_{Adj}\right) \nonumber \\
-3 & = & \frac{6^2}{3}\left(\frac{N_f}{12}-\frac{1}{2}\right) \nonumber \\
N_f & = & 3
\end{eqnarray}
Given that the dimensions of fundamental and adjoint are $27$ and $78$, respectively,
\begin{equation}
\Delta \hon = 78 - 3\times 27 -8= -3-8=-11
\end{equation}
Enhancing finally to $\ge_7$, we find
\begin{eqnarray}
\Sigma \cdot \Sigma &=& \frac{12^2}{3}\left(\frac{N}{24}-\frac{1}{6}\right) \nonumber \\
-3 &=& 2(N-4) \nonumber \\
N &=& \frac{5}{2}
\end{eqnarray}
which is possible because the fundamental ${\bf 56}$ of $\ge_7$ is
self-conjugate,  and
hence admits a half-hypermultiplet in six dimensions. This
contributes to $\hon$ in the amount $+133-\frac{5}{2}56=-7$, {\it
  i.e.} represents a shift of $-4$ subsequent to tuning an $\ge_6$, or
in total $\Delta \hon =-15$ from the generic case of $\gsu(3)$.

These calculations can be summarized simply as:
\begin{equation}
\begin{array}{cccccccc}
\gg & \gsu(3) & \gg_2 & \gso(7) & \gso(8) & \gf_4 & \ge_6 & \ge_7 \\
\hline
\Delta \hon & 0 & [-1] & -3 & -4 &  [-8] & -11 & -15
\end{array}
\end{equation}

\subsubsection{Spectrum and Hodge shifts from local geometry}

Now we would like to explain this from a more direct geometric
viewpoint. We will find that we must be careful to implement the most
generic tuning, which (when we consider monodromy) will not always be
obtained simply
by setting monomial coefficients to zero. We use a local
model that can be considered a convenient way to visualize the
monomials in $f$ and $g$ (in local coordinates); alternately, our
local models are simply concrete ways of generating the full set of
monomials consistent with equations \ref{eq:sheaves}.  Torically, the
self-intersection number of any toric divisor $\Sigma \leftrightarrow
v_i$ corresponding to $v_i$ in the fan can be determined by the
formula $v_{i-1}+v_{i+1}=-(\Sigma \cdot \Sigma) v_i$. Therefore, a
linear chain of $k$ rational curves with any specified
self-intersection numbers may be realized by a toric fan with $k+2$
rays, which corresponds to a non-compact toric variety. In this
example, we need three rays (corresponding to the $-3$ curve and its
neighbors). Without loss of generality, we take this fan to be
$(3,-1),(1,0),(0,1)$. Using the methods of section 2, we find that the
monomials of $-nK$ are determined to lie within (or on the boundary
of) a wedge determined by the conditions: $x\geq -n$,
$y\geq -n$, and $y \leq n+3x$. The first condition is automatically
satisfied when the latter two are.

\begin{figure}[!t]
\includegraphics[scale=0.7]{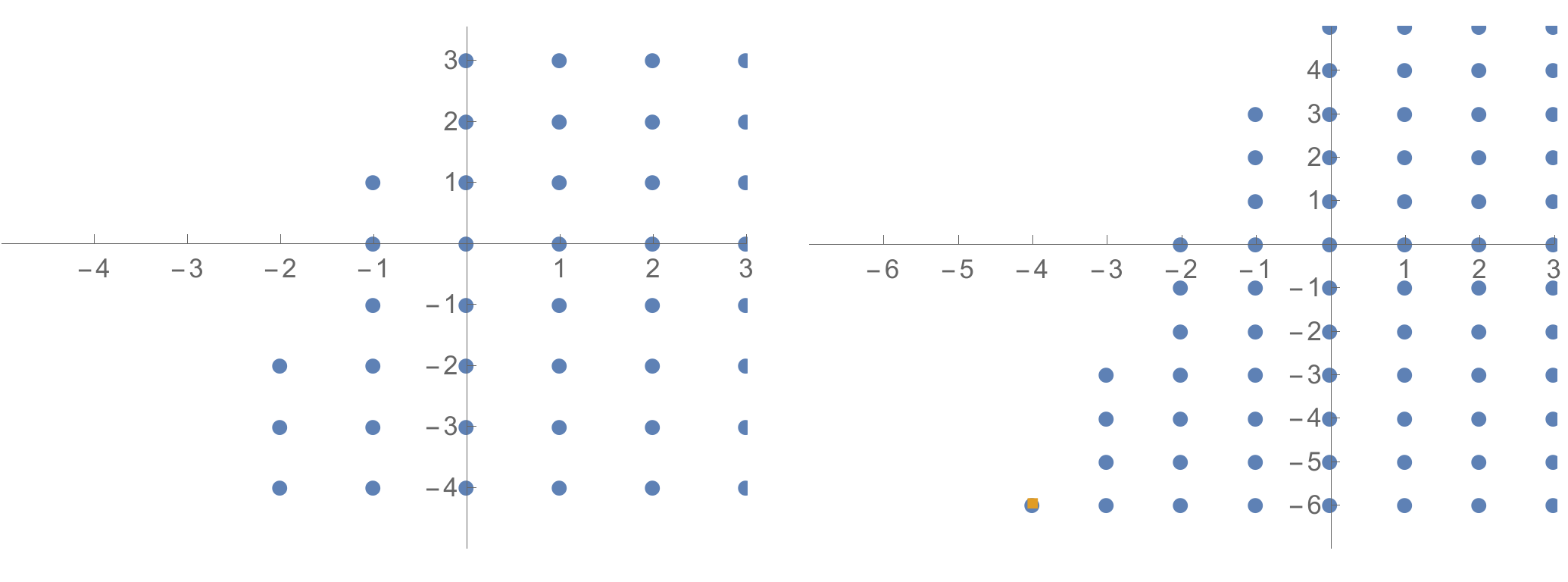}
\caption[x]{A representation of monomials in $-4K$ (left) and $-6K$
  (right) over the local model of a $-3$ curve $\Sigma$ and its two
  neighbors. Both sets of monomials should be considered as extending
  infinitely in the positive  $x$ and $y$ directions. To write these
  monomials explicitly, we may establish a local coordinate system
  $z$, $w$ such that $\Sigma=\{z=0\}$ (corresponding to the ray $(1,0)$ of the toric fan)
and one of its two neighbors
  $\Sigma'$ (corresponding to the ray $(0,1)$) is
  $\Sigma'=\{w=0\}$. Then a monomial $(a,b)\in -kK$ corresponds
  concretely to $z^{a+k}w^{b+k}$.} 
\end{figure}

In this description of the fan, the $-3$ curve $\Sigma$ corresponds to
the ray $v=(1,0)$. We will study the monomials of $f$ ($-4K$) and $g$
($-6K$) in order to determine the number of degrees of freedom
({\it i.e.}, complex structures) that must be removed in order to implement a
given tuning. We describe the different cases in order:\\

{\bf Case 1: $\gsu(3)$:} This is the untuned case, apparent from the
diagrams. If we put local coordinates $z$ and $w$ such that $\Sigma =
\{z=0\}$ and a neighboring curve (say corresponding to the ray $(0,1)$)
is $\{ w=0\}$, then the monomials in $-kK$ represented by points
$(a,b)$ correspond concretely to $z^{a+k}w^{b+k}$. Since the lowest-$x$
monomials in $f$ and $g$ are at $-2$ and $-4$ respectively, this
implies that the orders $(f,g)$ are $(2,2)$ on $\Sigma$. This places
us in Kodaira case $IV$. To determine whether this corresponds to
$\gsu(3)$ or $\gsu(2)$ requires testing a monodromy condition as
outlined in {\it e.g.} \cite{kmss-Tate}. To state this condition, let us expand
$f=\sum_i f_iz^i$ and $g=\sum_i g_i z^i$ as Taylor series in
$z$. Explicitly, then, the monodromy condition to check is whether
$g_2(w)$ is a perfect square:  if it is, the fibration corresponds
to an $\gsu(3)$; otherwise, it corresponds to $\gsu(2)$. In our case,
it is clear that $g_2(w)$ (being the constant polynomial in $w$) is a
perfect square. The properties of $-3$ curve geometry conspire to
force us into this usually non-generic branch of the Kodaira type IV
case. Having established $\gsu(3)$ as the base (untuned) case, let us
investigate tuned fibrations. Along the way, we will count how many
complex degrees of freedom (monomial coefficients) must be fixed in
order to tune a given model.\\

{\bf Case 2: $\gg_2$:} This is the generic case of an $(f,g)=(2,3)$
type singularity. To implement this tuning, then, all that is required
is that $g$ vanish to degree $3$, easily accomplished by removing the
single monomial in $g_2$. Hence, implementing this tuning
removes precisely one degree of freedom
\[ \Delta \hon =  [-1]
\]
as we had concluded earlier using anomaly calculations.\\

{\bf Case 3: $\gso(7)$: } We now encounter a more subtle issue of
counting. The monodromy conditions that distinguish the three gauge
algebras that can accompany a $(2,3)$ singularity are specified by the
factorization properties of the polynomial
\begin{eqnarray}
x^3+f_2 (w)x+g_3(w)   \,,
\end{eqnarray}
\begin{eqnarray}
x^3+Ax+B  \text{\ \ (generic)} & \Rightarrow & \gg_2 \nonumber \\
(x-A)(x^2+Ax+B) & \Rightarrow & \gso(7) \nonumber \\
(x-A)(x-B)(x+(A+B)) & \Rightarrow & \gso(8) 
\end{eqnarray}
The coefficients here are chosen in order to ensure that no quadratic
term appears in the total cubic polynomial. To obtain the second
condition ($\gso(7)$), we proceed by writing explicitly
\begin{eqnarray}
x^3+(f_{2,0}+f_{2,1}w+f_{2,2}w^2)x+(g_{3,0}+g_{3,1}w+g_{3,2}w^2+g_{3,3}w^3) \,.
\end{eqnarray}
This expression uses explicit knowledge of the
monomials. Recalling that the order in $w$ of a monomial $(a,b)$ in
$-4K$ is $b+4$, we may read off that the only monomials of $f_2$ are
$\{w^0,w^1,w^2\}$. Similarly, the only monomials available for $g_3$
are $\{w^0,w^1,w^2,w^3\}$. The seven coefficients above must
then be tuned to enforce the appropriate factorization. Expanding the
factorized version of the cubic (in $x$) polynomial, it is clear that
we must impose that the coefficient of $x$ be given by $B-A^2$ and
that of $x^0$ given by $-AB$. This can be minimally accomplished by
setting to zero the coefficients $c$ and $g$ above. 
More generally, $A$ and $B$ must be respectively linear and quadratic,
with 5 independent degrees of freedom.
This represents a
loss of two additional degrees of freedom (besides the first, which
represented tuning from $\gsu(3)$ to $\gg_2$). Hence
\begin{equation}
\Delta \hon =-3
\end{equation}
again in accordance with the anomaly results.\\

{\bf Case 4: $\gso(8)$:} Consulting the list above, to achieve $\gso(8)$, we must completely factorize the polynomial. Expanding yields the constraints
\begin{eqnarray}
a+bw+cw^2 & = & -A^2-AB  -B^2 \nonumber \\
d+ew+fw^2+gw^3 & = & AB (A + B)
\end{eqnarray}
This requires that now both $B$ and $A$ must be linear in $w$, so we
can for example simply set the $f$ coefficient to zero as well. This removes
an additional 1 degree of freedom (beyond the previously removed
three) leading to
\begin{equation}
\Delta \hon = -4
\end{equation}
as expected from anomaly results.\\

{\bf Case 5: $\gf_4$:} To tune to the $\gf_4$/$\ge_6$ case, we must
enhance the degrees of vanishing of $f$ and $g$ to $(3,4)$, which
requires that we eliminate \textit{all} $(a,b)\in -4K$ with $b\leq -2$
and $(c,d)\in -6K$ with $d\leq  -3$. 
The generic such tuning is an $\gf_4$ algebra.
Inspecting the monomial figure, we
find that from the initial (untuned) scenario, this requires us to
eliminate the leftmost column of $f$ (3 monomials) and the  leftmost
two columns of $g$ ($1+4$ monomials), so that in total
\begin{equation}
\Delta \hon= [-8]
\end{equation}
as expected from anomaly results.\\

{\bf Case 6: $\ge_6$:} In this case, the monodromy condition is
whether $g_4$ is a perfect square. Counting up from the left (as
before), the available monomials are $\{w^0,w^1,..., w^6\}$. 
We can make  this polynomial of degree 6 into a
square by restricting it to be of the form
\begin{equation}
g_4 (w)=(\alpha+\beta w+\delta w^2+\epsilon w^3)^2 \,,
\end{equation}
which clearly preserves $4$ of the $7$ original degrees of freedom.
This counting indicates that (as expected) we lose $3$ more degrees of freedom in
tuning from $\gf_4$ to $\ge_6$, for a total change of
\begin{equation}
\Delta \hon = -11
\end{equation}
from the original (untuned) $\gsu(3)$.
Note that this minimal tuning cannot be reached by simply setting to 0
three of the coefficients in $g_4$.

{\bf Case 7: $\ge_7$:} 
Enhancing finally to $\ge_7$ requires enhancing
the degrees of vanishing of $(f,g)$ to $(3,5)$. Up to this point, we
have already enhanced to $(3,4)$, so it remains only to eliminate the
remaining $7$ monomials of $g_4$, yielding a shift in $\hon$ of $-7$
in comparison to a tuning of $\gf_4$, or a shift of $-4$ subsequent to
tuning $\ge_6$, in accordance with anomaly calculations.
\vspace*{0.05in}

We thus have found that explicit computation of Weierstrass models
with monomials confirms that all tunings compatible with anomaly
cancellation on a $-3$ curve can be realized in F-theory, with the
proper number of degrees of freedom tuned in each case.

\subsection{Special case: tuning $\gso(N > 8)$}
\label{sec:8}

We have not so far explored the tuning of $\gso(N)$ with $N>8$, {\it
  i.e.} a Kodaira type $I_{m>0}^*$ singularity. Such gauge algebras
are non-generic in two ways: 1) they require a vanishing of $\Delta$
to order greater than the minimum
$\text{min}\{2\ \text{ord}(f),3\ \text{ord}(g)\}$, and 2) even and odd
$\gso(N)$ are distinguished by a subtle monodromy condition
\cite{Katz-Vafa}. Due to these complications, we have reserved the
treatment of these tunings to this section. The results to follow are
almost precisely in accord with Table \ref{t:isolated}, by applying
the rule that a tuning is not allowed when the formula for the
multiplicity of any of its representations becomes negative or
fractional (with fractions of $\frac{1}{2}$ allowed for real
representations). However, for clarity, we 
explicitly list the allowed tunings for $-2$, $-3$, and $-4$ curves in
 Table 
\ref{t:soN-matter}.

\begin{table}[h]
\begin{center}
\begin{tabular}{|c|c|c|c|c|}
\hline
Curve & $\gso(N)$ & $N_f$ & $N_s$ & $(\Delta h^{1,1},\Delta \hon)$ \\
\hline
-2 & 7 & 1 & 4 &  $(3,[-18])$ \\
& 8 & 2 & 4 & $(4,-20)$ \\
& 9 & 3 & 2 & $(4,[-23])$\\
& 10 & 4 & 2 & $(5,-27)$ \\
& 11 & 5 & 1 & $(5,[-32])$\\
& 12 & 6 & 1 & $(6,-38)$\\
& 13 & 7 & $\frac{1}{2}$ & $(6,[-45])$\\
\hline
-3 & 7 & 0 & 1 & $(1,-3)$ \\
& 8 & 1 & 2 & $(2,-4)$ \\
& 9 & 2 & 1 & $(2,[-6])$\\
& 10 & 3 & 1 & $(3,-9)$\\
& 11 & 4 & $\frac{1}{2}$ & $(3,[-13])$ \\
& 12 & 5 & $\frac{1}{2}$ & $(4,-18)$ \\
\hline
-4 & $N$ & $N-8$ & 0 & $(\lfloor (N-8)/2\rfloor, N\left(-\frac{N}{2}+\frac{15}{2}\right)+28)$ \\
\hline
\end{tabular}
\end{center}
\caption[x]{\footnotesize Allowed $\gso(N \geq 7)$ representations and associated matter. No $\gso(N)$ tuning is allowed on a curve of self-intersection $\leq -5$. All shifts are from the generic gauge algebras: $\emptyset$, $\gsu(3)$, and $\gso(8)$, respectively.}
\label{t:soN-matter}
\end{table}

The situations in which these complications 
arise on curves of negative self-intersection are quite limited. 
For instance, it is impossible for $\gso(N)$ to arise on a
curve of self-intersection $\leq -5$. This is a straightforward
consequence of the NHC classification, which dictates that such curves
must at least host algebras of $\gf_4$ or $\ge_{6,7,8}$. Because no $\gso(N)$
contains any exceptional algebra as a subalgebra, we can conclude that
these low self-intersection curves cannot be enhanced to any
$\gso(N)$. 
We focus here on
tunings only on individual curves of self-intersection $-4$, $-3$, and $-2$, excluding
tunings on clusters of more than one curve. 
These are the cases relevant for tunings on NHC's.
Indeed, it is easy to see,
upon consulting the analysis of $-2$ chains, that higher $\gso(N)$'s
are impossible on any chain of more than one $-2$ curve. Therefore
such tunings, if they occur, can do so only on isolated $-2$, $-3$, or
$-4$ curves. We proceed to analyze each case separately.
Tunings of $\gso(N)$ on individual curves of more general
self-intersection are treated in the following subsection.

As is our general strategy, we examine each potential tuning both
geometrically (in a local toric model) and from the standpoint of
anomaly cancellation. We again find that these methods agree.   
The structure of this section is
slightly different from previous ones: we first perform anomaly
calculations for curves of self-intersection $-4$, $-3$, and $-2$; and
then perform geometric calculations for these curves. We will see that
tunings are limited by the existence of spinor representations--they
become too large to satisfy anomaly cancellation. Geometrically, this
will manifest as a $(4,6)$ singularity at a codimension-2 ({\it i.e.}
dimension-0) locus in the base which corresponds to the location of
spinor matter. Let us now see how this unfolds explicitly.

\subsubsection{Spectrum and Hodge shifts from anomaly cancellation}
The anomaly calculation proceeds simply, but reveals an intricate
pattern of spinors depending upon the curve self-intersection,
matching {\it e.g.} \cite{Bershadsky-all}. First notice that the
adjoint representation has $C_{adj}=3$, and therefore the $C$
condition can be satisfied only for representations with $C \leq
3$. (There are no representations with negative $C$.) Given $\lambda =
2$ for $\gso(N)$, the $C$ condition takes the form
\begin{eqnarray}
\Sigma \cdot \Sigma = \frac{4}{3}\left( \sum_R C_R - 3 \right)
\end{eqnarray}
Indeed, since we consider only curves with $\Sigma \cdot \Sigma < 0$,
  the only representations that can cancel the $-3$  are those with $C
  < 3$. As discussed in \cite{finite}, the only such representations
    are the fundamental ($C=0$) and the spinor (for $N \leq
    13$). Given that the fundamental representation does not
    contribute at all to this condition, this equation uniquely fixes
    the number of spinor representations on a given curve. The results
    are summarized in table \ref{t:soN-matter}. It is important in
    implementing these conditions to recall that the $32$ and $64$
    dimensional spinor representations (of $\gso(11/12)$ and
    $\gso(13)$) are both self-conjugate. Therefore, half-hypermultiplets can
    transform in these representations. These hypers are counted with
    multiplicity $\frac{1}{2}$; if they were to be counted with
    multiplicity $1$, anomaly cancellation for $\gso(11/12)$ would be
    impossible on a $-3$ curve and anomaly cancellation for
    $\gso(13)$ would be impossible on a $-2$ curve. As an example,
    on a $-3$ curve, the $C$ condition reads $-\frac{9}{4}+3 =
    \frac{3}{4} = N_s C_s$. For $N = 9,10$, $C_s=\frac{3}{4}$, so
    there is one spinor rep. For $N=11,12$, $C_s=\frac{3}{2}$ and
    therefore a half-hyper spinor rep is required. Above this, we can
    see that $C_s = 3$, and even the smallest amount of spinor matter,
    a half-hyper, can no longer satisfy anomaly cancellation. As
    spinors were the only candidates to satisfy these conditions in
    the first place, we can decisively state that $\gso(N)$ tunings
    with $N>12$ are forbidden on $-3$ curves.  

As to the fundamental representations, their numbers can be determined
from the $A$ condition
\begin{eqnarray}
K \cdot \Sigma = \frac{1}{3} \left(\sum_R A_R-A_{adj} \right) 
\end{eqnarray}
Using $A_s=2^{\lfloor (N+1)/2 \rfloor - 4}$, $A_{adj}=N-2$, and
$A_f=1$, we can easily solve for $N_f$. For instance, on a $-3$ curve,
we have already determined that there is 1 spinor hyper  for
$\gso(9/10)$ and $\frac{1}{2}$ for $\gso(11/12)$. Since the left hand
side of the condition is $K \cdot \Sigma = -1$, we obtain
$-3=N_f+N_sA_s-(N-2)$. As an example, for $\gso(9)$ on a $-3$ curve,
this gives $N_f=7-2-3=2$. For $\gso(10)$, the only change is that
$A_{adj}=10-2=8$ increases by one, hence $N_f=3$. Similarly, for
$N=11,12$, although only a half-hyper transforms in the spin
representation, the coefficient $A_s$ doubles in comparison to the
previous cases. Thus again, the only numerical change is that
$A_{adj}$ increases by one as $N$ increases; $N_f=5,6$ for
$\gso(11/12)$. This one example of tunings on a $-3$ curve illustrates
the general pattern of matter representations on all three curves
considered; further calculations are entirely analogous and are
therefore omitted. 

One final remark is in order: we found that $\gso(13)$ was the largest
$\gso(N)$ that anomalies allow on a $-2$ curve and $\gso(12)$ was the
highest allowed on a $-3$ curve. One might expect this pattern to
continue, with {\it e.g.} $\gso(10)$ the highest possible tuning on a
$-4$ curve. But, at this very lowest self-intersection curve before
such tunings become impossible, they regain renewed vigor: all
$\gso(N)$'s are tunable on a $-4$ curve. It is straightforward to
verify that the matter in table \ref{t:soN-matter} for a $-4$ curve
satisfies anomaly cancellation. The reason there is no restriction is
simply that it was the spin representations that led to a problem
before, and on a $-4$ curve, there are no spin represenations: all
matter is in the fundamentals.

\subsubsection{Spectrum and Hodge shifts from local geometry}

Now we check the predictions from anomaly cancellation by
constructing, where possible, local models for the allowed $\gso(N)$
tunings and showing how the disallowed tunings fail. These local
calculations are subtler than others we have so far
encountered. Namely, we must impose a non-generic vanishing of
$\text{ord}(\Delta) > \text{min}\{ 3\ \text{ord}(f), 2\ \text{ord}(g)
\}$. Moreover, there are two distinct monodromy conditions
distinguishing $\gso(8+2m)$ from $\gso(7+2m)$ in the $I_m^*$ Kodaira
case, one condition each for $n$ even and odd. These conditions are
clearly stated in \cite{Grassi-Morrison-2, kmss-Tate}. For our
purposes, instead of using these results directly, we note that the
monodromy conditions can be summarized succinctly as follows. To be in
the generic Kodaira case $I_m^*$, {\it i.e.} $\gso(7+2m)$, requires that
$\Delta$ vanish to order $6+m$. However $\Delta_{6+m}$ must not
vanish; otherwise we would be in the next highest Kodaira case. All
monodromy conditions for $I_m^*$ can be summarized as the requirement
that $\Delta_{6+m}$ be a perfect square. When this is the case, the
resulting gauge algebra is enhanced $\gso(7+2m)\longrightarrow
\gso(8+2m)$.

For our local models, we use the fan $\{ (0,1), (1,0), (n,-1)\}$, where $n$ is (negative) the self-intersection number of the middle curve $\Sigma$ represented by the vector $(1,0)$ and assumes the values $2,3,4$. Because a $-3$ curve is the simplest example that captures the complexity of these tunings, we begin with it, moving then to $-2$ and finally $-4$ curves.

On a $-3$ curve, we will be able to tune up to but not including an $\gso(13)$. Let us see how this is possible. We have already seen how an $\gso(8)$ may be tuned, so let us investigate the first new case: $\gso(9)$, {\it i.e.} the generic $I_1^*$  singularity. To implement this tuning, we simply require
\begin{eqnarray}
\Delta_6 \propto f_2^3+g_3^2 = 0
\end{eqnarray}
In the above we suppress the coeffients of the separate terms (4 and
27, respectively) as they will play no role in determining whether
this quantity can be set to zero and (if this is possible) how many
degrees of freedom must be fixed to do so. In implementing this
condition, we must keep in mind that the orders of $f$ and $g$ must
remain $2$ and $3$, respectively. It is clear that such a tuning (on a
smooth divisor) will
be possible iff $f_2$ is a perfect square and $g_3$ is a perfect
cube. Indeed, expanding in a local defining coordinate $w$ for the
curve represented by $(0,1)$, we see that $f_n\sim w^{3n-4}$ and
$g_n\sim w^{3n-3}$, which we will use repeatedly. In particular,
$f_2 \sim w^2$ and $g_3 \sim w^3$,\footnote{The notation $\sim w^n$
  will be used throughout this section to denote a polynomial of
  degree $n$ in $w$ with arbitrary coefficients.} whence this
condition can be satisfied if $f_2 \propto \phi^2$, $g_3 \propto
\phi^3$ for an arbitrary linear term $\phi = a+bw$. Note that for
cancellation between these terms, the two coefficients $a$ and $b$ are
arbitrary but fixed between $f$ and $g$, and moreoever, once an
overall coefficient for $\phi^2$ is chosen for $f_2$, this fixes the
overall coefficient of $\phi^3$. Hence there are $2$ degrees of
freedom remaining, whereas we started with $3+4=7$ in arbitrary
quadratic and cubic polynomials. Therefore, we lose $5$ degrees of
freedom in this tuning, in precise accordance with the change $\Delta
\hon$ predicted from the matter determined by anomaly
cancellation. (NB: This change is counted from $\gg_2$, 
the generic $(2,3)$ singularity.) Comparing with the known change of
$\Delta \hon=-3$ in tuning $\gso(8)$ from $\gg_2$, we see that this
tuning represents a loss of $2$ additional degrees of freedom. (Or,
from the generic non-Higgsable algebra $\gsu(3)$, we have $\Delta
\hon=-6$). To reassure ourselves that this indeed works, we recall the
anomaly calculation: $\gsu(3)$ has no hypers, only a vector in the
adjoint, so its contribution to $\hon$ (in the terms $V-H_{\rm charged})$
is $+8$. For $\gso(9)$, we have both vectors and hypermultiplets, so
the contribution to $V-H_{\rm charged}$ is $+36 - 1 \cdot 16 - 2 \cdot 9 =
+2$, a loss of $6$ degrees of freedom.

To tune an $\gso(10)$ requires implementing a mondromy condition: $g_4+\frac{1}{3} \phi f_3$ must be a perfect square \cite{kmss-Tate}. This is indeed possible, and since $g_4 \sim w^6$, we lose $3$ degrees of freedom in requiring that it take the form $(\sim w^3)^2$. Again, we have a match with anomaly cancellation.

To tune an $\gso(11)$ we now require $\Delta_7 = 0$, namely
\begin{eqnarray}
0 & = & \Delta_7 \propto f_3f_2^2+g_3g_4 \nonumber \\
& = & f_3\phi^4 + \phi^3 g_4
\end{eqnarray}
Now this will be possible with $g_4 \propto \phi f_3$, which consumes all the degrees of freedom of $g_4$, {\it i.e.} all $7$ (or the $4$ that remain after tuning an $\gso(10)$); again, we have agreement with anomaly cancellation calculations.

To tune an $\gso(12)$, we now impose the more complicated monodromy condition that
\begin{eqnarray}
\mu = 4\phi (g_5+\frac{1}{3}\phi f_4) - f_3^2
\label{eq:12-monodromy}
\end{eqnarray} 
be a perfect square. This is also possible. We have  $10$ degrees
of freedom in $g_5 \sim w^9$ and $6$ in $f_3 \sim w^5$. By requiring
$g_5 \propto \phi  \tilde{g}_5,  \tilde{g}_5
 \sim w^8$ and $f_3 \propto \phi \tilde{f}_3, \tilde{f}_3 \sim w^4$, we
can factor out $\phi^2$ from $\mu$; we can then tune
$\tilde{g}_5$ so that $\mu/\phi^2$ is a perfect square, so that we
satisfy
the monodromy condition (\ref{eq:12-monodromy}). This fixes $1+4=5$ degrees of freedom,
consistent with anomaly cancellation.

What goes wrong at the crucial case $\gso(13)$? To implement this tuning, we must set to zero $\Delta_8$:
\begin{eqnarray}
0 = \Delta_8 & \propto & f_3^2f_2+f_2^2f_4+g_3g_5+g_4^2 \nonumber \\
& = & f_3^2 \phi^2 + \phi^4 f_4 + \phi^3 g_5
\end{eqnarray}
We combined the first and last terms upon recalling that $g_4
 \propto
\phi
f_3$. For this quantity to be zero, $\phi$ must divide $f_3^2$, which
implies in this case that $\phi$ divides $f_3$ because $\phi$ cannot
be a perfect square. This leads to an unacceptable singularity on the
curve $C_{\phi} = \{ \phi = 0 \}$! This arises because, now, $\phi^2 |f_2$ and
$\phi |f_3$, so that $f$ vanishes to order
$4$ on  $C_{\phi}$. Meanwhile, we have already found that $\phi^3
|g_3$, 
and now
$ \phi^2 |g_4$ and $\phi |g_5$. This
leads directly to a $(4,6)$ singularity on $C_{\phi}$.

On a $-2$ curve, the story is similar but slightly more complicated:
we can tune up to but not including $\gso(14)$, in agreement with
anomaly cancellation.\footnote{Thanks to Nikhil Raghuram and Yuji
  Tachikawa for illuminating discussions on this point.}
Geometrically, we will see that the $\gso(14)$ tuning fails for
reasons similar to but slightly different from the failures we have
previously encountered. To begin investigating these tunings, note
that on this geometry, $f_n$, $g_n\propto w^{2n}$. On an isolated $-2$
curve, we have $f_2 \sim w^4$, $g_3 \sim w^6$, which implies that an
$I_1^*$ can be tuned by taking $\phi = a+bw+cw^2$ to be an arbitrary
quadratic. In the process, $5+7-3=9$ degrees of freedom are
fixed. This is to be interpreted as a change from $\gg_2$, the generic
$(2,3)$ gauge algebra; or since $\gso(8)$ can be tuned from $\gg_2$ by
fixing $6$ degrees of freedom, this is a change of $3$ additional
degrees of freedom from $\gso(8)$, consistent with anomaly
cancellation calculations. (Recall that anomaly cancellation predicts
$\Delta \hon = -14$, $-20$, $-23$ for $\gg_2$, $\gso(8)$, and
$\gso(9)$, respectively.) We can continue the analysis exactly as
before, and no subtleties arise in counting degrees of freedom. Let us
jump, then, to the case of $\gso(13)$, the generic $I_3^*$
singularity. Again we must require
\begin{eqnarray}
0 = \Delta_8 & \propto & f_3^2 \phi^2 + \phi^4 f_4 + \phi^3 g_5
\end{eqnarray}
but in this case, $\phi$ is quadratic and can therefore be chosen to
be a perfect square. Hence we can satisfy the required condition $\phi
| f_3^2$ while also maintaining $\phi \nmid f_3$. Let us denote
$\psi^2 = \phi$. Making this choice, namely fixing a quadratic to be
the square of a linear function eliminates one degree of freedom. Also,
factoring $f_3=\psi \bar{f}_3$, for generic degree $5$ $\bar{f}_3$, we
lose one more degree of freedom. Finally, setting
$g_5=\bar{f}_3^2+\phi f_4$ fixes all $11$ degrees of freedom of $g_5
\sim w^{10}$. This completes the required cancellation, fixing
$1+1+11=13$ degrees of freedom in the process. This matches with the
shift from $\gso(11)$ expected from anomaly cancellation.

It is not possible to tune $\gso(14)$. The appropriate monodromy
condition is to require that $\Delta_9$ be a perfect square.
\begin{eqnarray}
\Delta_9 & = & \psi^3\left(-\frac{1}{2}\bar{f}_3^3-18\psi^2 \bar{f}_3f_4+108\psi^3\left( g_6 - f_5 \psi^2 \right) \right)  
\end{eqnarray}
As the $\bar{f}^3_3$ term has lowest order (in $\psi$), it cannot be
cancelled by any other term unless $\psi | \bar{f}_3^3$, hence $\psi |
\bar{f}_3$. But now $\psi | g_5 = \bar{f}_3^2+\psi^2 f_4$, and one can
check that our previous constraints have likewise made all lower order
terms in $f$ and $g$ divisible by $\psi$
to sufficiently high order that a $(4,6)$ singularity
at $\{ \phi = 0\}$ is  inevitable, and we conclude that $\gso(14)$
cannot be tuned on a $-2$ curve. 
This conclusion was also reached in \cite{global-symmetries} based on
the analysis of \cite{Grassi-Morrison-2}, and matches the expectation
from anomaly cancellation.

On a $-4$ curve, the discussion is completely analogous, save for one
crucial difference: $f_2, \ g_3\sim z^0$ are constants, therefore
$\phi$ is a constant. It is always true that a constant divides higher
order terms in $\Delta$, and this condition therefore places no
restrictions on the tunings. On an isolated $-4$ curve, there is no
apparent restriction on tuning $\gso(N)$'s. In SUGRA models, of
course, large $N$ will eventually cease to be tunable because
$h^{2,1}$ is finite and there will not be sufficiently many complex
degrees of freedom to implement the tuning. Such failures result from
global properties of the base. From an anomaly cancellation
standpoint, this failure eventually results from an inability to
satisfy gravitational anomaly cancellation. 
As we discuss in \S\ref{sec:generalized-e8}, a local bound on $N$ can
also be imposed by other curves of non-positive self intersection that
intersect a $-4$ curve supporting an $\gso(N)$, though such bounds are
not fully understood in the low-energy theory.
In SCFT's, there is no
reason to expect that the series of $\gso(N)$ tunings will ever terminate
at any $N$.

\subsection{Tuning on rational curves of self-intersection $n \geq -2$}
\label{sec:general-curves}

In this section we consider the possible tunings on an isolated curve
of self-intersection $n \geq -2$.  For such curves, it is
straightforward to use anomaly analysis along the lines of the
preceding section to confirm that in general the generic matter
spectrum is that given by Table~\ref{t:isolated} for each of the
groups listed.  Since in some cases the set of possible tunings is
unbounded given only local constraints, a case-by-case analysis is
impossible.  Fortunately, beginning at $n = -2$, we can systematically
organize the computation easily as a function of $n$.  For the
exceptional Lie algebras ($\ge, \gf, \gg$) we can check that the
tunings are possible using Weierstrass, and explicitly check the Hodge
numbers.  For the classical Lie algebras ($\gsp, \gsu$) we use Tate
form.
For $\gso(N)$ the analysis closely parallels that of the previous
section, and we compare the Tate and Weierstrass perspectives.

\subsubsection{Tuning exceptional algebras on $n \geq -2$
curves}

For simplicity we begin with $-1$ curves, then generalize.  From the
local toric analysis, we have an expansion of $f, g$ in polynomials
$f_k, g_k$ in a local coordinate $w$, where deg ($f_k$) = $4 + k$,
deg ($g_k$) = $6 + k$.  The number of independent monomials in $f_0,
f_1, \ldots$ and $g_0, g_1, \ldots$ are thus $5, 6, \ldots$ and $7, 8,
\ldots$ respectively.

To tune an $\ge_7$ on the $-1$ curve, we must set $f_0 = f_1 =f_2 =
g_0 = \cdots g_4 = 0$.  This can clearly be done by setting $63$
independent monomials to vanish.  Thus, we can tune the $\ge_7$ and we
confirm the Hodge number shift in Table~\ref{t:isolated}.  A similar
computation allows us to tune $\gf_4$ by the same tuning but leaving
$g_4$ generic (11 monomials), giving the correct Hodge shift of $52$.
For $\ge_6$ we have the monodromy condition that $g_4$ is a perfect
square, so we get the correct shift by $57$.  Finally, for $\gg_2$ we
leave $f_2, g_3$ generic ($7 + 10 = 17$ monomials), for a correct
Hodge shift of $35$.  We have the usual caveat regarding the $\gf_4$
and $\gg_2$ and rank-preserving breaking.

We can generalize this analysis by noting that on a curve of
self-intersection $n \geq -2$, the local expansion gives
\begin{equation}
{\rm deg} (f_k) = 8 + n (4-k), \;\;\;\;\;
{\rm deg} (g_k) = 12 + n (6-k) \,.
\label{eq:fg-degrees}
\end{equation}
Tuning any of the exceptional algebras on any such curve then is
possible, since the degree 
is nonnegative for $k \leq 4, 6$ for $f, g$ respectively.  The number
of Weierstrass monomials that must be tuned can easily be computed in
each case and checked to match with Table~\ref{t:isolated}.
In this comparison it is important to recall from (\ref{eq:h-neutral})
that each $-2$ curve contributes an additional neutral scalar field,
while each curve of self-intersection $n \geq 0$ contributes $n + 1$
to the number of automorphisms, effectively removing Weierstrass
moduli from the neutral scalar count.  For example, for
$\ge_7$ the number of monomials tuned is
\begin{equation}
((9 + 4n) + \cdots + (9 + 2n)) + ((13 + 6n) + \cdots (13 + 2n))
=  92 + 29n \,.
\end{equation}
The number of neutral scalars removed by this tuning is then
\begin{equation}
92 + 29n-(n +1) = 91+28n \,,
\end{equation}
in perfect agreement with the last line of Table~\ref{t:isolated}.
The shifts for the other exceptional groups can similarly be computed
to match the anomaly prediction.

\subsubsection{Tuning $\gsu(N)$ and $\gsp(N)$
on $n \geq -2$
curves}
\label{sec:classical-tuning}

We now consider the classical Lie algebras, beginning with $\gsu(N)$.
These tunings are more subtle for several reasons.  First, the tunings
involve a cancellation in $\Delta$ that is not automatically imposed
by vanishing of lower order terms in $f, g$ so the computation of such
tunings is more algebraically involved.  Second, these tunings can
involve terms of arbitrarily high order in $\Delta, f, g$, and can be
cut off when higher order terms do not exist in $f, g$, even in a
purely local analysis.

To illustrate the first issue consider the tuning of $\gsu(2)$ on a
$-1$ curve.  The analysis of the general $\gsu(N)$ tuning through
Weierstrass was considered in \cite{mt-singularities}.  For $\gsu(2)$,
this tuning involves setting $f_0 = -3 \phi^2, g_0 = 2 \phi^3$ to
guarantee the cancellation of $\Delta_0$, and then solving the
condition $\Delta_1 = 0$ for $g_1$.  On a $-1$ curve this amounts to
replacing the $5 +7 + 8 = 20$ monomials in $f_0, f_1, g_1$ with $3$
monomials in a quadratic $\phi$.  The shift in $\hon$ is therefore
by 17, in agreement with Table~\ref{t:isolated}.  As $N$ increases,
the explicit tuning of the Weierstrass model in this way becomes
increasingly complicated.  For $N \geq 6$, there are multiple
branches, including those with non-generic matter contents, even for
smooth curves; we return to this in \S\ref{sec:matter}.  A systematic
procedure for tuning $\gsu(N)$ for arbitrary $N$ through explicit
algebraic manipulations of the Weierstrass model is not known.  Thus,
in these cases rather than attempting to explicitly compute the
Weierstrass model to all orders we simply use the Tate approach
described in \S\ref{sec:Tate}.

Already from Table~\ref{t:isolated}, we can see that as $n$ increases,
the bound of allowed values on $N$ so that the number of fundamental
representations is nonnegative decreases.  We then wish to determine
which values of $N$ can be realized
using the Tate description and compare with this bound from
anomalies.  For $n = -2, -1, 0$, there is no bound on $N$ from
anomalies.  To analyze the Tate forms, we determine the degrees of the
coefficients in an expansion
$a_i = \sum_{k} a_{i (k)} z^k$ to be, in parallel to (\ref{eq:fg-degrees}),
\begin{equation}
{\rm deg} (a_{i (k)}) = 2i + n (i-k) \,.
\label{eq:a-degrees}
\end{equation}
To tune $\gsu(N)$, we must tune in Tate form $a_2$ to vanish to order
1, $a_3$ to vanish to order $\lfloor n/2\rfloor$, etc..  This can
clearly be done for any $N$ on $n = -2, -1, 0$ curves, so there is no
problem with tuning any of these groups, consistent with the absence
of a constraint from anomalies.

Now, however, consider for example a curve of self-intersection $+
1$.  From anomalies we see that the number of fundamental matter
representations is $16 + (8-N) n = 24 -N$, which becomes negative for
$N > 24$.  So we want to check which values of $N \leq$ 24 admit a
Tate tuning of $\gsu(N)$.  For $n = 1$, the maximum degree possible of
the $a_i$'s is
\begin{equation}
{\rm deg} (a_1, a_2, a_3, a_4, a_6) = (3, 6, 9, 12, 18).
\end{equation}
For each $a_i$, this is the largest value of $k$ such that
(\ref{eq:a-degrees}) is nonnegative.  To tune a Tate $\gsu(24)$, we
need the $a$'s to vanish to orders $(0, 1, 12, 12,  24)$.  This can be
achieved by setting $a_3 = a_6 = 0$ and leaving arbitrary the largest
terms in $a_{4}$ ({\it i.e.}, $a_{4 (12)}$).  So we can tune through
Tate an $\gsu(24)$.  
Tuning a higher $\gsu(N)$ would require the vanishing also of $a_4$ to
all orders, which would produce a singular Weierstrass model with
$\Delta = 0$ everywhere, consistent with the anomaly constraint.
This is not the end of the story, however.  To
tune a Tate $\gsu(23)$ requires the $a$'s to vanish to orders $(0, 1,
11, 12, 23)$.  But since there is no order 11 term in $a_3$ or order
23 term in $a_6$, this drives the Tate model automatically to
$\gsu(24)$.  Thus, there is no Tate tuning of $\gsu(23)$ on a $+1$
curve.  Similarly, there is no Tate tuning of $\gsu(21)$, although $\gsu(22)$
may be tuned; and $\gsu(19)$ can also be tuned without obstruction.  Precisely this same
pattern was encountered in \cite{mt-singularities} when an attempt was
made to tune these groups
 directly in the Weierstrass model over a $+1$ curve, although in that
 context a particular simplification was made and there was no
 complete proof that there was no more complicated construction of
 these algebras.  The upshot, however, is that on a curve of
self-intersection $+ 1$, there is a slight discrepancy between the
anomaly constraints and what we have been able to explicitly tune
through Tate or Weierstrass.  We have an almost exact agreement, but
the gauge algebras $\gsu(21)$ and $\gsu(23)$ lie in the ``swampland''
of models that seem consistent from low-energy conditions but cannot
at this time be realized in any known version of string theory.

We can perform a similar analysis for the $\gsu(N)$ groups on other
curves of positive self intersection;
the results of this analysis  are tabulated in Table~\ref{t:su}.
Several other curve types have similarly missing $\gsu(N)$ groups in
the Tate analysis.  For $+ 2, + 3$ curves it is impossible to tune
$\gsu(15), \gsu(13)$ respectively using the Tate form, an explicit
attempt to construct Weierstrass models showed a similar obstruction
(with some simplifying assumptions made) in \cite{KMT}.  It is
interesting to note that in most of the cases where the Tate analysis does not
provide an $\gsu(N -1)$ but does allow for tuning an algebra
$\gsu(N)$, the $\gsu(N)$ theory always has either zero or one
hypermultiplets in the fundamental ${\bf N}$ representation, so that
there is no direct Higgsing to $\gsu(N -1)$.  
One might think that in the two cases ($n = 1, N = 22$ and $n = 7, N =
10$) there should be two fundamentals, so that the
theory might be Higgsable to the missing model.  
In the $\gsu(22)$ case, for example however, this tuning also
forces\footnote{Thanks to Nikhil Raghuram for pointing this out.}
an additional $\gsu(2)$ to arise in some cases on a curve that
intersects $\Sigma$; this would absorb the two fundamentals into a
single bifundamental, so that there may generally be no direct Higgsing to $\gsu(21)$.
It may also be relevant
that in the explicit tuning of $\gsu(24)$ on a $+ 1$ curve in
\cite{mt-singularities}, the resulting gauge group had a global
discrete quotient, so that the precise gauge group is SU($24$)$/\Z_2$,
not SU($24$).
In any case, it seems likely that there is no Weierstrass model
corresponding to these configurations that cannot be constructed using
Tate, and this would give a self-consistent picture with the other
results in this paper, but we do not have a complete proof of that statement.

\begin{table}
\begin{center}
\begin{tabular}{| c | c | c |c | c |}
\hline
$n$ &  anomaly bound on N & Tate realizations & swamp\\
\hline
$ + 1 $ & 24 & $\ldots, 20, 22, 24$ & 21, 23\\
$+ 2 $ & 16 & $\ldots, 14, 16$& 15\\
$+ 3 $ &  13 & $\ldots, 12$& 13\\
$+ 4 $ &  12 & $\ldots, 10, 12$& 11\\
$+  5 $ &   11 & $\ldots, 10$& 11\\
$+ 6 $ &    10 & $\ldots, 10$& $\cdot$\\
$+ 7, + 8 $ &    10 & $\ldots, 8, 10$& 9\\
$+ 9, \ldots, + 16 $ &     9 & $\ldots, 8$& 9\\
$> + 16$ & 8 & $\ldots, 8$ & $\cdot$\\
\hline
\end{tabular}
\end{center}
\caption[x]{\footnotesize Table of $\gsu(N)$ algebras that can be
  realized on a curve of positive self-intersection $+ n$ using Tate,
  compared to bounds from anomaly cancellation.  The last column are
  cases in the ``swampland''.}
\label{t:su}
\end{table}

For tunings of $\gsp(N)$, the story is similar but simpler.  Anomaly
cancellation shows that $\gsp(N)$ can only be tuned on curves of
self-intersection $n \geq -1$.  From the Kodaira conditions it is
immediately clear that $\gsp(N)$ cannot be tuned on a curve of
self-intersection $-3$ or below.  For a curve of self-intersection
$-2$, the monodromy condition that distinguishes $\gsp(N)$ from
$\gsu(N)$ automatically produces an $\gsu(N)$ group, since the
condition is that $f_0 = \phi^2$ where $\phi$ itself is a perfect
square, and since $f_0$ is a constant on a $-2$ curve, it is always a
perfect square.
Just as for $\gsu(N)$ we can use Tate to determine when $\gsp(N)$ can
be tuned on a given curve of self-intersection $n \geq -1$.  In this
case there are no inconsistencies between anomaly conditions and the
tuning possibilities; the swampland in this case is empty, and all
possibilities in Table~\ref{t:isolated} that have nonnegative matter
content are allowed.

\subsubsection{Tuning $\gso(N)$ on 
 $n \geq -2$ curves}

Finally, we consider $\gso(N)$ on curves of self-intersection $n \geq
-2$.  Complementing the analysis of \S\ref{sec:8}, we see what the
Tate analysis has to say about these cases.  It is straightforward to
check that there is no problem with tuning up to $\gsu(12)$ using Tate
for a local analysis around any curve of self-intersection $n \geq
-2$.  We simply cancel according to the rules in Table~\ref{t:Tate} and
we get Weierstrass models that provide the desired group.  The Tate
procedure breaks down, however, at $\gso(13)$.  To tune this algebra
the $a$'s must be taken to vanish to orders $(1, 1, 3, 4, 6)$.  Taking
the Tate form
\begin{equation}
y^2 + z \tilde{a}_1 yx + z^3\tilde{a}_3y
 = x^3 + z\tilde{a}_2x^2 +  z^4\tilde{a}_4x + z^6 \tilde{a}_6 \,,
\end{equation}
and converting to Weierstrass form we find that $\tilde{a}_2$ divides
all coefficients in $f$ and $g$ up to $f_4, g_6$.  This is the $\phi$
that played a key role in the analysis of \S\ref{sec:8}.  Unless $\phi
= \tilde{a}_2$ is a constant, the Tate tuning of $\gso(13)$ and beyond
gives a problematic Weierstrass model.  For $-4$ curves alone, $\phi$
is a constant, so the Tate form breaks down for all other curves.
Note that one might try to set $\phi$ to a constant, even though it
has monomials of higher order.  This leads to a problem at the
coordinate value $w = \infty$ on the curve where the group is tuned.

Analysis of the anomaly equations and the properties of the $\gso(N)$
spinor representations as discussed above indicates that the anomaly
conditions are satisfied for $\gso(13)$  on a curve of
self-intersection $n$ if and only if $n$ is even.  While the Tate
analysis is problematic in these cases, the Weierstrass analysis of
\S\ref{sec:8} easily generalizes to arbitrary $n \geq -4$.  As long as
the degree of $\phi$ is even, which occurs when the degree of $f_2$ is
a multiple of four, we can decompose $\phi = \psi^2$ and find a
Weierstrass solution for $\gso(13)$.  This occurs precisely when $n$
is even, so the Weierstrass analysis shows that all $\gso(N)$ gauge
groups with $N \leq 13$ allowed by anomaly cancellation can be tuned
on a single smooth rational curve in a local analysis.  In the same
way that $\gso(14)$ develops a (4, 6) singularity on a $-2$ curve as
described in \S\ref{sec:8}, the same occurs on any curve of
self-intersection $-2 + 4m, m > 0$.  

The only remaining situation is $\gso(N)$ on an isolated $-4$
curve where $N > 13$. In this case, there is no constraint from anomalies as the
number of fundamental matter fields is $N -8$
and there are no spinors.  Similarly, there is no
constraint from Tate for any $N$.  So in this case, everything
allowed from anomalies can be tuned in F-theory.

This completes our analysis of tunings of all gauge groups on rational
curves on the base using only constraints from the local geometry.

\subsection{Higher genus curves}
\label{sec:higher-genus}

In the discussion in this section so far we have focused on curves of
genus 0.  For tuning toric curves on toric bases, or for 6D SCFT's,
this is all that is necessary.  For tuning more general curves on
either toric or non-toric bases for general F-theory supergravity
models, however, we must consider tuning gauge groups on curves of
higher genus.  For example, we could tune a gauge group on a cubic on
the base $\P^2$; such a curve has genus one.

For a smooth curve of genus $g$, the matter content includes $g$
matter hypermultiplets in the adjoint representation of the group, and
the rest of the matter content is determined accordingly from the
anomaly cancellation condition.  The generic
matter content and Hodge number shifts for tunings over a curve of
general genus $g$ are given in Table~\ref{t:tunings-g}.  Unlike the
genus 0 cases, where we have performed explicit local analyses in each
case  (except those of large $N$ for the classical groups), in this Table we have simply given the results expected from
anomaly cancellation.  
In each case, the matter content is uniquely determined from the
anomaly cancellation conditions (\ref{eq:hv}--\ref{eq:bb-condition})
with $\Sigma\cdot \Sigma= n$ and $(K + \Sigma)\cdot \Sigma= 2g-2$,
given the constraint that only the adjoint and generic matter types
({\it {\it e.g.}} the fundamental and two-index antisymmetric
representations for $\gsu (N)$) arise.
\begin{table}[h]
\begin{center}
\begin{tabular}{|c|c|c|}
\hline
$\gg$ & matter & $\Delta(h^{1,1},\hon)$ \\
\hline
$\gsu (2)$ &$ (6n + 16-16g) {\bf 2} + (g) {\bf 3} $ & $(1, -12n-29 (1-g)) $\\
$\gsu (3)$ &$ (6n + 18-18g) {\bf 3} + (g) {\bf 8} $ & $(2, -18n-46 (1-g)) $\\
$\gsu(N)$ & $((8-N) n + 16 (1-g)){\bf N}$
\hspace*{0.7in} & $(N-1, 
-(\frac{15 N
  -N^2}{2}) n-(15 N + 1) (1-g)) $
\\
&\hspace*{0.1in}
$+(n+2 -2g)\frac{{\bf N}({\bf
    N-1})}{{\bf 2}} + (g) ({\bf N^2 -1})$ &\hspace*{0.1in}
\\
$\gsp(N/2)$ & $((8-N)n + 16 (1-g)){\bf N}  $
\hspace*{0.9in}
 & $\left(N/2,
\left[-(\frac{15 N
  -N^2 -2}{2}) n-
\left( \frac{
31 N -N^2}{2} \right)  (1-g)
\right]\right)$ \\
& \hspace*{0.1in}
$+ (n+1 -g)\left(\frac{{\bf N}({\bf N}-1)}{2} -1 \right)+ 
(g) {\bf \frac{N  (N + 1)}{2} } $ &\\
$\gso(N)$ & $(n + (N -4) (1-g)){\bf N}
$\hspace*{0.9in}  & $\left( \lfloor N/2\rfloor,-
(N+ 16)n -(\frac{N^2 -7 N + 128}{2})(1-g)\right)$\\
& \hspace*{0.1in}
$+(n+4 -4g)2^{5-\lfloor (N+1)/2\rfloor} {\bf S} 
+ (g){\bf \frac{N (N -1)}{2} }$&\\
$\gg_2$ & $(3n+10-10g){\bf 7}
+ (g) {\bf 14}$ & $(2,[-7(3n+8 -8g)])$\\
$\gf_4$ & $(n+5-5g){\bf 26} + (g) {\bf 52}$ & $(4,[-26(n+3 - 3g)])$\\
$\ge_6$ & $(n+6-6g){\bf 27}  +
(g) {\bf 78}$ 
& $(6,-27n-84 (1-g))$\\
$\ge_7$ & 
$(4-4g+n/2) {\bf 56} + (g) {\bf 133}$ & $(7,-28 n -91 (1-g))$\\
\hline
\end{tabular}
\end{center}
\caption[x]{\footnotesize  Possible tunings on a curve
  $\Sigma$ of
genus $g$ and self-intersection $n$, together with matter and shifts
  in Hodge numbers. 
Note that $\gsu (2)$ and $\gsu (3)$
are listed separately; the antisymmetric in  the
case of $\gsu (2)$ is a singlet, and does not contribute to the Hodge number shift,
so this case differs slightly from the general $\gsu (N)$ formula.
The $\gsu (3)$ case, which also lacks a quartic Casimir, is also
listed explicitly for convenience.
} 
\label{t:tunings-g}
\end{table}

\section{Classification II: multiple-curve clusters}
\label{sec:C-II}

It is  useful to break our analysis of allowed tuned gauge
symmetries into tunings on isolated curves and on multiple-curve
clusters. First, these multiple-curve NHCs already arise in regular
patterns in bases; and second, the possible tunings on multiple-curve
NHCs are very tightly constrained and thus represent a very small
subset of combinations of {\it a priori} allowed tunings on each curve
within these clusters. For clusters containing a $-3$ curve, some
possibilities that satisfy anomaly cancellation (and all rules to be
discussed later) are not possible; these therefore deserve special
attention and cannot be treated except as individual cases. For $-2$
chains, on the other hand, a very distinctive ``critical'' structure
appears, which is also best highlighted by examining this class
individually.

\subsection{The clusters $(-3,-2,-2)$, $(-2,-3,-2)$, and $(-3,-2)$}
\label{sec:tuning-clusters}

Multiple-curve NHCs containing a $-3$ curve present examples of an interesting
phenomenon: although the calculations proceed similarly to the above
(and can be found in appendix \ref{sec:appendix}), we will pause to
highlight this phenomenon. Simple anomaly cancellation and
geometry-based arguments both immediately show that the NHC
$\gg_2\oplus \gsu(2)$ cannot be enhanced to more than
$\gso(8)\oplus\gsu(2)$. From the geometry point of view, this
restriction arises because the next Kodaira singularity type 
beyond $\gso(8)$
is
$\gf_4$, which would lead to a $(4,6)$ singularity at the intersection
between the $-2$ and $-3$ curves. (A more detailed analysis shows that
an attempt to enhance $\gsu(2)$ to $\gsu(3)$ by tuning monodromy will
also force a $(4,6)$ singularity.) This leaves two possible
enhancements, both of which satisfy anomaly cancellation:
$\gso(7)\oplus \gg_2$ and $\gso(8)\oplus\gsu(2)$. In no case can
$\gso(8)$, however, be realized. The allowed tunings are presented in
table \ref{t:multiple}.

\begin{table}
\begin{center}
\begin{tabular}{|c|c|c|c|}
\hline
cluster & $\gg$ & $(\Delta h^{1,1},\Delta \hon)$ & matter \\
\hline
(-3,-2) & $\gg_2\oplus\gsu(2)$ & (0,0) & $({\bf 7},\frac{1}{2}{\bf 2})+\frac{1}{2}{\bf 2}$ \\
& $\gso(7)\oplus\gsu(2)$ & (1,-1) & $({\bf 8}_s, \frac{1}{2}{\bf 2})+{\bf 8}_s $ \\
\hline
(-3,-2,-2) & $\gg_2\oplus\gsu(2)$ & (0,0) & $({\bf 7},\frac{1}{2}{\bf 2})+\frac{1}{2}{\bf 2}$ \\
\hline
(-2,-3,-2) & $\gsu(2)\oplus \gso(7)\oplus \gsu(2)$ & (0,0) &
$(\frac{1}{2}{\bf 2}, {\bf 8_s}, \cdot)+(\cdot, {\bf 8_s} ,\frac{1}{2}{\bf 2})$ \\ 
\hline
\end{tabular}
\end{center}
\caption[x]{\footnotesize {\it A priori} possible tuned gauge
  algebras, together with matter and Hodge shifts, on the NHCs with
  multiple divisors. (Tunings on $-2$ curves can be found in the separate Table \ref{t:2chains} in the following subsection.)}
\label{t:multiple}
\end{table}

This curious fact was first derived in \cite{4D-NHC, SCFT-II}, where
it  was
shown generally that a Kodaira type $I_0^*$ meeting type $IV$ can only
be consistently implemented when the $I_0^*$ is $\gg_2$; when meeting
type $III$, it can only be implemented as a $\gg_2$ or $\gso(7)$. Our
local analysis simply confirms these results while also explicitly
constructing local models for those cases that are allowed. Although
these facts are mysterious from the standpoint of anomaly
cancellation, some progress has been made to explain this discrepancy
solely in the language of field theory (in particular global
symmetries \cite{Tachikawa}).

\subsection{$-2$ clusters}\label{sec:-2}

The final cluster type to consider is a configuration of intersecting
$-2$ curves.  We begin by discussing linear chains of $-2$ curves
connected pairwise by simple intersections, focusing on tunings of
$\gsu(N)$ gauge algebras.  We then comment on $-2$ clusters with more
general structure, and discuss the small number of specific possible
tunings with larger gauge algebras.

Several interesting phenomena arise in the study of tunings on
clusters of $-2$ curves. It seems that as far as tunings are
concerned, $-2$ is a ``critical'' value of the self-intersection
number; first of all, these curves and clusters are the lowest in
self-intersection number to admit a null tuning. They form the only
unbounded family of NHC's for 6D F-theory models, at least insofar as
they arise in non-compact toric bases for F-theory
compactifications. Second, tunings of $\gsu(N)$ on $-2$ chains are
also critical, in that precisely {\it all} matter transforming under a given $\gsu(N)$ can be shared with neighboring $\gsu(N)$'s.\footnote{When extending our analysis
  to non-compact bases, it is also of interest that $-2$ curves
  present the main ingredient in constructing the ``end-points''
  crucial to the study of SCFT's in \cite{SCFT-II}.}  Finally, certain
combinations of $-2$ curves can form degenerate elliptic curves
associated with an elliptic curve in the base itself.  We will
proceed to analyze these chains both by local models, general
geometric arguments, and anomaly cancellation arguments. The results
of this analysis are summarized in table \ref{t:2chains}.

\begin{table}
\begin{center}
\begin{tabular}{|c|c|c|}
\hline
cluster & $\gg$ & $(\Delta h^{1,1},\Delta \hon)$ \\
\hline
22 & $\gg_2\oplus \gsu(2)$ & $(3,[-12])$ \\
& $\gso(7)\oplus \gsu(2)$ & $(4,[-15])$ \\
\hline
222 & $\gsu(2)\oplus \gg_2\oplus \gsu(2)$ & $(4,[-8])$ \\
& $\gsu(2)\oplus\gso(7)\oplus\gsu(2)$ & $(5,[-12])$ \\
& $\gg_2\oplus\gsu(2)\oplus \cdot $ & $(3,[-11])$ \\
\hline
2222 & $\gsu(2)\oplus\gg_2\oplus\gsu(2)\oplus\cdot$ & $(4,[-8])$ \\ 
\hline
22222 & $\cdot \oplus \gsu(2)\oplus \gg_2\oplus \gsu(2)\oplus \cdot$ &
$(4,[-8])$\\
\hline
$2_1\cdots 2_{k>5}$ & (Only $\gsu(n)$'s; see \S\ref{sec:-2}) & 
see (\ref{eq:2-shift}) \\
\hline
\end{tabular}
\end{center}
\caption[x]{\footnotesize Table of possible tunings on $-2$
  chains. (Chains are listed with self-intersections sign-reversed.)
  Because matter is very similar between these cases, we do not list
  it explicitly, preferring to display the shift in Hodge numbers
  resulting from that matter. For convenience, we summarize the
  relevant matter content here: $4 \times {\bf 2}$ for $\gsu(2)$, $4
\times {\bf 7}$
  for $\gg_2$, $4 \times {\bf 8_s}+{\bf 7}$ for $\gso(7)$ and $2
\times {\bf
    8_s} + 2 \times {\bf 8_c}
+2 \times {\bf 8_f}$ for $\gso(8)$, where the ${\bf 8_s}, {\bf 8_c}$
matter representations are functionally equivalent. $\gsu(2)$ shares a
  half-hypermultiplet with all groups but itself, where it shares a
  whole hyper; with $\gso$'s, it is the spinor representation which is
  shared. } 
\label{t:2chains}
\end{table}

\subsubsection{Linear $-2$ chains}

Rather than using local toric models, 
in this section we primarily use a simple
local feature of the tuning over $-2$ curves to simplify the
analysis.
This feature, which can be understood both geometrically and from
anomaly cancellation, gives a simple picture of the structure of $-2$ cluster tuning
that avoids detailed technical analysis.  The conclusions of this
simple analysis can then be checked using local models, which we do in
part later in this section and in part in the following section.
To see the basic structure of tuning over $-2$ curves,
recall that the Zariski analysis leads
to inequality \ref{eq:avg}, which constrains the minimum order of a
section of $-nK$ given its orders on neighboring curves. The feature
of interest appears when this formula is applied to $-2$
curves. Indeed, letting $k$ denote the order of vanishing of any section (of {\it
  any} $-nK$) on a $-2$ curve $\Sigma$ and $k_R$ and $k_L$ denote the
orders of vanishing of the section on the neighbors of $\Sigma$, \ref{eq:avg} becomes
\begin{eqnarray}\label{eq:avg-2}
k & \geq & \frac{k_L+k_R}{2} 
\end{eqnarray}
This feature and some of its consequences
was used in \cite{large-h21}, and was dubbed the ``convexity
condition'' in \cite{SCFT-II}. 
More generally, if the $-2$ curve $\Sigma$ intersects $j$ other curves
$\Sigma_i, i = 1, \ldots, j$, then the order of vanishing on $\Sigma$
satisfies $k \geq (\sum_{i}k_i)/2$.  The consequence for $\gsu(N)$
tunings on a $-2$ curve connected to a set of other $-2$ curves with
tuned gauge algebras $\gsu(M_i)$ is that
\begin{equation}
2N   \geq \sum_{i}M_i\,.
\label{eq:2-local}
\end{equation}
This condition follows immediately from anomaly cancellation, since at
every intersection there is a hypermultiplet of shared matter in the
$({\bf N}, {\bf M_i})$ representation, and $\Sigma$ only carries $2 N$
matter fields in the fundamental representation.  Thus, this simple
convexity condition naturally captures the constraints of anomaly cancellation.

Comparison to harmonic functions yields some immediate insight. For
instance, on a closed or infinite
chain of $-2$ curves, a $(0,0,n)$ tuning on any divisor forces a
$(0,0,n)$ tuning on every divisor. 
More generally, imagine that a curve $\Sigma$ supports a tuned
$\gsu(n)$ gauge algebra, associated with vanishing orders of $(f, g,
\Delta)$ of $(0,0,n)$.  Now consider a linear
chain of $k$ $(-2)$-curves
connected in sequence to $\Sigma$,  with curves labeled by 
$\Sigma_1,\cdots, \Sigma_k$, with $\Sigma_k=\Sigma$. 
The order of vanishing of $\Delta$ on $D_l$  then satisfies 
$n_l\geq \lceil n\frac{l}{k} \rceil$.  We
can recover
from this rule the infinite case.
Note that in some cases the inequality cannot be saturated .

The local rule (\ref{eq:2-local}) gives a clear bound on possible
tunings of $-2$ curves combined in an arbitrary cluster.  In the
following section we prove using Tate tunings that, at least at the
local level of pairwise intersections, every tuning of a combination
of $\gsu(N)$ algebras on intersecting divisors that is allowed from
(\ref{eq:2-local}) can be realized through a Weierstrass construction,
so at least locally there is a perfect match between the constraints
of the low-energy field theory and F-theory.
Here we proceed simply using (\ref{eq:2-local}) to make some
observations about possible tunings of $\gsu(N)$ combinations on $-2$
clusters.

The local rule (\ref{eq:2-local}) has simple consequences for tunings
over any linear chain of $-2$ curves.  In particular, on a linear
chain of $-2$ curves $\Sigma_i$, the sequence of gauge algebras
$\gsu(N_i)$ must be convex, with each $N_i$ greater or equal to the
average of the adjoining $N_{i-1}, N_{i + 1}$.  This constraint gives
a systematic framework for analyzing local tunings on any linear chain
of $-2$ curves.  Note, however, that the set of possible tunings even
on a single isolated $-2$ is {\it a priori} infinite, when no further
constraints from neighboring divisors are taking into account.  The
finite bound on possible tunings of curves of self-intersection $-2$
or above is discussed in the following section and
\S\ref{sec:algorithm}, in the context of intersection with other
curves in the base.  For 6D supergravity models with a compact base,
the actual number of possible tunings is always finite, while for 6D
SCFT's the family of possible tunings is infinite.  Similarly, for 6D
SCFT's there is no bound on the number of possible $-2$ curves that
can be combined in a chain, while for compact supergravity models
there is a finite bound.

Given this structure, we can simply classify all $\gsu(N)$ tunings
over clusters of $-2$ curves.
The tunings allowed are precisely those that satisfy
(\ref{eq:2-local}).  If we have a set of $l$ curves $\Sigma_i$ that carry
gauge algebras $\gsu(N_i)$, with intersection numbers $I_{ij} \in\{0,
1\}$, then the total gauge algebra is
\begin{equation}
\oplus_i \gsu(N_i) \,,
\end{equation}
the matter content is
\begin{equation}
\sum_{i, j:I_{ij} = 1}({\bf N_i},  {\bf N_j}) \,,
\end{equation}
and the shift in Hodge numbers is
\begin{equation}
\Delta (h^{1, 1},h^{2, 1}) =
\left(\sum_{i}(N_i -1), \sum_{i} (-N_i^2 -1) + \sum_{i, j:I_{ij} = 1}(N_i
N_j)\right) \,.
\label{eq:2-shift}
\end{equation}

The ``critical'' nature of $-2$ chains is particularly apparent when
the inequality (\ref{eq:2-local}) is saturated.  In this case, all of
the $2N$ fundamental matter fields on $\Sigma$ are involved in
bifundamental matter fields.  An interesting feature of this is that
there is an almost perfect cancellation between the number of vector
and hyper multiplets.  In particular, for a closed chain of $-2$
curves, with $\gsu(N)$ tuned on each, we have a contribution to
$H_{\rm charged} -V$ of precisely 1 for each $-2$
curve.  This
interesting possibility is discussed further in a related set of
circumstances in the following
subsection.

\subsubsection{Nonlinear $-2$ clusters}
\label{sec:nonlinear-2}

We can use (\ref{eq:2-local}) to describe $\gsu(N)$ tunings on more
complicated configurations of $-2$ curves, which may include branching
or loops.  Remarkably, this simple averaging rule strongly constrains
the kinds of clusters that can support tunings, revealing a
potentially very interesting structure.

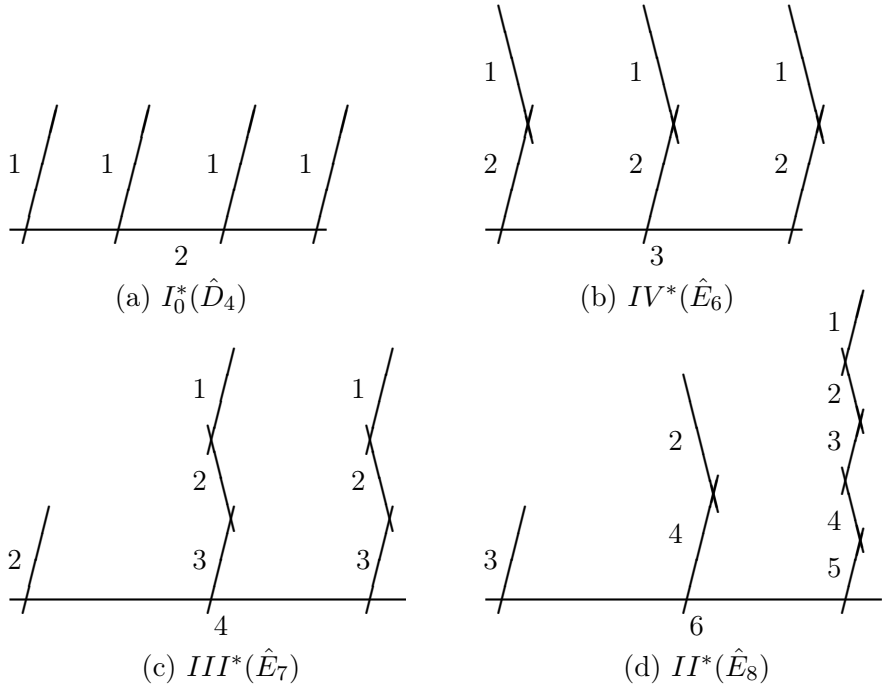
\begin{figure}
\begin{center}
\begin{picture}(200,200)(- 100,- 200)
\thicklines
\put(-150, -40){\line(1, 0){120}}
\put(-85,-65){\makebox(0,0){(a) $I_0^* (\hat{D}_4)$}}
\put(-85,-50){\makebox(0,0){2}}
\put(-145, -45){\line(1, 4){13}}
\put(-148, -15){\makebox(0,0){1}}
\put(-35, -45){\line(1, 4){13}}
\put(-38, -15){\makebox(0,0){1}}
\put(-110, -45){\line(1, 4){13}}
\put(-113, -15){\makebox(0,0){1}}
\put(-70, -45){\line(1, 4){13}}
\put(-73, -15){\makebox(0,0){1}}

\put(30, -40){\line(1, 0){120}}
\put(95,-65){\makebox(0,0){(b) $IV^* (\hat{E}_6)$}}
\put(95,-50){\makebox(0,0){3}}
\put(35, -45){\line(1, 4){13}}
\put(35, 45){\line(1, -4){13}}
\put(32, -15){\makebox(0,0){2}}
\put(32, 20){\makebox(0,0){1}}
\put(90, -45){\line(1, 4){13}}
\put(90, 45){\line(1, -4){13}}
\put(87, -15){\makebox(0,0){2}}
\put(87, 20){\makebox(0,0){1}}
\put(145, -45){\line(1, 4){13}}
\put(145, 45){\line(1, -4){13}}
\put(142, -15){\makebox(0,0){2}}
\put(142, 20){\makebox(0,0){1}}

\put(-150, -180){\line(1, 0){150}}
\put(-70,-205){\makebox(0,0){(c) $III^* (\hat{E}_7)$}}
\put(-70,-190){\makebox(0,0){4}}
\put(-145, -185){\line(1, 4){10}}
\put(-148, -165){\makebox(0,0){2}}
\put(-75, -185){\line(1, 4){10}}
\put(-75, -125){\line(1, 4){10}}
\put(-75, -114){\line(1, -4){10}}
\put(-78, -165){\makebox(0,0){3}}
\put(-78, -135){\makebox(0,0){2}}
\put(-78, -100){\makebox(0,0){1}}
\put(-15, -185){\line(1, 4){10}}
\put(-15, -125){\line(1, 4){10}}
\put(-15, -114){\line(1, -4){10}}
\put(-18, -165){\makebox(0,0){ 3}}
\put(-18, -135){\makebox(0,0){2}}
\put(-18, -100){\makebox(0,0){1}}

\put(30, -180){\line(1, 0){150}}
\put(110,-205){\makebox(0,0){(d) $II^* (\hat{E}_8)$}}
\put(110,-190){\makebox(0,0){6}}
\put(35, -185){\line(1, 4){10}}
\put(32, -165){\makebox(0,0){3}}
\put(105, -185){\line(1, 4){13}}
\put(105, -95){\line(1, -4){13}}
\put(102, -155){\makebox(0,0){4}}
\put(102, -120){\makebox(0,0){2}}
\put(165, -185){\line(1, 4){8}}
\put(162, -168){\makebox(0,0){5}}
\put(165, -130){\line(1, -4){8}}
\put(162, -150){\makebox(0,0){4}}
\put(165, -140){\line(1, 4){8}}
\put(162, -120){\makebox(0,0){3}}
\put(165, -85){\line(1, -4){8}}
\put(162, -102){\makebox(0,0){2}}
\put(165, -95){\line(1, 4){8}}
\put(162, -75){\makebox(0,0){1}}
\end{picture}
\end{center}
\caption[x]{\footnotesize
Some configurations of $-2$ curves associated
  with Kodaira-type surface singularities associated with degenerate
  elliptic fibers.  
The numbers given are the weightings needed to give an elliptic curve
with vanishing self-intersection.
Labels correspond to Kodaira singularity type and associated Dynkin diagram.}
\label{f:2-curves}
\end{figure}

First, consider a $-2$ curve $\Sigma$ that is connected to $c$ linear
chains of $-2$ curves of length $l_i -1$. Assume that $\gsu(N)$ is
tuned on $\Sigma$.  Then from the above analysis, each of the curves
connected to $\Sigma$ from the linear chains must support at least a
gauge factor $\gsu(\lceil N (l_i -1)/l_i\rceil)$.  From
(\ref{eq:2-local}), however, the sum of the resulting $M_i$ has an
upper bound of $2 N$.  This immediately bounds the types of chains
that can be connected to $\Sigma$.  A chain of length 1 contributes at
least $N/2$ to $\sum_{i}M_i$, a chain of length 2 contributes at least
$2 N/3$, {\it etc.}.  Thus, we can have at most four chains connected
to $\Sigma$, and this is possible only for chains of length one.  If
we have 3 chains, it is straightforward to check that the allowed
lengths are $(1, 1, l -1)$ for arbitrary $l$, $(1, 2, l-1)$ for $l
\leq 6$, $(1, 3, 3)$,
and $(2, 2, 2)$.  Thinking of these configurations as Dynkin
diagrams, the extremal cases in this enumeration correspond precisely
with the classification of degenerate elliptic curves associated with
affine Dynkin diagrams $\hat{D}_4$, $\hat{D}_{l + 1}$, and $\hat{E}_n$
\cite{bhpv}.
Examples of these degenerate elliptic curves are illustrated in 
Figure~\ref{f:2-curves}.
All of these nontrivial $-2$ curve
configurations can be realized, for example on
rational elliptic surfaces \cite{Persson,  Miranda}.  Specific
examples of such realizations were encountered in the classification
of $\C^*$ bases in \cite{martini}.  

In the extremal cases, we have a situation
where combinations of $\gsu(N)$ can be tuned on these divisors with a
contribution to $H_{\rm charged}-V$ that is independent of $N$.  For
example, on the $(2, 2, 2)$ configuration corresponding to
$\hat{E}_6$, we can tune a gauge algebra $\gsu(3N)$ on $\Sigma$,
$\gsu(2N)$ on the components of the chains that intersect $\Sigma$,
and $\gsu(N)$ on the terminal links in the chains.
This presents an apparent puzzle, since in a compact base only a
finite number of tunings are possible and we would expect higher-rank
tuning to require more moduli.

Weierstrass models for the extremal $-2$ clusters can be analyzed
using the methodology  used in \cite{martini}.  For example, for the
case of a $-2$ curve $\Sigma$ intersecting four other $-2$ curves, in local
coordinates where the other curves intersect $\Sigma$ at $w = 0, 1, 2,
\infty$, the generic Weierstrass model takes the form
\begin{eqnarray}
f & = &  f_0 + f_2 \zeta z^2+ f_4 \zeta^2 z^4 + \cdots \label{eq:Weierstrass-elliptic}\\
g & = &  g_0 + g_2 \zeta z^2+ g_4 \zeta^2 z^4 + \cdots \,,
\end{eqnarray}
where $\zeta = w (w-1) (w-2)$.  We can set $\Delta = {\cal O} (z^2)$
by tuning $g_0$ to cancel in the leading term.  This gives a gauge
algebra of $\gsu(2)$ on $\Sigma$ and no gauge algebra on the other
curves.  
The shift in $H_{\rm \neutral}$ is then given by $\Delta H_{\rm
  \neutral} = V-H_{\rm charged} = -5$.  This appears surprising as we
have only tuned one modulus, but keeping careful track of extra moduli
from $-2$ curves this is correct; once we have done this tuning, the
discriminant identically vanishes on the four additional curves, so
they are no longer counted as contributing to $N_{-2}$ as discussed in
\S\ref{sec:Calabi-Yau}.  We can further tune
$g_2$ so that the next term in the discriminant vanishes.  This then
gives an $\gsu(4)$ on $\Sigma$ and an $\gsu(2)$ on the other four
curves.  We now have $V-H_{\rm charged} = -5$ again, but we have
nonetheless tuned a modulus.  Repeating this, we use one modulus each
time we increase the algebra on $\Sigma$ by $N \rightarrow N + 1$.
This represents an apparent disagreement between the moduli needed for
Weierstrass tuning and anomaly cancellation.  

Some insight can be gleaned into what is transpiring in these
situations by observing that these $-2$ configurations are essentially
degenerate genus one curves that satisfy $\Sigma \cdot \Sigma = -K
\cdot \Sigma = 0$.  On a smooth curve of this type, the only matter
would be a single adjoint representation of $\gsu(N)$, giving $H_{\rm
  charged} -V = 1$, independent of $N$. When the smooth genus one
curve degenerates into a combination of $-2$ curves, the resulting
configuration of $\gsu(N)$ groups is precisely that realized on the
extremal $-2$ clusters, with no matter in the fundamental, and
bifundamental matter at the intersection points.
Indeed, the multiplicity of $N$ that is tuned on each $-2$
curve for each extremal cluster associated with a degenerate elliptic
curve is precisely the proper multiplicity to give the elliptic
curve.  Thus, this can be thought of in each of these cases as tuning
an $\gsu(N)$ on the elliptic curve and taking a degenerate limit.

The tuning of a single degree of freedom for each increase in $N$ can
be understood as the motion of a single seven-brane in the transverse
direction to the genus one curve $\Sigma$. 
The fact that $N$ can only be tuned to a certain maximum value for the
smooth genus one curve $\Sigma$ with $\Sigma \cdot \Sigma = 0$
 follows from the fact that $\Delta = -12K$; there is always some
 maximum $N$ such that $-12K-N  \Sigma$ is effective.
In a compact base with an extremal $-2$ cluster, 
this corresponds to the fact that
the Weierstrass
expansion terminates and only a finite number of tunings are
possible.  
For example, tuning two of these clusters, with the sets of
four additional $-2$ curves in each cluster connected pairwise by $-1$
curves gives a base studied in \cite{martini}, which has the unusual
future of supporting a generic U(1) factor associated with a higher
rank Mordell-Weil group.  This structure may be a clue to the
anomalous anomaly behavior.  In this compact base, $f, g$ in
(\ref{eq:Weierstrass-elliptic}) only go out to order $z^8, z^{12}$
respectively, giving a bound on the gauge group that can be tuned.

A closed loop of $-2$ curves fits into this framework as the affine
$\hat{A}_{l-1}$ Dynkin diagram/degenerate elliptic curve.
We comment on this further in \S\ref{sec:generalized-intersection}.
Note also that in some situations these degenerate genus one curve
configurations of $-2$ curves can be blown up further, giving more
complicated configurations with related properties \cite{mpt}; in
general such configurations, once blown up, have curves of
self-intersection $-5$ or below, and do not admit infinite tunings.
Blowing up a loop of $-2$ curves to form a loop of alternating $-1,
-4$ curves is an exception, as mentioned in
\S\ref{sec:generalized-intersection}.

\subsubsection{Tuning other groups on $-2$ curve clusters: special
  cases}
\label{sec:special-cases}

So far all tunings we have discussed on clusters with multiple $-2$
curves have involved only $\gsu(N)$. It happens that only a handful of
other algebras can can be tuned on $-2$ clusters. For instance,
$\gsp(N)$ cannot be tuned on any $\leq -1$ curve. Also, it is
straightforward to see that the $\gf$ and $\ge$ algebras cannot arise
on multiple $-2$ curve clusters. (Simply observe that a $(3,4)$
singularity on one $-2$ curve must force at least a $(2,2)$
singularity on any intersecting $-2$ curve, hence these exotic
algebras would share matter. In section
\ref{sec:intersecting-constraint}, we discuss why this is always
impossible both for geometric and anomaly-cancellation reasons.) This
analysis leaves only the case $I^*_n$.

In this subsection we go through the explicit analysis of the various
cases of $-2$ curve clusters that support algebras other than
$\gsu(2)$, looking at Weierstrass models and the corresponding anomaly
conditions.  While some aspects of this analysis are essentially
covered by the rules of tunings on intersecting brane combinations in
the following section, it is worth emphasizing one subtlety, which is
related to the fact that certain algebras such as $\gsu(2)$ can be
tuned in several different ways, either as a type $I_2$ or as a type
$III$ or $IV$.  We have for the most part not emphasized this
distinction as it is not relevant for minimal tunings in most cases,
and is not easy to understand in terms of the low-energy theory,
however it is relevant in these cases, which serve as an illustration
of how these distinctions are relevant in special cases.

In more detail, the Kodaira cases to consider are combinations of
$I_0^*$ tunings with $\gsu(2)$ tuned as either type $III$ or $IV$. Let
us see why this is. To understand why $I_{\geq 1}^*$ cannot be tuned,
suffice it to say that geometrically, it forces a $(4,6)$ singularity at
the intersection point with any other $-2$ curve. Although this could
be demonstrated directly, we note here that it will follow from our
geometric arguments that even $\gso(8)$ cannot be tuned on any $-2$
cluster; the monomials in $f$ and $g$ that must be set to zero to
achieve an $\gso(8)$ are a strict subset of those which must be set to
zero in tuning $\gso(9)$. From the anomaly cancellation standpoint,
this statement is perfectly consistent, because we expect all matter
to be $8$ half-hypermultipets of an $\gsu(2)$ forced on an adjacent
curve, and therefore the dimension of the matter shared with $\gsu(2)$
cannot exceed $8$, but of course would be for $\gso(\geq 9)$. In fact, it
is impossible for an $I_0^*$ to appear next to anything but
$\gsu(2)$. This is because an $I_0^*$ $(2,3)$ singularity forces at
least a $(1,2)$ singularity on any intersecting $-2$ curve. Attempting
to tune $\gsu(3)$ with a $IV$ singularity requires that $g_2$ be a
perfect square. However, a toric analysis for two intersecting $-2$
curves reveals that $g_2$ has a lowest order term of order $w^3$,
which cannot be the lowest order term of a perfect square. Eliminating
this monomial directly produces a $(4,6)$ singularity at the
intersection point. This conclusion is also reasonable from the field
theory point of view: the fundamental of $\gsu(3)$ is not self-conjugate,
and therefore the matter shared with it can be at most
$6$-dimensional.

In summary, the only tunings on $-2$ curve clusters that contain
algebras other than $\gsu(N)$ are combinations of $I_0^*$ and
$III$/$IV$ $\gsu(2)$'s. In fact, the averaging rule implies that
tunings containing one $I_0^*$ component may only occur in chains with
$\leq 5$ curves. (On larger clusters, the averaging rule implies that
the $(2,3,6)$ singularity would persist to at least the nearest
neighbor, immediately yielding a $(4,6)$ singularity at the
intersection point.) Therefore, our task is to classify the allowed
combinations of $\gsu(2)$, $\gg_2$, $\gso(7)$, and $\gso(8)$. We will
find that only $\gg_2$ and $\gso(7)$ can be realized, but not
$\gso(8)$.\footnote{This matches with the low-energy constraint from
  global symmetries \cite{Tachikawa}, as discussed in more detail in
  \S\ref{sec:generalized-e8}.} As discussed above, an $I_0^*$ tuning
necessarily forces at least a $III$ singularity on an intersecting
$-2$ curve, so we will not encounter any isolated $I_0^*$
tunings. (This geometrical constraint is not yet well characterized in
terms of the low-energy field theory; see \S\ref{sec:generalized-e8})

Let us now classify these tunings: namely, a single $\gg_2$,
$\gso(7)$, or $\gso(8)$ together with its neighbors, which must be
$\gsu(2)$'s. This must occur on a chain of length $\leq 5$. We will
construct these models in order of increasing length of chain,
starting on a configuration $(\Sigma_1, \Sigma_2)$ of two intersecting
$-2$ curves, and progressing to a chain $(\Sigma_1,\cdots,\Sigma_5)$
of five $-2$ curves in a linear configuration. For each of these four
configurations, we tune a $\gg_2$, then attempt to enhance to
$\gso(7)$ or $\gso(8)$. Because we will immediately find that
$\gso(8)$ cannot be tuned, we will not consider it on any
configuration but the first. (Configurations with additional $-2$
curves have strictly fewer monomials, so any obstruction to tuning on
smaller configurations will apply trivially to larger configurations
as well.) To set notation, we will use $\{z=0\}$ to be a local
defining equation for the curve $\Sigma$ on which the type $I_0^*$
singularity is tuned; any intersecting curve of interest we will take
to be defined as $\{ w=0\}$. We will consider
$f=\sum_{i,j}f_{i,j}z^iw^j$ and similarly for $g$; when discussing
orders of vanishing on {\it e.g.} $\Sigma=\{ z=0\}$, we will use terms
$f_{i,\cdot}$ and $g_{i,\cdot}$ with one blank subscript to refer to
functions of $w$ in an implicit expansion $f=\sum_i
f_{i,\cdot}(w)z^i$. Similarly $f_{\cdot,j}$ refers to an implicit
expansion $f = \sum_i f_{\cdot,j}(z)w^j$.
While we only discuss linear chains here explicitly, a similar analysis
governs the tuning of an $I_0^*$ factor on a small branched $-2$
cluster; details are left to the reader.

{\bf Case 1 (a): $\gg_2$ on $(-2,-2)$.} To implement the tuning
$\gg_2$ on $\Sigma_1$, we must impose
$(\text{ord}_{\Sigma_1}f,\text{ord}_{\Sigma_1}g)=(2,3)$, so for $f$ we
eliminate $1+2$ degrees of freedom in setting $f_{0,\cdot}$ and
$f_{1,\cdot}$ to zero, while for $g$ we eliminate $1+2+4$ degrees of
freedom in setting $g_{0,\cdot}$, $g_{1,\cdot}$, and $g_{2,\cdot}$ to
zero. Now let us examine $f_{2,\cdot}=\sum_{i=1}^4f_{2,i}w^i$ and
$g_{3,\cdot}=\sum_{i=2}^6 g_{3,i} z^i$. By inspection, the geometry
does not force a non-generic factorization, so we have indeed tuned
$\gg_2$, and not one of the other $I_0^*$ cases. The removal of these
monomials forces
$(\text{ord}_{\Sigma_2}f,\text{ord}_{\Sigma_2}g)=(1,2)$, so that we
obtain an $\gsu(2)$ neighbor. As a check, note that implementing this
tuning fixes $10$ monomials as well as two $-2$ curve moduli, which
yields a shift in $\hon$ of $-12$, consistent with anomaly
cancellation. No further complications have arisen in this instance.

{\bf (b): $\gso(7)$} Enhancing this tuning to $\gso(7)$, we impose
\begin{eqnarray}
f_2 & = & B-A^2 \\
g_3 & = & -AB
\end{eqnarray}
which is most generically achieved by setting $A=A_1w+A_2w^2$ and
$B=B_1w+B_2w^2+B_3w^3+B_4w^4$, for a total loss of $9-6=3$ degrees of
freedom. Note that all coefficients $f_{2,i}$ and $g_{3,i}$ will
generally remain nonzero. This implies that the orders of $(f,g)$ on
$\Sigma_2$ remain $(1,2)$, so that the neighboring $\gsu(2)$ remains
type $III$. It is worth mentioning that these calculations agree with
the anomaly calculations of shifts in $\hon$ when one tunes the
$\gg_2\oplus \gsu(2)\longrightarrow \gso(7)\oplus \gsu(2)$
combination, provided that it is one of the four ${\bf 8_s}$
representations of $\gso(7)$ that is also charged as a fundamental
under $\gsu(2)$--\textit{not} the ${\bf 7_f}$. This bifundamental
matter is consistent with the process of Higgsing back to $\gg_2$ as
it leaves the ${\bf 7}$ unharmed to play the role of the ${\bf 7}$
fundamental of $\gg_2$. Moreover, this matches with
the global symmetry analysis of
\cite{Tachikawa}, which indicates that the $\bf{8_s}$ representation must be
shared instead of the ${\bf 7}$ representation.

{\bf (c): $\gso(8)$: Forbidden} This tuning is impossible, as
mentioned above; hence all $\gso(8)$ tunings will be impossible on
larger $-2$ chains. To see this in this context, recall that to
enhance the tuning to $\gso(8)$, we require the more stringent
factorization condition
\begin{eqnarray}
f_2 & = & AB-(A+B)^2 \\
g_3 & = & AB(A+B)
\end{eqnarray}
which is most generically achieved by setting $A=A_1w+A_2w^2$, $B=B_1w+B_2w^2$. Notice that this removes all monomials $z^aw^b$ in $g$ with $b \leq 3$, hence $a+b \geq 6$. Therefore the order of $g$ at the intersection point between the $-2$ curves is at least $6$. Notice as well that this tuning removes the order $z^2$ term in $f$. Combining these facts, this tuning attempt would yield a $(4,6)$ singularity where the $-2$ curves meet.

{\bf Case 2 (a,i): $\gg_2$ on $\Sigma_1$ of $(-2,-2,-2)$.}  In this cluster we may tune either on the first curve $\Sigma_1$ or the middle curve $\Sigma_2$. The former presents distinct differences, which we will discuss first, after which we will move on to discuss the tuning on $\Sigma_2$, which proceeds analogously to the tuning on $(-2,-2)$.

Let us now implement the tuning on $\Sigma_1$. The additional
complication will be the possibility of a forced gauge algebra on the
final curve $\Sigma_3$. On $\Sigma_2$, the effect is a type $IV$
singularity with $g_2$ consisting of a single monomial. In fact, this
monomial is $z^3$ in a defining coordinate $z$ for $\Sigma_2$, so we
will never encounter the issue of an $\gsu(3)$ on $\Sigma_2$. On
$\Sigma_3$, there is only an $II$ type singularity, which does not produce a gauge algebra. This tuning has
proceeded without obstruction.

{\bf (a,ii): $\gso(7)$: Forbidden} Since $f_{2,\cdot}=f_{2,1}w+f_{2,2}w^2$ and $g_{3,\cdot}=g_{3,2}w^2+g_{3,3}w^3+g_{3,4}w^4$, the required factorization condition is satisfied by the choice $A=A_1w$, $B=B_1w+B_2w^2$. Note that this requires the $w^4$ term in $g_{3,\cdot}$ to vanish, in addition to imposing a relation among the remaining coefficients. We lose $2$ degrees of freedom, as expected from the point of view of anomaly calculations above. This factorization constraint removes the single monomial of in $g$ that is order $2$ over $\Sigma_3$ , leading to a $III$ singularity on $\Sigma_3$. An $\gsu(2)$ on $\Sigma_3$ would have to share more matter than it carries in the first place; therefore this enhancement, as well as subsequent ones, are inconsistent. In more geometric terms, a $(4,6)$ singularity appears at $\Sigma_1\cdot \Sigma_2$.

{\bf (b,i) $\gg_2$ on $\Sigma_2$} Attempting to tune $I_0^*$ on the middle curve $\Sigma_2$ proceeds without difficulty, in complete analogy to tuning on the $\Sigma_1$ in the configuration $(-2,-2)$ of case $1$. Implementing a tuning of $\gg_2$ on the middle curve $\Sigma_2$, we investigate the forced tunings on its neighbors $\Sigma_1$ and $\Sigma_3$. (By symmetry, it suffices to consider only $\Sigma_1$, which incurs an $(f,g)=(1,2)$ type $III$ singularity.) One can confirm that there is generic factorization on $\Sigma_2$, so we are indeed in the $\gg_2$ case.

{\bf (b,ii): $\gso(7)$}  Since (on $\Sigma_2$) $f_{2,\cdot}=f_{2,1}w+f_{2,2}w^2+f_{2,3}w^3$ and $g_{3,\cdot}=g_{3,2}w^2+g_{3,3}w^3+g_{3,4}w^4$, the relevant factorization condition can be achieved with $A=A_1w$, $B=B_1w+B_2w^2+B_3w^3$, for a loss of $2$ degrees of freedom. This is consistent with anomaly calculations, and moreover the singularities on the adjacent curves remain type $III$.

{\bf Case 3 (a,i): $\gg_2$ on $\Sigma_1$ of $(-2,-2,-2,-2)$} The novel contribution to this cluster is the possibility of tuning on the initial curve $\Sigma_1$. However, this is impossible. The pathology of this attempted tuning is visible even without investigating monomials. A $(2,3,6)$ singularity on $\Sigma_1$ will, by the averaging rule, immediately produce at least a $(2,3,6)$ singularity on $\Sigma_2$, leading to an unacceptable $(4,6)$ singularity at $\Sigma_1\cdot \Sigma_2$. Indeed, this same logic shows that {\em no} tuning of $I_0^*$ is possible on a chain of $\geq 6$ $-2$ curves, as there is no curve in this chain that with have fewer than $3$ additional $-2$ curves to one side. We reemphasize: even $\gg_2$ cannot be tuned here.

{\bf (b,i): $\gg_2$ on $\Sigma_2$ of $(-2,-2,-2,-2)$.} In this case,
we will tune on $\Sigma_2$, which is quite analogous to tuning on
$\Sigma_1$ of the configuration $(-2,-2,-2)$. We  find no
additional restrictions, but we  the presence of the $-2$
curve $\Sigma_1$ leads to the presence of another (type $III$)
$\gsu(2)$ neighbor. Examining first the effect on $\Sigma_1$, a
$\gg_2$ produces the expected $III$ or $(f,g)=(1,2)$ singularity. On
$\Sigma_3$, the effect is a type $IV$ singularity with $g_{\cdot,2}$
consisting of a single monomial, and on $\Sigma_4$, there is
only a type $II$ type singularity, as in case (a, i) above.

{\bf (b,ii): $\gso(7)$: Forbidden} This is already assured from the
analysis of case $2$, as we have merely added another $-2$ curve,
which can only add constraints.

{\bf Case 4 (a): $\gg_2$ on $(-2,-2,-2,-2,-2)$.} The previous analysis
already shows that we cannot tune an $I_0^*$ singularity anywhere but
the middle curve; otherwise, there would be a string of $\geq 3$ $-2$
curves to one side of the $I_0^*$, which would force a $(4,6)$
singularity. Thus, our goal is simply to verify that a $\gg_2$ can be
tuned on the middle curve $\Sigma_3$, in precise analogy to tuning on
$\Sigma_2$ of case $3$. We already know that tuning $\gso(7)$ and
$\gso(8)$ tunings are forbidden in this context, because this case is
obtained from the previous one by an additional blowup at the endpoint
$-1$ curve of the local model. The task, then, is simply to verify
that the generic $I_0^*$ $\gg_2$ singularity can be consistently
imposed. By investigating the monomials, one can implement this
tuning, finding that on $\Sigma_3$, $f_{2,\cdot}=f_{2,2}w^2$ and
$g_{3,\cdot}=g_{3,2}w^2+g_{3,3}w^3+g_{3,4}w^4$ (in a defining
coordinate $w$ for $\Sigma_2$), which implies that no factorization is
generically forced, so this tuning belongs in the $\gg_2$ subcase of
$I_0^*$ tunings, as desired. Moreover, on $\Sigma_2$, there is a
forced $IV$ singularity, for which $g_{\cdot,2}= g_{3,2}z^3$, yielding
$\gsu(2)$. Similarly for $\Sigma_4$. This ensures an $\gsu(2)$ is
adjacent to the tuned $\gg_2$ on either side. As to $\Sigma_{1,5}$,
each carries a type $II$ singularity--in other words, no tuned
algebra. This ensures that such a $\gg_2$ tuning can in fact be
realized, completing the desired classification.

\section{Classification III: connecting curves and clusters}\label{sec:C-III}

At this point, we have investigated tunings over individual curves or
non-Higgsable clusters.  To go further, we would like to determine
constraints on what groups can be tuned over intersecting divisors.
In particular, low-energy anomaly cancellation conditions and corresponding
F-theory geometric conditions impose clear constraints on what groups
can be tuned over intersecting divisors.  At this point we are unaware
of any specific constraints on global models that go beyond conditions
that can be expressed in terms of gauge groups tuned on a single
divisor $\Sigma$ and its immediate neighbors ({\it i.e.}, divisors
intersecting $\Sigma$).  Thus, it may be that determining local
constraints on such configurations may be sufficient to determine the
full set of global tunings that is possible.  We do not attempt to
prove the completeness of local conditions here, but focus in this
section on various conditions that constrain tunings that are possible
on multiple intersecting curves.

In \S\ref{sec:intersecting-constraint} we give a simple set of
arguments that show that there are only 5 (families of) pairs of gauge
groups that can be tuned on a pair of intersecting divisors.  In
\S\ref{sec:intersecting-2}, we determine constraints on these families
in terms of the self-intersections of the curves involved and the
group types.
In \S\ref{sec:generalized-e8} we consider a more general set of
constraints on a curve $\Sigma$ that intersects with two or more other curves
supporting gauge groups, including generalizations of the $E_8$ rule
for curves $\Sigma$ that do not themselves support a gauge group.

In this analysis we continue to focus on divisors with single pairwise
intersections.  A few comments on more general intersection
possibilities are made in \S\ref{sec:generalized-intersection}

\subsection{Types of groups on intersecting divisors}
\label{sec:intersecting-constraint}

We begin by giving some simple arguments that rule out all but five
possible combinations of (families of) pairs of algebras supported on
divisors $\Sigma_1, \Sigma_2$ that intersect at a single point.  
The allowed combinations, determined from
anomaly cancellation conditions, are listed in
Table~\ref{t:intersecting-possibilities}.

\begin{table}
\begin{center}
\begin{tabular}{| c | c | c | c |c |}
\hline
$\gg_n$ & $\gg_m$ & 
anomaly constraints on $n, m$ & matter & $\Delta \hon$\\
\hline
 $\gsu(N)$ &  $\gsu(M)$ &  $\lfloor N/2 \rfloor \leq 8 + (4-\lfloor M/2 \rfloor)m,
\lfloor M/2 \rfloor \leq 8 +  (4-\lfloor  N/2 \rfloor)n$  &
 ${\bf (N, M)}$ &  $+ NM$ \\
$\gsp(j)$ &  $\gsp(k)$  &
$j \leq 8 + (4-k)m,
k \leq 8 +  (4-j)n$ &{\bf (2j, 2k)} & $+ 4jk$\\
$\gsu(N)$ &  $\gsp(k)$  &
$\lfloor N/2 \rfloor \leq 8 + (4-k)m,
k \leq 8 +  (4-\lfloor  N/2 \rfloor)n$ 
 &{\bf (N, 2k)} &  $+ 2 N k$\\
$\gso(N)$ &  $\gsp(k)$  & 
$k \leq n + N -4,
N \leq 32 + (16-4k)m$
& $\frac{1}{2} {\bf (N, 2k)} $ & $ + N k$\\
& & $N = 7: k \leq 8 + 2n, 2 \leq 8 + (4-k)m$  &
$\frac{1}{2}{\bf (8_s, 2k)}$ & + 8k\\
$\gg_2$ &  $\gsp(k)$  & 
$k \leq 3n + 10,7 \leq 32 + (16-4k)m$
&$ \frac{1}{2} {\bf (7, 2k)} $& $+ 7k$\\
\hline
\end{tabular}
\end{center}
\caption[x]{\footnotesize  Algebras of
product groups that can be tuned on a pair
  of intersecting curves of self-intersection $n, m$, and constraints
  from anomaly conditions.  Further constraints from Tate tuning
  conditions are discussed in text.  Shift to Hodge
  numbers is relative to the shift of the individual group tunings.}
\label{t:intersecting-possibilities}
\end{table}

From the field theory point of view, the possibilities of the groups
that are tuned is constrained from the anomaly equation \ref{eq:bij-condition}
\begin{equation}
\Sigma_1\cdot \Sigma_2 = \lambda_1\lambda_2\sum_{R_{1,2}}A_{R_1}A_{R_2}x_{R_{1,2}}
\end{equation}
In words, the shared matter, weighted with the product of its $A$
coefficients and its multiplicity, must equal $\Sigma_1\cdot \Sigma_2$
(which is one or zero in the cases we studied here). 

From this low-energy constraint it is clear that bi-charged matter is
quite difficult to achieve for any algebras other than $\gsu(N)$ and
$\gsp(N)$.  For these algebras $\lambda = 1$.  For $\gg_2$ and
$\gso(N)$, $\lambda = 2$.  For all the other algebras, $\lambda > 2$.
  For all matter representations that appear in generic F-theory
  models, and all known matter representations that can arise in
  F-theory, the coefficients $A_R$ are integers.
We assume that this is generally the case though we have no completely
general proof.
  Thus, we can only have $\Sigma_1 \cdot \Sigma_2 = 1$ when both
  factors are either $\gsu(N)$ or $\gsp(N)$ and $ A_1 = A_2 =x = 1$,
  such as for a situation where there is a full matter hypermultiplet
  in the bifundamental representation, or when one factor is $\gg_2$
  or $\gso(N)$ and the other is $\gsu(N)$ or $\gsp(N)$ and we have a
  half-hypermultiplet in the bifundamental representation.
(Note that the $\gso(N)$ fundamental can be replaced by a spinor when
  $N = 7, 8$ and the anomaly conditions are unchanged.)
While the fundamental $2 N$ of $\gsp(N)$ is self-conjugate
(pseudoreal), only for the special case $SU(2) = Sp(1)$ is the
fundamental of $\gsu(n)$ self-conjugate.  Thus, field theory
considerations seem immediately to restrict to the 5 possibilities in
Table~\ref{t:intersecting-possibilities}.

We can show directly in F-theory using a local monomial analysis that
indeed the five possibilities in
Table~\ref{t:intersecting-possibilities} are the only combinations of
algebras on intersecting divisors that admit a tuning in the
Weierstrass model.  We begin by showing that $\gf_4$ cannot live on a
curve that intersects another curve supporting any nontrivial
algebra.  To begin, let us
label the curves of $\gf_4$ and its neighbor $\gg$ as $\Sigma_1$ and
$\Sigma_2$ respectively. Now notice that $\gf_4$ corresponds to a
$(3,4)$ singularity. If a non-trivial algebra other than one in the
$I_n$ series appeared on $\Sigma_2$, it would have to be at least a
$(1,2)$ singularity--immediately leading to a $(4,6)$ singularity at
$\Sigma_1\cdot \Sigma_2$. Thus, such tunings are inconsistent. It
remains only to prove that a $I_{\geq 2}$ algebra cannot appear on
$\Sigma_2$, for which it will be sufficient to rule out just $I_2$,
{\it i.e.} an $\gsu(2)$ tuned by $(0,0,2)$. By thinking in toric
monomials, it is easy to see why this is inconsistent with a $(3,4,8)$
tuning on $\Sigma_1$. Let $w$ be a local coordinate on $\Sigma_1$ such
that $\Sigma_2=\{w=0\}$, and vice versa for $z$ and
$\Sigma_1=\{z=0\}$. Then we can expand $f$, $g$, and $\Delta$ in
Taylor series in $z$ and $w$. By hypothesis, $f$ contains no monomials
with powers of $z$ lower than $z^3$, and $g$ contains none with powers
lower than $z^4$. To tune $\Delta$ to order $2$ on $\Sigma_2$, we must
require the vanishing of both $\Delta_0$ and $\Delta_1$ in the
expansion $\Delta = \sum_{i=0}\Delta_iw^i$. For $\Delta_0 \sim
f_0^3+g_0^2$, this implies $f_0\propto \phi^2$ and $g_0\propto \phi^3$
for some expression $\phi$. Immediately we see that $f_0$ must be a
perfect square, which forces us to exclude the monomial $w^0z^3$;
moreover, for $g_0$ to be a perfect cube, the coefficients in $g$ of
both $w^0z^4$ and $w^0z^5$ must be zero.

Now let us impose the constraint $\Delta_1 \sim f_0^2f_1 + g_0g_1 =
0$. The lowest order (in $z$) term in $f_0^2f_1$ is now $z^{11}$
($4+4+3$) whereas the lowest order term in $g_0g_1$ is $z^{10}$
($6+4$). The required cancellation  between these terms can only occur,
then, if $g_1$ vanishes to order at least $5$ instead of $4$. But now
this putative tuning is in serious trouble. We have removed from $f$
the single monomial $w^0z^3$ with $m+n < 4$ (where
  $w^mz^n$). Moreover, we have removed from $g$ the three monomials
  with $m+n < 6$. This guarantees a $(4,6)$ singularity at
    $\Sigma_1\cdot \Sigma_2$, as we expected to find.
This proves that no gauge algebra can be tuned on a divisor intersecting a
divisor carrying a singularity of order $(3, 4)$ or higher.

A similar analysis shows that divisors intersecting other divisors
carrying the groups with algebras $\gso(n), \gg_2$ can only have
$\gsp(n)$ algebras.  If we assume $\Sigma_1$ carries a gauge algebra
with $(f, g)$ vanishing to orders $(2, 3)$, we similarly analyze $f_0,
g_0$ at leading orders in $z$, {\it etc.}.  A second curve $\Sigma_2$
intersecting $\Sigma_1$ cannot carry a $(2, 3)$ singularity or we
immediately have a $(4, 6)$ singularity at the intersection point.  We
thus need only consider gauge algebras $\gsu(n), \gsp(n)$ on
$\Sigma_2$.  We consider $I_n$ type singularities.  As above, we have
$f_0 \propto \phi^2, g_0 \propto \phi^3$.  For an $\gsu(n)$ algebra
the split condition dictates that we must have $\phi$ a perfect square
$\phi = \phi_0^2$.  But then $ z | \phi_0$, $z^4| f_0$ and similarly
$z^6 | g_0$.  To tune an $\gsu(3)$ we have \cite{mt-singularities}
$f_1 \sim \phi_0 \psi_1, g_1 \sim \phi f_1, g_2 \sim \psi_1^2 + \phi
f_2$, and since $g_2$ and $\phi f_2$ scale at least as $z^3, z^4$,
$z^2 |\psi_1$, which means $z^3 | f_1, z^5 |g_1, z^4 | g_2$, so we get
a (4, 6) singularity at the intersection.  A similar effort to tune an
$\gsu (3)$ through a type IV singularity would give a term $z^3$ in
$g_2$, which implies that $g_2$ is not a perfect square so the
monodromy gives an $\gsu(2)$ algebra on any curve with a type IV
singularity intersecting a singularity of order $(2, 3)$.

This completes the demonstration that the only possible pairs of
nontrivial algebras that can be realized on intersecting curves are
those in Table~\ref{t:intersecting-possibilities}.  Note that the
analysis here was independent of the dimension of the base, so the
same result holds for 4D F-theory compactifications.

\subsection{Constraining groups on intersecting divisors}
\label{sec:intersecting-2}

We now consider the possible combinations of gauge groups that can
actually be realized for the five possible pairings from
Table~\ref{t:intersecting-possibilities}.  In each case we compare the
constraints from anomaly cancellation to a local Tate analysis, as was
done for single curves in \S\ref{sec:general-curves}.
We take the self-intersections of the two curves to be $\Sigma_1 \cdot
 \Sigma_1 = n$, $\Sigma_2 \cdot
 \Sigma_2 =  m$, and we are assuming that $\Sigma_1 \cdot
 \Sigma_2 = 1$.
We consider the various cases in turn, indicating  potential swampland
contributions in each case.
\vspace*{0.05in}

\noindent
${\bf \gsp(j) \oplus \gsp(k)}$ (no swamp):

We begin with the case $Sp(j = N/2) \times Sp(k=M/2)$, where the
analysis is simplest.  In this case, we expect a single (full)
bifundamental hypermultiplet in the ${\bf (N, M)} = {\bf (2j, 2k)}$
representation.  The number of fundamentals on each of the two curves
is, from Table~\ref{t:isolated}, $16 + (8-2j)n, 16 + (8-2k) m$
respectively.
We therefore have the constraints from anomaly cancellation
\begin{eqnarray}
j & \leq & 8 + (4-k)m \,,   \nonumber\\
k &\leq &  8 +  (4-j)n \,.\label{eq:sp-sp-constraints}
\end{eqnarray}
Here, the self-intersections satisfy $n, m \geq -1$, since $\gsp(k)$
cannot be tuned on a $-2$ curve.

Now let us consider the Tate model.  
Tuning $\gsp(j)$ on a curve of
self-intersection $n$ requires tuning
the $a$ coefficients $(a_1, a_2, a_3, a_4, a_6)$
to vanish to orders $(0, 0, j, j, 2j)$.  The weakest constraint comes
from the $a_4$ condition.
From (\ref{eq:a-degrees}), we see that the degrees of the coefficients
in $a_4$ are ${\rm deg} (a_{4 (s)}) = 8 + n (4- s)$.  Imposing both
constraints, we see that $a_4$ can be written in terms of monomials
$z^s w^t$ subject to the conditions that $t \leq 8 + (4- s)n, s \leq 8 +
(4- t)m$.  Tuning the algebras $\gsp(j) \oplus \gsp(k)$ on our curves
of self-intersections $n, m$ requires having a monomial in $a_4$ with
degrees $ s = j,  t =k$, and we see that such a monomial exists if and
only if the constraints (\ref{eq:sp-sp-constraints}) are satisfied.
This shows that in a local model, using the Tate construction, all
possible $\gsp(j)\oplus \gsp(k)$ algebras consistent with anomaly
constraints can be tuned on a pair of intersecting curves.

\vspace*{0.05in}

\noindent
${\bf \gsu(2j) \oplus \gsu(2k)}$ (no swamp):

A similar analysis can be carried out in the other cases
of Table~\ref{t:intersecting-possibilities}.
We next consider SU($N$) $\times$ SU($M$).  If $N, M$ are both even,
with $N = 2j, M = 2k$, the tuning is precisely like that of the $Sp(j)
\times Sp(k)$ case just considered, except for the tuning of $a_2$ to
first order.  The $a_2$ tuning is always possible, so it cannot affect
the conclusion, and so for even $N, M$ everything that is allowed from
anomalies can be realized using Tate.  
Note that for these algebras, we can have in particular $n = m = -2$.

\vspace*{0.05in}

\noindent
${\bf \gsu(2j + 1) \oplus \gsu(2k)}$ (apparent swamp):

Next consider the case $N = 2j
+ 1, M = 2k$.  In this case, the constraint  from $a_4$ is just as
(\ref{eq:sp-sp-constraints}) but with $j$ replaced by $j + 1$.  But
there must also
{\it either} be at least one monomial in $a_3$ of order at least
$j$ in $z$ or a monomial in $a_6$ of order at least $2j + 1$,
or else the symmetry automatically enhances to SU($N + 1$).  The
conditions that must be satisfied are then
($a_4$ and  ($a_3$ or $a_6$)), where
\begin{align}
a_4: \hspace*{0.1in} &
j  \leq   7 + (4-k)m \,,  \hspace*{0.1in} \nonumber
k \leq 8 + (3-j) n\\
a_3: \hspace*{0.1in} &
j \leq 6 + (3-k)m \,,
\hspace*{0.1in}
k \leq  6 +  (3-j)n  \label{eq:even-odd-constraints}\\
a_6: \hspace*{0.1in} &
2j + 1 \leq 12 + (6-2k)m \,, \hspace*{0.1in}
2k \leq 12 + (5-2j) n
\,.  \nonumber
\end{align}
Some of these conditions imply others.  In particular, the $a_6$
condition on $j$ is always stronger than the $a_4$ condition on $j$,
and the $a_3$ condition on $k$
always implies the $a_4$ condition on $k$.  Nonetheless, the analysis
is a bit subtle as different combinations are ruled in or out in
different ways. For example, $\gsu(3) \oplus \gsu(6)$ violates the
$a_3$ condition but satisfies the $a_6$ condition, while $\gsu(9)
\oplus \gsu(2)$ satisfies the $a_3$ condition but violates the $a_6$
condition.

Let us consider some specific cases of even-odd SU($N$) $\times$
SU($M$).  First, we note that when $m = -2$, the $a_6$ condition on
$j$ is weaker than the $a_3$ condition, and equal to the $a_4$
condition as well as to the anomaly condition.  And when $n = -2$, the
$a_6$ condition is again equivalent to the $a_4$ condition and the
anomaly condition, and all of these are in this case stronger than the
$a_3$ condition.  It follows that when $n = m = -2$ a Tate tuning is
possible precisely when the anomaly conditions are satisfied, and
there is no swampland contribution.

Now, however, we consider the case $ n = -1, m = -2$.  In this case,
the $a_6$ condition on $k$ is $2k \leq 7 + 2j$; this is stronger than
the $a_4$ constraint and weaker than the $a_3$ constraint so it must
be satisfied for a Tate tuning.  But this condition is also stronger
than the anomaly cancellation condition $2k \leq 9 + 2j$.  For $j =
1$, there is a potential swamp contribution at $k = 5$, and more
generally the algebras $\gsu(2j + 1) \oplus \gsu(2j + 8)$ will be
allowed by anomalies but not by Tate.  This represents a simple family
of cases that either should be shown to be inconsistent in the
low-energy theory or realized through Weierstrass if possible.  These
cases are of particular interest since they are relevant for 6D SCFT's
as they can be realized on intersecting $-1, -2$ curves that can be
blown down to give a decoupled field theory.  The simplest of these
examples is $\gsu(3) \oplus \gsu(10)$ where the $\gsu(10)$ has 20
hypermultiplet matter fields in the fundamental representation and
$\gsu(3)$ has 12 matter fields in the fundamental representation (of
which one is technically distinct as it lies in the anti-fundamental,
affecting the global symmetry group even though the content is
identical as 6D hypermultiplets include a complex degree of freedom in
a representation $R$ and a matching complex degree of freedom in the
conjugate representation).

Next, consider $m = n = 1$, where two curves of self-intersection 1
are intersecting.  In this case the anomaly constraint says that $N +
M \leq 24$, and the even-odd Tate constraints impose the condition $N
+ M \leq 19$.  This is similar in spirit to the results of
\S\ref{sec:general-curves}, and can be related explicitly in some
circumstances.  For example, on $\P^2$ a pair of lines supporting
gauge groups SU($N$), SU($M$) can be tuned to be coincident to reach a
gauge group SU($N + M$) on a single line.  Thus, both the upper bound
and the ``swamp'' of models where $N + M = 21, 23$ are consistent
between these pictures.

\vspace*{0.05in}

\noindent
${\bf \gsu(2j + 1) \oplus \gsu(2k + 1)}$ (apparent swamp):

Finally, we consider the odd-odd case $N = 2j + 1, M = 2k + 1$.  In
this case the constraints are similar to
(\ref{eq:even-odd-constraints}), with appropriate replacement of $k
\rightarrow k + 1$ in the $a_4$ constraint and $2k \rightarrow 2k + 1$
in the $a_6$ constraint.  Again we must satisfy $a_4$ and either $a_3$
or $a_6$.  As in the even-odd cases, once again
in the special case $n = m = -2$, this set of constraints
again leads to no conditions
beyond those imposed by anomalies.  For other combinations we have
further contributions to the potential swampland from tunings that are
not possible in Tate.  For $n = m = 1$, where the anomaly
constraint is $N + M \leq 24$, the  odd-odd Tate conditions
 impose the stronger condition $N + M
\leq 20$, so odd-odd combinations with $N + M = 22, 24$ cannot be
tuned by Tate.

We thus see that the Tate approach only gives a subset of the $\gsu(N)
\oplus \gsu(M)$ models that anomaly cancellation suggests should be
allowed on a pair of intersecting divisors, giving some apparent
additional contributions to the ``swampland''.  The number of cases
with no known F-theory construction is relatively large, and it would
be nice to understand whether these admit Weierstrass constructions or
are somehow inconsistent due to low-energy constraints, or neither.

We summarize some of the apparent swampland contributions where
Tate tuning is not possible in Table~\ref{t:uu-swamp}
The fact that there is no swamp when $n = m = -2$ indicates that, at
least locally, the convexity condition used in the preceding section
is the only constraint on tuning product groups on any $-2$ cluster
that need be considered.

\begin{table}
\begin{center}
\begin{tabular}{| c | c | c |}
\hline
$n$ & $m$ & swamp contribution $(N, M) \rightarrow (\gsu(N) \oplus \gsu(M))$\\
\hline
\hline
$-2$ & $-2$ & no swamp\\
\hline
$-2$ & $-1$ & (10, 3), (11, 3), (12, 5), (13, 5), \ldots\\
\hline
$-1$ & $-1$ &  (3, 10), (3, 11), (5, 12),
(5, 13), $\ldots (+ \leftrightarrow)$\\
\hline
$0$ & $0$ &(2, 15), (3, 14), (3, 15),(3, 16), (4, 15), (5, 14), $\ldots
(+ \leftrightarrow) $\\
\hline
1 & 1 & (2, 19), (2, 21), (3, 18), (3, 19), (3, 20), (3, 21), $\ldots
(+ \leftrightarrow)$\\
\hline
\end{tabular}
\end{center}
\caption[x]{\footnotesize  
Some
combinations of $\gsu(N) \oplus \gsu(M)$ algebras that appear to be in
the ``swampland'' as they
cannot be tuned
on intersecting curves of self-intersection $n, m$ through Tate, but
satisfy anomaly cancellation.
}
\label{t:uu-swamp}
\end{table}

\vspace*{0.05in}

\noindent
${\bf \gsu(N) \oplus \gsp(k)}$ (apparent swamp):

We now consider $\gsu(N) \oplus \gsp(k)$.  From the above analysis, the
anomaly and Tate constraints are identical to the case $\gsu(N)$
$\oplus$  $\gsu(2k)$. When $N$ is even there are no swamp contributions.
When $N$ is odd and $ n > -2$, there are additional potential swamp
  contributions.

\vspace*{0.05in}

\noindent
${\bf  \gg_2 \oplus \gsp(k)}$ (apparent swamp):

Next consider $\gg_2 \oplus \gsp(k)$.  The anomaly constraints dictate
$k \leq 3n + 10, 7 \leq 32 + (16-4k)m$.  The primary constraining Tate
coefficient is again $a_4$.  To have a $\gg_2$ on the $n$-curve, $a_4$
must have a coefficient proportional to $z^2$. This imposes the
constraint $2 \leq 8 + (4-k)m$, equivalent to the second anomaly
constraint.  On the other hand, the Tate constraint on $k$ is $k \leq
8 + 2n$.  For $n = -2$ this is equivalent to the anomaly constraint
($k \leq 4$); for larger $n$, however, the Tate constraint is
stronger.
There is also a further constraint from the condition that $a_6$ must
have a coefficient proportional to $z^3$, or the $\gg_2$ will be
enhanced to $\gso(7)$ or greater.  This condition implies that $2k
\leq 12 + 3n$, which is substantially stronger than the anomaly
conditions.  For $n = -2$, we have $k \leq 3$, for $n = 1$ we have $k
\leq 4$,  for $n = 0$ we have $k \leq 6$, {\it etc.}.  Thus, there is
an apparent tuning swampland of models that contains, for example,
 $\gg_2 \oplus \gsp(4)$ when the
$\gg_2$ is on a $-2$ curve, and has four fundamental hypermultiplets.
This is true in particular when the $\gsp(n)$ factor is on a $-1$
curve, in which case this should be an SCFT, so this represents
another potential contribution to the 6D SCFT swampland.  It is not
clear whether there can or cannot be a Weierstrass model in this case
as Weierstrass and Tate are not necessarily equivalent for $\gsp(4)$,
though as mentioned earlier it seems likely that no non-Tate Weierstrass model
can be realized for this algebra without involving exotic matter.
Further elements of the swampland in this case include $\gg_2 \oplus
\gsp(5)$ through $\gg_2 \oplus \gsp(7)$ when the $\gg_2$ is on a $-1$
curve, $\gg_2 \oplus \gsp(7)$ through $\gg_2 \oplus \gsp(10)$ when the
$\gg_2$ is on a $0$ curve, {\it etc.}.

\vspace*{0.05in}

\noindent
${\bf \gso(N) \oplus \gsp(k)}$ (apparent swamp):

Finally, we turn to the cases $\gso(N) \oplus \gsp(k)$.  These are the
most delicate cases; we consider the small values of $N$ explicitly.
We begin with $\gso(8) \oplus \gsp(k)$.  In this case, the three
different eight-dimensional representations ${\bf 8_v, 8_s, 8_c}$ are
equivalent under anomalies, and physically related through triality
symmetry.  Anomaly constraints limit $k \leq 4 + n, 2 \leq 8 +
(4-k)m$.  
For a Tate tuning the constraints  associated with upper bounds on the
size of the group are, similar to the $\gg_2$ case, $2 \leq 8 +
(4-k)m$, again like the second anomaly constraint, and again the
constraint $k \leq 8 + 2n$ which now matches the first anomaly
constraint.  So it seems that the anomaly and Tate constraints are
consistent.  There is however a subtlety here associated with the
monodromy condition for $\gso(8)$.  This imposes the condition that
the order $z^4$ term in $a_2^2 -4a_4$ must vanish \cite{kmss-Tate}.
In the special case that the $\gsp(k)$ has $k= 1$, and $n = -2$, the
only possible monomial in $a_4$ of order $w$ is $wz^2$.  But this
cannot be part of a perfect square, so the $\gso(8)$ monodromy
condition cannot be satisfied.  Thus, in this case the tuning is not
possible.  Furthermore, in this case even a Weierstrass tuning is not
possible.  This fact was mentioned in \S\ref{sec:tuning-clusters}, and
we elaborate further here.
It was shown in \cite{4D-NHC} that $\gso(8) \oplus \gsu(2)
= \gso(8) \oplus \gsp(1)$ cannot be tuned on any pair of intersecting
divisors where the second factor is realized through a Kodaira type
III or IV singularity; the argument there was given in the context of
threefold bases, but holds for bases of any dimension.  The argument
given there shows in this context that in 6D, $\gso(8) \oplus \gsu(2)$
cannot be tuned on any pair of intersecting divisors where the second
divisor is a $-2$ curve.  It was also
shown in \cite{SCFT-II}, in the context of
6D SCFTs, that an $\gso(8) \oplus \gsu(2)$ cannot be
realized on a pair of intersecting $-3, -2$ curves.  
The upshot of this analysis is that an $\gsu(2)$ on any $-2$ curve
cannot intersect an $\gso(8)$ on any divisor (not just a $-3$
curve)\footnote{Thanks to Clay Cordova for discussions on this
  point}.  This apparent element
of the tuning swampland was shown to be inconsistent at the level of
field theory in \cite{Tachikawa}; actually, the argument there
demonstrated this result only at the superconformal point, but the
same result should hold for a general 6D F-theory supergravity model,
since locally the $-2$ curve can generally be contracted to form an
SCFT.  This is an interesting example of a case where an apparent
element of the tuning swampland is removed by realization of a new
field theory inconsistency.

We turn now to $\gso(7) \oplus \gsp(k)$. The anomaly constraints are
similar but now depend on whether the bi-charged matter is in the
spinor (${\bf 8_s}$) or fundamental ({\bf 7}) representation.  In all
cases, the second constraint $2 \leq 8 + (4-k)m$ agrees between
anomalies and Tate.  The anomaly conditions for spinor matter are $k
\leq 8 + 2n$ and for fundamental matter are $k \leq 3 + n$.
Performing a generic Tate tuning, the bound is $k \leq 8 + 2n$.  This
suggests that the general Tate tuning gives matter must be in the
spinor-fundamental representation.  This matches with the known
examples of the $-2, -3, -2$ non-Higgsable cluster, which carries
spinor matter for an $\gso(7)$ on the $-3$ curve, and the results of
\cite{Tachikawa} that the four $\gsu(2)$ fundamental matter fields on
an $-2$ curve transform in the spinor representation of an $\gso(7)$
flavor group.  That result implies that for an $\gsu(2)$ on a $-2$
curve there cannot be a bifundamental with $\gso(7)$, but a remaining
open question is whether there can be an explicit Weierstrass tuning
of an $\gso(7) \oplus \gsp(k)$ algebra on a more general pair of
divisors with bifundamental matter, and if this is indeed impossible
what the field theory reason is.

The other $\gso(N)\oplus \gsp(k)$ tunings can be analyzed in a similar
fashion, though the analysis is simpler since anomalies show that
spinors cannot appear at the intersection point.  For $\gso(9)$ the
Tate condition computed using the Tate form from Table~\ref{t:Tate}
appears to give the bound $k \leq 6 + n$ from $a_6$.  This cannot be
correct, since the anomaly bound gives $k \leq 5 + n$, which seems to
allow Tate constructions of models that violate anomaly cancellation.
In fact, the maximal case $k = 6 + n$ actually gives $\gso(10)$.  This
can be understood if we carefully impose the proper additional
monodromy condition.  In terms of the Tate form of $\gso(9)$ from
Table~\ref{t:Tate}, the algebra is actually $\gso(10)$ if $(a_3^2 + 4
a_6)/z^4$ is a single monomial when evaluated at 
$z = 0$, and thus a perfect square, which
occurs for the generic Tate form in the current context precisely when
$k = 6 + n$.  So this Tate condition and the corresponding anomaly
condition for $\gso(9)$ match perfectly.  The Tate bound on $k$ from
$m$, $4 \leq 12 + (6-k)m$, is stronger than the anomaly bound, $9 \leq
32 + (16-4k)m$, but both are satisfied for all compatible values of
$m, k$ from \S\ref{sec:classical-tuning}, so there is no swampland.

For $\gso(10)$, we can enforce the monodromy condition that $(a_3^2 +
4a_6)/z^4$ is a perfect square on $z = 0$ by setting $a_6$ to vanish
to order $z^5$ instead of $z^4$.  In this class of tunings,
the anomaly condition $k \leq 6 + n$ matches the $a_3$
Tate condition $k \leq 6 + (3-2)n$, while the $m$ condition $10 \leq
32 + (16-4k)m$ from anomalies is slightly weaker than the Tate
condition $2 \leq 6 + (3-k) m$, leaving in the tuning swampland for
example $\gso(10) \oplus \gsp(k)$ for $k = 8, 9$ when $m = + 1$ (and
necessarily $n \geq 2, 3$).  For $\gso(11)$, the Tate condition
without considering monodromy is $k \leq 8 + n$, which is again weaker
than the anomaly condition $k \leq 7 + n$.  The discrepancy can again
be corrected by the monodromy condition that for Tate $\gso(11)$ as in
table~\ref{t:Tate}, we have $\gso(12)$ when $(a_4^2 -4a_2a_6)/z^6$ is
a perfect square on $z = 0$.
This monodromy condition is stated in \cite{kmss-Tate} for $\gso(4n +
4)$ with $n \geq 3$, but the analysis here indicates that it must also
hold at $n = 2$.  
With this monodromy, the first conditions agree; 
the other conditions, 
$11 \leq 32 + (16-4k)m$ and $2 \leq 8 + (4-k)m$ also agree.
There is an exact matching between anomaly
conditions and Tate conditions for the cases $\gso(12)$.

For tunings of $\gso(N) \oplus \gsp(k)$ where the $\gso(N)$ is on a
$m =-4$ curve, there is no {\it a priori} upper bound on $N$, and the
pattern continues in a similar way as for
$\gso(9)$-$\gso(12)$.  For $\gso(N =4j)$, the $a_4$ Tate bounds $k \leq 8 + (4
-j) m = N-8$, $j \leq 8 + (4-k)m$ precisely match the anomaly bounds.
For $\gso(N = 4j + 2)$, the $a_3$ Tate bound $k \leq 6 + (3-j)m = N
-8$ matches the anomaly bound and $j \leq  6 + (3-k)m \Rightarrow N
\leq 26 + (12-4k)m$ is stronger than the anomaly bound $N \leq 32 +
(16-4k)m$, leaving a small swampland contribution.  Similarly for
$\gso(4j + 1), \gso(4j + 3)$ when the proper monodromy conditions are
incorporated as for $\gso(9), \gso(11)$.
The upshot of this analysis is that $\gso(N) \oplus \gsp(k)$ tunings
have a few swampland contributions, but not many.

\subsection{Multiple curves intersecting $\Sigma$}
\label{sec:generalized-e8}

Having analyzed the combinations of algebras that can be tuned on a
pair of intersecting curves, we can consider the more general class of
local constraints associated with a single curve $\Sigma$ that
supports a gauge algebra $\gg$, and which
intersects $k$ other curves $\Sigma_i, i = 1, \ldots, k$, with each
curve having a fixed self-intersection.  In principle such geometries
can be analyzed using the same methods used in the preceding section
for a pair of curves.  A more general structure relevant for this
analysis is related to the global symmetry of the field theory over
the curve $\Sigma$.  Such global symmetries were recently analyzed in
the context of 6D SCFTs in \cite{global-symmetries}.
From the field theory point of view, the global symmetry can be
determined by the nature of the matter transforming under $\gg$.  For
example, the fundamental representations of $\gsu(N)$ are complex, and
on a curve carrying $M$ such representations, there is a global
symmetry $\gsu(M)$ that rotates the representations among themselves.
In general, the direct sum of the algebras $\gg_i$ supported on the
$k$ curves $\Sigma_i$ that intersect $\Sigma$ must be a subalgebra of
the global algebra of $\Sigma$.  The global symmetries for curves of
negative self intersection were computed in \cite{global-symmetries},
and these are included in the Tables in the Appendix of information
about tunings of groups on curves of fixed self-intersection.  A
similar computation can be carried out for curves of nonnegative self
intersection; indeed, the computations in the preceding section are
closely related to the computation of the global symmetry, though for
the global symmetry the constraint associated with the curve
intersecting the desired curve would be dropped.  Note that in
\cite{global-symmetries}, only global symmetries associated with
generic intersections were incorporated, more generally, for example,
there could be a component of the global symmetry group associated
with antisymmetric representations of $\gsu(N)$, which can appear in
more complicated bi-charged matter configurations \cite{agrt}.
Note also that in considering situations where multiple curves
intersect a given curve $\Sigma$ that supports a gauge algebra $\gg$,
the distinction between different realizations of $\gg$, such as
between type $I_2$ and $III, IV$ realizations of $\gsu(2)$, are
important.  These distinctions are relevant for instance for the cases
in \S\ref{sec:special-cases}, and are
tracked in \cite{global-symmetries}.  A complete analysis of all local
rules for a single curve intersecting multiple other curves would need
to distinguish these cases.

In general,
in the situation where multiple curves $\Sigma_i$ intersect a single
curve $\Sigma$ supporting a gauge algebra $\gg$, there are constraints
on the gauge algebras that can be tuned over the $\Sigma_i$ coming
from the pairwise constraints determined in the preceding subsection,
and a further overall constraint associated with the global symmetry
on $\Sigma$.  Every configuration that satisfies these constraints
automatically satisfies the local anomaly conditions.
It is natural to expect that perhaps all possibilities
compatible with these two conditions can actually be realized in
F-theory.  In principle this could be investigated systematically for
all possible combinations.  We do not do this here, but to illustrate
the point we consider a couple of specific examples; in one case this
hypothesis holds, and in the other case it seems not to and there are
further contributions to the swampland. We leave a detailed analyses
of all the cases for this story to future work.

Consider in particular the case where $\Sigma$ has self-intersection
$n$ and supports a gauge algebra $\gsu(2N)$.  The global symmetry in
this case associated with the $16 + (8- 2 N)n$ matter fields in the
fundamental representation of $\gsu(2N)$ is $\gsu(16 + (8-2 N)n)$.  In
a local model around $\Sigma$ we have the usual Tate expansion.  Let
us ask what gauge groups $\gsu(2M_i)$ can be tuned on the intersecting
divisors $\Sigma_i$.  We take all the $2 N, 2M_i$ to be even for
simplicity; as discussed above in the odd cases there will be some
anomaly-allowed models that cannot be tuned by Tate.  Without imposing
any constraints from the self-intersections of the $\Sigma_i$, anomaly
constraints impose the condition $\sum_{i}M_i \leq N$.  This is also
the condition imposed by the global symmetry.  Looking at the
constraining term $a_4$ in the Tate expansion, we see that as above
${\rm deg} (a_{4 (s)})= 8 + n (4-s)$.  The gauge group on $\Sigma$
indicates that we set to vanish all coefficients in $a_4$ up to $s=
N$.  The leading coefficient is then of degree $8 + n (4-N)$.  To tune
gauge groups $\gsu(2M_i)$ at points $w = w_i$ on $\Sigma$, we must
then take $a_{4 (N)} = \prod_{i}(z-z_i)^{M_i}$.  This can precisely be
done for all sets $\{M_i\}$ that satisfy the condition.  Thus, in this
case all possible tunings are possible that are compatible with
anomaly constraints, which are the same as the tunings obeying the
pairwise and global symmetry constraints.

Another interesting class of cases arises when we consider a $-1$
curve intersecting with two $-4$ curves.  In this case, with a
$\gsp(k)$ on the $-1$ curve, anomaly cancellation and the global
symmetry group suggest that it should be possible to tune $\gso(N),
\gso(M)$ on the two $-4$ curves as long as $ N + M \leq 16 + 4k$.
This does not always seem to be the case, at least with Tate tunings,
even when each intersection is pairwise allowed.  For example, while
$\gso(11) \oplus \gsp(1) \oplus \gso(9)$ and $\gso(11) \oplus \gsp(2)
\oplus \gso(13)$ can be tuned in Tate, $\gso(15) \oplus \gsp(2) \oplus
\gso(9)$ cannot.  Thus, it seems there is a further component of the
tuning swampland associated with cases allowed by the global group and
pairwise intersections that cannot be realized as three-divisor
tunings.

\subsection{No gauge group on $\Sigma$ (generalizing the ``$E_8$ rule'')}
\label{sec:e8-rule}

We can also consider situations where $\Sigma$ carries no gauge group
and intersects a set of other curves $\Sigma_i$.  Although anomaly
cancellation does not give any apparent constraint in such a
situation, F-theory geometries are still constrained.  An example of
this is
the $E_8$ rule that has been mentioned above, which from the SCFT
point of view can be viewed as a
generalization of the above arguments regarding global
symmetries. When a rational curve $E$ is a generic exceptional divisor $E\cdot E = -1$, the
analysis of {\it e.g.} \cite{phase-transitions} establishes that in
the limit in which the curve shrinks to zero size in a non-compact
geometry, the resulting SCFT has a global $E_8$ symmetry. 
Therefore it is
natural to expect that $\gg_1\oplus \gg_2 \subseteq \ge_8$
for gauge algebras on a pair of curves $\Sigma_1, \Sigma_2$ that
intersect $\Sigma$, or  more generally that the sum of algebras over
any set of  curves that intersect $\Sigma$ is contained in $\ge_8$. 
The $E_8$ rule  holds in the case of NHCs, as discussed in
Appendix C of \cite{SCFT-I}; the full set of rules for NHCs that can
intersect a $-1$ curves is given in \cite{clusters}. 

It is natural to conjecture that the $E_8$ rule holds for all
tunings on any set of curves $\Sigma_i$ that intersect a $-1$ curve
that does not support a gauge algebra.  A consequence of this would be
that any tunings over $\Sigma$ and the $\Sigma_i$ that could be
Higgsed to break all gauge factors over $\Sigma$ would also lead to a
configuration that satisfies the $E_8$ rule.  

A slightly stronger version of the $E_8$ rule would be that any
tuning allowed by the $E_8$ rule and anomaly cancellation should be
possible.  
We have used Tate tunings to investigate the validity of the tuned
version of the $E_8$ rule, both in the weaker form and the stronger
form.  It is straightforward to check, given a pair of algebras $\aa, \bb$, what the consequences are of the Tate tuning of these
algebras on a pair of curves intersecting a $-1$ curve $\Sigma$.  In
analogy with the rule  (\ref{eq:avg-2}), from the Zariski
decomposition (or a local toric analysis), it follows that if a $-1$
curve $\Sigma$ intersects other curves $\Sigma_i$ on which $a_n
\in{\cal O} (-n K)$ vanishes to orders $k_i$, then $a_n$ must vanish
to order $k \geq 0$ where
\begin{equation}
k \geq -n + \sum_{i}k_i \,.
\end{equation}
Thus,
for example, if we  try to perform a Tate tuning of $\gsu(5)$ on each
of two divisors $\Sigma_1, \Sigma_2$ that intersect the $-1$
curve $\Sigma$, since for $\gsu(5)$ we have ${\rm ord} (a_1, a_2, a_3,
a_4, a_6) = (0, 1, 2, 3, 5)$, it follows that
${\rm ord}_\Sigma (a_1, a_2, a_3,
a_4, a_6)\geq (0, 0, 1, 2, 4)$, which forces an $\gsu(3)$ on
$\Sigma$.  In fact, even trying to tune $\gsu(5) \oplus \gsu(4)$ leads
to an $\gsu(2)$ on $\Sigma$.  This suggests that the stronger version
of the $E_8$ rule may fail.   As another check on this, we can
consider a form of
Weierstrass analysis.  As discussed previously, the Weierstrass and
Tate formulations are equivalent for tunings of $\gsu(N), N \leq 5$
on a smooth irreducible divisor.   We
can write the general Weierstrass form for $\gsu(5)$ on  such an
irreducible divisor in the form
\cite{kmss-Tate, mt-singularities}
\begin{eqnarray}
f &= & -\frac{1}{3} \Phi^2 +
\frac{1}{2}\phi_0 \psi_2 \sigma^2 +
\tilde{f}  \sigma^3   \label{eq:tw-5}\\
g & = &  -\frac{1}{3} \Phi f
-
\frac{1}{27}  \Phi^3 +\frac{1}{4} \psi_2^2 \sigma^{4} +\tilde{g}\sigma^{5} \,,
\nonumber\\
\Phi & = & \phi_0^2/4 + \tilde{\Phi} \sigma
\end{eqnarray}
and the resulting discriminant is of the form
\begin{equation}
\Delta = \frac{1}{16} ( \tilde{g}  \phi_0^6
-\tilde{f} \psi_0^5 \psi_2 + \phi_0^4 \tilde{\Phi} \psi_2^2) \sigma^{5} +{\cal O}
(\sigma^{6}) \,.
\end{equation}
We can try using this form for Weierstrass to tune an $\gsu(5) \oplus
\gsu(5)$, by considering this as a single $\gsu(5)$ on a reducible
divisor.  We thus consider the discriminant now in terms of a local
Weierstrass analysis on $\Sigma =\{z = 0\}$.  The term $\tilde{f}$
multiplies $\sigma^3$, giving a section of $-4K$.  From a local toric
analysis like those we have been doing, it follows that $\tilde{f}
\sigma^3$ (essentially $a_4$ in the Tate analysis) must vanish at
least to order $z^2$.  Similarly, $\psi_2 \sigma^2$ $(\sim a_3)$ is a
section of $-3K$, which must vanish to at least order $z$, and
$\tilde{g} \sigma^5 (\sim a_6)$ must vanish to order $z^3$.  It
follows that this form of the Weierstrass model has at least a Kodaira
$I_2$ singularity on $\Sigma$ that supports an $\gsu(2)$ when we tune
$\gsu(5) \oplus \gsu(5)$ on a pair of curves intersecting $\Sigma$.
While this is not the most general form of Weierstrass model that
might be considered\footnote{Thanks to David Morrison and Tom Rudelius
  for discussions related to this point, which has been corrected from
  v1 of this paper.}, it illustrates the challenge of
this kind of tuning.

This shows that the tuned version of the $E_8$ rule may fail, in the
sense that there are some configurations that this rule apparently
would accept from the low-energy point of view, which we do not have a
method for realizing
in F-theory.  We can view this as part of the current swampland, assuming that
the justification of the $E_8$ rule from field theory holds for an
arbitrary  $-1$ curve holds, that is that there is always a limit
where the curve shrinks to an SCFT with global symmetry $E_8$.
Similar considerations show that other subgroups of $E_8$ suffer from
similar tuning issues, in particular this occurs for $\gsu(9)$.  
While it is possible, for example, that even though there is no Tate
realization of $\gsu(9)$ next to a $-1$ curve carrying no gauge group
there may be a non-Tate Weierstrass realization of such an $\gsu(9)$,
such a realization seems likely to have exotic matter
or other unusual features so that there is
potentially no standard realization of $\gsu(9)$ with generic matter even
though it is a subalgebra of $\ge_8$.
It
does seem, on the other hand, that all tunings that go beyond the
groups contained in $E_8$ are disallowed, at least at the level of
Tate tunings.  A summary of this analysis is given in
Table~\ref{t:e8-analysis}.

\begin{table}
\begin{center}
\begin{tabular}{| l | l |}
\hline
Algebras $\aa \oplus\bb \subseteq \ge_8$ &
 $\ge_8 \oplus \cdot, \ge_7 \oplus \gsu(2), \ge_6 \oplus \gsu(3),
\gf_4 \oplus \gg_2, \gso(8) \oplus \gso(8),$
\\
that can be tuned in F-theory &
$\gso(8) \oplus \gsu(4),
\gsu(4) \oplus \gsu(4), \gsu(6) \oplus \gsu(2),  $
\\
& $\gsu(5) \oplus \gsu(3),
\gsu(8) \oplus \cdot,
\gso(16) \oplus \cdot$
\\
\hline
Algebras $\aa \oplus\bb \subseteq \ge_8$ &
$\gsu(5) \oplus \gsu(5), \gsu(4) \oplus \gsu(5), \gsu(3) \oplus
\gsu(6),$
\\
with no known F-theory tuning &
$
\gsu(2) \oplus\gsu(7), \gsu(9) \oplus \cdot$
\\
(``$E_8$ (Tate) swamp'')&$ \gso(7) \oplus \gso(9), 
\gsp(2) \oplus \gso(11)$\\ 
\hline
Algebras $\aa \oplus\bb$
not in $\ge_8$ &
$\ge_8 \oplus \gsu(2), \ge_7 \oplus \gsu(3), \ge_6 \oplus \gg_2, \ge_6
\oplus \gsu(4), \gf_4 \oplus \gsu(4)$
\\
that cannot be tuned in F-theory &
$\gsu(4) \oplus \gso(9),
\gso(17) \oplus \cdot$
\\
\hline
\end{tabular}
\end{center}
\caption[x]{\footnotesize Tunings that do and do not satisfy the $E_8$
  rule, which states that any pair of algebras tuned on curves
  intersecting a $-1$ curve not supporting an algebra must have a
  combined algebra that is a subalgebra of $\ge_8$.  Tunings in the
  swamp are only tested using Tate, and may admit Weierstrass
  realizations.  Algebras not in $\ge_8$ that cannot be tuned in
  F-theory are largely checked using both Tate and Weierstrass.  This
  list is not comprehensive but illustrates the general picture.  }
\label{t:e8-analysis}
\end{table}

In regard to the failure of tuning $\gsu(9)$ using Tate on a divisor
intersecting a $-1$ curve, a further comment is in order.
Note that the analysis in \cite{kmss-Tate, mt-singularities,
  agrt} gives a systematic description of all $\gsu(N)$ tunings for $N
< 9$, but that there is as yet no completely systematic description of
  $\gsu(9)$ tunings. In fact,
a similar issue has been encountered in tuning an
$\gsu(9)$ algebra on a divisor to attain a non-generic
triple-antisymmetric matter field at a singular point on the divisor
that would need to have an $\ge_8$ enhancement \cite{mt-singularities,
  agrt}.  It may be
that the unusual way in which $\ge_8$ contains $\gsu(9)$ as a subgroup
may act as some kind of obstacle to F-theory realizations of $\gsu(9)$
in contexts where other subgroups of $\ge_8$ are allowed.

Further investigation of the $E_8$ rule in the context of tunings,
particularly trying
to understand whether any of the subgroups that cannot be realized
using Tate constructions have a Weierstrass construction, either with
generic or exotic matter,
may provide
fruitful insight into the connection of F-theory and low-energy
supergravity theories.

We can also ask about the analogue of the $E_8$ rule for curves of
higher self-intersection.  We consider here the situation of a 0-curve
$\Sigma$
intersecting two or more other curves $\Sigma_i$.  For all the
exceptional groups (including $\gg_2, \gf_4$), it is clear that any
pair of groups can be tuned on a pair of divisors $\Sigma_1,
\Sigma_2$, so for example we can tune $\ge_8 \oplus \ge_8$ on the two
divisors $\Sigma_i$.  There are, however, constraints on what
classical groups can be tuned on the $\Sigma_i$.  Most simply, it is
clear from the Tate construction, since ${\rm deg}\ a_{4 (s)}= 8$, that
using Tate to produce any combination of algebras $\gsu(2M_i)$ is only
possible if $\sum_{i} M_i \leq 8$, so that the total algebra is always
a subalgebra of $\gsu(16)$.  Similarly, an $\ge_8$ tuned on one side
can be combined through Tate with an $\gsu(8)$  on the
other side, but not with $\gsu(N), N > 8$.
We can also tune  $\gso(32)$ on one side, or {\it e.g.} $\ge_8 \oplus
\gso(16)$.
It is tempting to speculate that the consistency condition is related
to the weight lattice being a sublattice of one of the even self-dual
dimension 16 lattices $\ge_8 \oplus \ge_8, {\rm spin}(32)/\Z_2$.
It is also possible that an $\gsu(17)$ algebra may be realizable using
a Weierstrass construction  \cite{Raghuram-17}; in any case, it seems that
the rank of the algebra realized must be 16 or less.

Note that there are some analogues of the $E_8$ rule for $-2$ curves,
as discussed in \S\ref{sec:C-II}, which depend upon details of the
geometry that are not easily understood from the low-energy point of
view.  For example, when a $-2$ curve $\Sigma$ is connected to two
other $-2$ curves, it is not possible to tune an $\gsu(2)$ on one of
them and not on $\Sigma$.  This phenomenon is not currently
understood from the low-energy point of view.

Although one could imagine an extension of the $E_8$ rule and the
corresponding rule for a 0-curve to a
curve of positive self-intersection, the primary constraints on gauge
groups tuned on other divisors intersecting such a curve seem to come
from the connection between the other curves and the remaining
geometry.  
We leave further investigation of such
non field-theoretic constraints for further work.

\subsection{More general intersection structures}
\label{sec:generalized-intersection}

We have focused here on situations where multiple curves intersect a
single curve $\Sigma$ each at a single point.  More complicated
possibilities can arise geometrically.  For example, the curve
$\Sigma$ can intersect itself at one or more points, or acquire a more
complicated singularity.  In such situations, a gauge
group on $\Sigma$ will either require an adjoint representation (when
the tuning is in Tate form on a singular divisor) or a
more exotic ``higher genus'' matter representation when the tuning is
in non-Tate form; such configurations were discussed in
\S\ref{sec:matter}.

Another interesting situation can arise when two curves $\Sigma,
\Sigma'$ intersect at multiple points.  In principle such geometries
can be analyzed using similar methods to those used here.  We point
out, however, one case of particular interest.  If two $-2$ curves
intersect at two points, or more generally if $k$ $-2$ curves
intersect mutually pairwise in a loop, then if we were to be able to
tune an SU($N$) group on each curve there would be bifundamental
matter on each pair of curves, and the shift in Hodge number $h^{2,
  1}(X)$ would be the same for every $N$.  This would appear to give
rise to an infinite family of theories with a finite tuning.  This
example, along with a handful of other similar situations, was
encountered in the low-energy theory in \cite{Schwarz-infinite}, and
shown to be impossible in any supergravity theory with $T < 9$ in
  \cite{finite, KMT-II}.  
Closed loops of this kind are also encountered in the context of
F-theory realizations of little string theories \cite{little}.
From the F-theory point of view the
  possibility of an infinite family of tunings is incompatible with
  the proof of finiteness for Weierstrass models in \cite{KMT-II}.  We
  know, however, that such $-2$ curve configurations are
  possible\footnote{Thanks to Yinan Wang for discussions on this
    point}, for example in certain rational elliptic surfaces that can
  act as F-theory bases.  In fact, these configurations are another
  example of degenerate elliptic curves, like those discussed in
  \S\ref{sec:nonlinear-2}, but in this case associated with the affine
  Dynkin diagram $\hat{A}_{k -1}$.  As in those cases, we expect that
  the tuning of a single modulus will increase $N$ by one, and that
  there is an upper bound on $N$ associated with the maximum value
  such that $-12K-N \Sigma$ is effective.  Note, however, that from
  the low-energy point of view this constraint is not understood.
  This issue is discussed further in \S\ref{sec:conclusions}.

Note that similar closed cycles of curves can occur for alternating
$-4, -1$ sequences, with alternating $SO(2k), Sp(k)$ gauge groups.
These can be thought of as arising from blowing up points between
every pair of $-2$ curves, and again correspond to degenerate genus
one curves with $\Sigma \cdot \Sigma = K \cdot \Sigma = 0$, and the
explanation for finite tuning is again similar.

\section{Tuning exotic matter}
\label{sec:matter}

Up to this point we have focused on classifying the gauge groups that
can be tuned over the various effective divisors in the base through
tuning codimension one singularities in the Weierstrass model.  In
many circumstances, the gauge group content and associated
Green-Schwarz terms, which are fixed by the divisors supporting the gauge group,
uniquely determine the matter content of the theory.  In other cases,
however, there is some freedom in tuning codimension two singularities
that can realize different {\it anomaly-equivalent} matter
representations that can be realized in different F-theory models
associated with distinct Calabi-Yau threefold geometries.  The
existence of anomaly-equivalent matter representations for certain
gauge groups was noted in \cite{Bershadsky-all, Sadov,
  mt-singularities, Grassi-Morrison-2}, and explicit Weierstrass
models for some non-generic matter representations were constructed in
\cite{mt-singularities, ckpt, agrt, Klevers-WT}.
Tuning a non-generic matter representation without changing the gauge
group in general involves passing through a superconformal fixed point
(SCFT) \cite{agrt}.  At the level of the Calabi-Yau threefold, such a
transition leaves the Hodge number $h^{1, 1}(X)$ invariant (since the
rank of the gauge group and the base are unchanged), but generally
decreases $h^{2, 1}(X)$ as generic matter representations are
exchanged for a more exotic singularity associated with non-generic matter.

A systematic approach to classifying  possible exotic matter
representations that may arise was developed in
\cite{kpt, mt-singularities}.  Associated with each representation $R$
of a gauge group $G$ is an integer
\begin{equation}
g_R = \frac{ \lambda}{12}\left(2 \lambda C_{\mathbf{R}} + B_{\mathbf{R}} - A_{\mathbf{R}} \right) \,.
\label{eq:genus-contribution}
\end{equation}
The number $g_R$ corresponds to the arithmetic genus of a singularity
needed in a curve $C$ to support the representation $R$, in all cases
with known Weierstrass realizations; it was argued in
\cite{kpt} that this relationship should hold for all
representations.  For example, antisymmetric $k$-index representations
of $\gsu(N)$ have $g_R = 0$ and can be realized on smooth curves, while
the symmetric $k$-index representation of $\gsu(2)$ has  $g_k =k (k
-1)$, and for $k =  3$ a half-hypermultiplet of this representation
is realized on a triple point singularity in
a base curve having arithmetic genus 3 \cite{Klevers-WT}.

The exotic codimension two tunings
of exotic matter on a single gauge factor that have been shown explicitly to
be possible through construction of Weierstrass models are listed in
Table~\ref{t:exotic-matter}, along with the corresponding shifts in
the Hodge number $h^{2, 1}(X)$.  The change in $h^{2, 1}(X)$ corresponds
to the number of uncharged hypermultiplets that enter into the
corresponding matter transition, {\it {\it i.e.}} to the number of moduli
that must be tuned to effect the transition.  This list may not be
complete; it is possible that other exotic matter representations may
be tuned through appropriate Weierstrass models.  This subject is
currently an active area of research.    Nonetheless,  if
there are one or more other exotic representations possible that are
not listed in the table, for a given tuning of gauge groups from
codimension one singularities on a given base, the generic matter
content is finite.  Each divisor supporting a gauge group has a finite
genus, and there are a finite number of states in each of the generic
representations.  In each specific case, anomalies in principle
constrain the number of possible transitions to exotic matter to a
finite set, so that the evaluation of a superset of the set of
possible codimension two tunings is a finite process; the remaining
uncertainty is whether each of those models with exotic matter not
listed in Table~\ref{t:exotic-matter} has an actual realization in
F-theory.
At the time of writing this is not determined for all
possible matter representations, but further progress on this in some
cases will be reported elsewhere.
Note that certain related exotic multi-charged representations have
also been identified
\cite{agrt}, such as the $({\bf 2}, {\bf 2}, {\bf 2})$
of $\gsu(2)$ , the $({\bf 6}, {\bf 2})$ of
$\gsu(4) \oplus \gsu(2)$, {\it etc.}, which can arise from Higgsing of
the exotic matter listed in Table~\ref{t:exotic-matter}; while not
listed explicitly in the table, such multi-charged exotic matter
should also be considered in possible tunings.

A further issue is that in some cases there are combinations of exotic
and conventional matter multiplets
that appear from low-energy anomaly cancellation considerations to be
possible but that cannot be realized in F-theory.  Thus, certain
transitions that appear to be possible may be obstructed in F-theory.
As an example, at least for the method of constructing Weierstrass
models developed in \cite{agrt}, an $\gsu(8)$ theory with some
{\bf 56} multiplets must also have at least one {\bf 28} multiplet.
This gives another class of situations where the finite enumeration of
tunings gives a superset of the set of allowed Weierstrass models, of
which some may not represent consistent F-theory constructions.

\begin{table}
\begin{center}
\begin{tabular}{| c | c | c | c |c |c |}
\hline
$\gg$ & $R$ & $g_R$ & initial matter & tuned matter & $\Delta h^{2, 1}$\\
\hline
$\gsu(N)$ & ${\bf N (N + 1)/2}$ & 1 &
$ ({\bf N^2 -1})\oplus {\bf 1} $ &
$({\bf N (N + 1)/2}) \oplus ({\bf N (N -1)/2}))$ &-1\\
$\gsu(2)$ & {\bf 4} & 3 & 
$3 \times {\bf 3} \oplus 7 \times {\bf 1} $&$
\frac{1}{2}{\bf 4}  \oplus 7 \times {\bf 2}$ & -7\\
$\gsu(6)$ & {\bf 20} & 0 & ${\bf 15} \oplus {\bf 1}  $&$\frac{1}{2}
{\bf 20} \oplus {\bf 6}  $ & -1\\
$\gsu(7)$ &  {\bf  35} & 0 & $3 \times {\bf 21} \oplus 7 \times {\bf 1}
$&$ {\bf  35} \oplus 5 \times {\bf 7} $ & -7\\
$\gsu(8)$ & {\bf 56} & 0 & $4 \times {\bf 28} \oplus 16 \times {\bf 1}
$&$ {\bf 56} \oplus 9 \times {\bf 8}$ & -16\\
  $\gsp(3)$ & {\bf 14'} & 0 &
${\bf 14} \oplus 2 \times {\bf 1}$&
$\frac{1}{2}{\bf 14'} \oplus\frac{3}{2} {\bf 6}$ & -2\\
$\gsp(4)$ & {\bf 48} & 0 &
$4 \times {\bf 27} \oplus 20 \times {\bf 1} $&$
{\bf 48} \oplus 10 \times {\bf 8}$ & -20\\
\hline
\end{tabular}
\end{center}
\caption[x]{\footnotesize  Known exotic matter representations that
  can be realized by tuning codimension two singularities over a
  divisor in the base, including the tuning of Weierstrass moduli that
  impose singularities in a
  generically smooth divisor.
Half hypermultiplets are indicated explicitly in the transitions; when
the result of the transition is a half-hypermultiplet, the genus given
is that of the half-hyper.
Hodge number shifts are indicated in each case.  Other exotic matter
representations may be possible that have not yet been realized
explicitly through Weierstrass models.
Note that the first example listed, of transitions from an adjoint to
symmetric + antisymmetric matter, has only been explicitly realized so
far in Weierstrass models for SU(3).
}
\label{t:exotic-matter}
\end{table}

\section{Tuning abelian gauge factors}
\label{sec:abelian}

In the analysis so far we have focused on tuning nonabelian gauge
factors, which are determined by the Kodaira singularity types in the
elliptic fibration over each divisor in the base.  Abelian gauge
factors are much more subtle, as they arise from nonlocal structure
that is captured by the Mordell-Weil group of an elliptic fibration
\cite{Morrison-Vafa}.  There has been substantial work in recent years
on abelian factors in F-theory, which we do not attempt to review
here.  While there are still open questions related to abelian
constructions, particularly those of high rank, the general
understanding of these structures has progressed to the point that a
systematic approach can be taken to organizing the tuning of abelian
factors in F-theory models over a generic base.  We describe here how
this can be approached  in the context of the general
tuning framework of this paper, beginning with a single abelian factor
and then considering multiple abelian factors and discrete abelian
groups.

\subsection{Single abelian factors}

A general form for a Weierstrass model with rank one Mordell-Weil
group, corresponding to a single U(1) factor, was described by
Morrison and Park \cite{Morrison-Park}.  Over a generic base, such a
Weierstrass model takes the form
\begin{equation}
 y^2 = x^3+ (e_1e_3- \frac{1}{3}e_2^2 -b^2e_0) x
+ (-e_0e_3^2 +\frac{1}{3}e_1e_2e_3 - \frac{2}{27}e_2^3 + \frac{2}{3}b^2e_0e_2
 -\frac{1}{4}b^2e_1^2) \,.
\label{eq:Abelian-1}
\end{equation}
Here, $b$ is a section of a line bundle ${\cal O} (L)$, where $L$ is
effective, and $e_i$ are sections of line bundles $\mathcal{O}((i - 4)
K+ (i -2) L)$, where $K$ is the canonical class of the base.
This provides a general approach to tuning a Weierstrass model with a
single U(1) factor.  One chooses the divisor class $L$, and then
curves $b, e_i$ in the corresponding classes, which define the
Weierstrass model.

A simple conceptual way of understanding this construction and the
associated spectrum comes from the observation, developed in
\cite{Morrison-Park, mt-sections}
(see also \cite{Klevers-3, Grimm-kk}), 
that when $b
\rightarrow 0$ (\ref{eq:Abelian-1}) becomes the generic form of a
Weierstrass model from the SU(2) Tate form, where $e_3$ is the divisor
supporting the SU(2) gauge group. The divisor class $[e_3] = -K + L$
always has positive genus, since $L$ is effective.  Thus, the
resulting SU(2) group has some adjoint representations.  The U(1)
model (\ref{eq:Abelian-1}) can be found as the Higgsing of the SU(2)
model on an adjoint representation, and has the corresponding
spectrum.  While in some situations the enhanced SU(2) leads to a
singular F-theory model, the spectrum can still be analyzed
consistently from this point of view.

Thus, the generic tuning of a U(1) factor on an arbitrary base can be
carried out by choosing a curve $e_3$ of genus $g > 0$.  From
Table~\ref{t:tunings-g}, and the usual rules of Higgsing an SU(2) to a
U(1), we see that the resulting matter spectrum consists of
\begin{equation}
{\rm generic\  } U(1): \hspace*{0.2in}
{\rm matter} = (6n + 16-16g) ({\bf \pm 1}) + (g -1) ({\bf \pm 2})
\end{equation}
where $n =[e_3] \cdot[e_3]$ is the self-intersection of the curve $e_3$.
Since the Higgsing introduces one additional modulus for each
uncharged scalar, the change in
Hodge numbers for this tuning is $g$ less than that for the SU(2) model:
\begin{equation}
{\rm generic\  } U(1): \hspace*{0.2in}
\Delta (h^{1, 1},h^{2, 1}) = 
(1, -12n+30 (g-1)+ 1)\,.
\end{equation}
As the simplest example, choosing $e_3$ to be a cubic on $\P^2$ with
$n = 9$ gives $g = 1$, and a matter content of 108 fields of charge
$\pm 1$ under the U(1), while the Hodge numbers are (3, 166).  Note
that here the Hodge number $h^{2, 1}(X)$ is determined from the
low-energy theory using the rules of Higgsing and anomaly
cancellation; directly computing the number of independent moduli in
the Weierstrass model (\ref{eq:Abelian-1}) is tricky due to possible
redundancies, and has not yet been carried out in general, to the best
knowledge of the authors.

This approach allows for the tuning of a generic U(1) on an arbitrary
base.  The spectrum will become more complicated when $e_3$ intersects
other divisors that carry gauge groups, and must be analyzed in a
parallel fashion to other intersecting divisors that each carry
nonabelian gauge factors.
A recent systematic analysis of this type, for example, was carried
out in
\cite{Lawrie-sw}.
When the U(1) derives from the Higgsing of an SU(2), however, the
matter follows directly from the Higgsing process and can be tracked
in the low-energy theory.
Note also that the curve $e_3$ can be reducible, in which case the
corresponding SU(2) model will arise on a product of irreducible
factors, with bifundamental matter in place of adjoint matter; such
situations are discussed in some detail in \cite{ckpt}.

The U(1) factors tuned in this way will have only the generic types of
U(1) matter, associated with charges $q =\pm 1, \pm 2$.  It is known,
however, that U(1) models with higher charges such as $q = \pm 3$ can
be found, see {\it {\it e.g.}} \cite{Klevers-3}.  As described in
\cite{Klevers-WT}, such U(1) models can be described as arising from
the Higgsing of SU(2) models with exotic matter, such as in the
three-index symmetric matter representation.  From the point of view
of the unHiggsed nonabelian model, the SU(2) factor on $e_3$ can
acquire exotic matter through a transition such as the one described
in the second line of Table~\ref{t:exotic-matter} where 3 adjoints of
SU(2) and seven neutral fields are exchanged for a half-hypermultiplet
in the {\bf 4} representation and seven fundamentals, with a shift in
$h^{2, 1}$ of -7.  If the resulting theory has at least one further
adjoint, on which the SU(2) is then Higgsed, this gives a U(1) factor
with charge $q = \pm 3$.  By systematically constructing all SU(2)
models with exotic matter we thus should in principle be able to
construct all U(1) models with higher charges.  There are a number of
questions here that are still open for further work, however.  We
summarize the situation briefly.  As discussed in \cite{mt-sections},
when the U(1) is unHiggsed to SU(2) it may introduce (4, 6)
singularities at codimension two or even at codimension one.  Thus, a
systematic analysis of all U(1) models with exotic matter through
Higgsing might require constructing classes of singular SU(2) models.
A more direct approach would be to consider how the SU(2) transition
to exotic matter is inherited in the U(1) theory, where it should
correspond to a direct transition of U(1) models exchanging standard
matter types for $q = \pm 3$ or higher matter.  From the general
analysis of \cite{agrt}, such abelian transitions
should pass through superconformal fixed
points of the theory.  A direct construction of these transitions in
the U(1) theory has not yet been completed.  It is also not known in
principle whether all U(1) models with charge 3 matter can be
constructed in this fashion through Higgsing of models with the same
or higher rank nonabelian symmetry.  Further work is thus needed to
complete the classification of non-generic F-theory models with a
single U(1) factor and higher charged matter over a given base.  One
way to frame this question is in the context of 6D string universality
\cite{universality}; from the low-energy point of view, for a given
gauge group and associated divisor class, we can classify the finite
set of matter representations that are in principle allowed.  
For a theory with abelian factors, some progress was made in
classifying the allowed spectra in \cite{Erler, Park-Taylor, Park-abelian}, but a
complete classification has not been made.
The  open
question is whether in all cases with abelian factors,
all anomaly-allowed matter representations can
be realized by explicit Weierstrass models in F-theory.  
In particular,
for generic
SU(2) models with {\bf 4} or higher matter or the corresponding
Higgsed U(1) models with matter of charges $q = \pm 3$ or higher,
while some F-theory examples have been constructed others have not,
and the universality question is still open.

\subsection{Multiple U(1)'s}

As the rank of the abelian group rises, the algebraic complexity of
explicit constructions increases substantially.  A general rank two
U(1) $\times$ U(1) Weierstrass model was constructed in \cite{ckpt}
(less general $U(1)^2$ models were constructed in
\cite{Mayrhofer:2012zy,Braun:2013yti,Borchmann:2013jwa,Cvetic-Klevers-1,
Cvetic:2013uta,Borchmann:2013hta}).
The general $U(1)^2$ model can be understood in a similar fashion to
the U(1) models just described, in terms of unHiggsing to a nonabelian
model.  In general, there is a divisor class associated with each U(1)
factor; most generally, these divisor classes can be reducible, and
there may be some overlap between the curves supporting the resulting
SU(2) factors, in which case the rank is increased to two over the
common divisor.  Specifically, if we denote by $AC, BC$ the curves on
which the ``horizontal'' U(1) divisors become vertical, with $C$ the
common factor, as described in \cite{ckpt}, the unHiggsed gauge group
will be $SU(2) \times SU(2) \times SU(3)$, with the factors supported
on curves $A, B, C$, each of which may be further reducible in which
case the gauge group acquires multiple factors accordingly.  The
charges of the general U(1) $\times$ U(1) model can then be understood
simply from the Higgsing of the appropriate nonabelian model, in
parallel to the construction described above for a single U(1) factor
from the unHiggsed SU(2) spectrum.  The details become somewhat more
complicated, and we do not go into them explicitly here (see
\cite{ckpt}), but the tuning and full spectra of a generic U(1)
$\times$ U(1) model can in principle be described using this analysis.

As in the case of a single U(1), the generic U(1) $\times$ U(1)
spectrum described here consists of only the most generic types of
charged matter under the two abelian factors.  Note, however, that
this includes matter charges associated with a symmetric
representation of SU(3), which can be realized for example, from the
Higgsing of an SU(3) on a singular curve with double point
singularities and a non-Tate model with a symmetric representation and
at least one additional adjoint, as constructed in \cite{ckpt, agrt}.
For more exotic matter representations, there is as yet no general
understanding, and many of the open questions described above, such as
the explicit realization of abelian matter transitions, and the
existence and Higgsing of appropriate rank two nonabelian models with
exotic matter, are relevant here as well.

Constructing models with more U(1)'s becomes progressively more
difficult.  One class of $U(1)^3$ models was constructed in
\cite{Cvetic-Klevers-2}, though these models do not capture many of the
spectra that could result from Higgsing nonabelian rank 3 models, and
are certainly not general.  The construction of a completely general
abelian model with $U(1)^k$ where $ k > 2$ is still an open problem.
Nonetheless, from the point of view of geometry and field theory a
general approach was outlined in \cite{ckpt} that in principle gives
an approach to the tunings that gives what should be
a superset of the set of allowed
possibilities, in the spirit of this paper.
The idea is that each U(1) factor should come from a divisor $C_i$,
and these divisors can be reducible, with separate components in
principle for each subset of the divisors, generalizing the $AC, BC$
rank two construction described above.  To proceed then, we consider
all possible divisor combinations that can support an unHiggsed rank
$k$ nonabelian group.  The possible rank $k$ abelian group constructions,
and the corresponding charges, can then be determined from Higgsing
the corresponding nonabelian model.  In each case, the specific
spectrum and anomaly cancellation conditions allow us to compute the
potential shift in Hodge numbers.  This gives in principle a finite
list of possibilities that would need to be checked for the existence
of an explicit Weierstrass model.  In practice, the fact that no
generic way is known to implement Higgsing in the Weierstrass context
makes the explicit check impossible with current technology, even for
generic types of matter.  Here we also in principle would need to deal
with exotic matter contents.

Another approach to constructing higher-rank abelian models proceeds
through constructing fibrations with particular special fiber types
that automatically enhance the Mordell-Weil rank, see {\it e.g.}
\cite{Klemm-lry, Klevers-3}.  It is not clear, however, how this
approach can be used in the systematic construction of models,
particularly through the perspective of tuned Weierstrass models as we
have considered here.  Nonetheless, this approach may provide a useful
alternative perspective on the systematic construction of higher-rank
abelian theories.

To summarize, for rank one U(1) models with generic matter, we have a
systematic approach to constructing all tunings.  When considering
more exotic matter or higher rank nonabelian groups, we have a
systematic algorithm for constructing a finite set of possibilities
along with the Hodge numbers of the elliptic Calabi-Yau threefold, but
the technology does not yet exist to explicitly check all
possibilities.  Note that there is also as yet no proof that the
general higher rank Higgsing strategy will give all possible
higher-rank abelian spectra, unlike for rank one and two with generic
matter, where the results of \cite{Morrison-Park, ckpt} represent
general constructions, all of which are compatible with the Higgsing
approach.

\subsection{Discrete abelian gauge factors}

Finally, we describe briefly the possibility of systematic tunings of
discrete abelian gauge factors.  Such discrete factors, and
corresponding matter, have been the subject of substantial recent work
\cite{mt-sections, Mayrhofer:2014opa, Anderson:2014yva,
  Garcia-Etxebarria:2014qua, Mayrhofer:2014haa, Mayrhofer:2014laa,
  Cvetic:2015moa, Grimm:2015ona, Oehlmann:2016wsb}.  As described in \cite{mt-sections},
one systematic way to approach discrete abelian factors is through the
Higgsing of continuous abelian U(1) factors on states of higher
charge.  For example, Higgsing a U(1) theory with matter of charge
$\pm 1$ on a field of charge $+ 2$ gives a theory with a discrete
abelian $\Z_2$ symmetry and matter of charge $1$ under the $\Z_2$.  In
the context of (\ref{eq:Abelian-1}), this Higgsing can be realized by
transforming $b^2$ into a generic section $e_4$ of the line bundle
${\cal O} (2L)$.  This gives an explicit approach to constructing the
simplest class of discrete abelian gauge models, those with $\Z_2$
gauge group and charges $1$.  On a generic base, choosing $e_3$ to be
a curve of genus $g$ and self-intersection $n$, the resulting spectrum
and Hodge shifts should be
\begin{equation}
{\rm generic\ } \Z_2:
{\rm matter} = (6n + 16-16g) ({\bf \pm 1}), \;\;\;\;\;
\Delta (h^{1, 1},h^{2, 1}) = 
(0, -12n+32 (g-1)) \,.
\end{equation} 
A construction of a theory with a discrete abelian $\Z_3$ symmetry
from a U(1) factor with matter of charge $q = \pm 3$ was given in
\cite{Cvetic:2015moa}.
While constructions of models with more complicated groups and/or
matter over generic bases have not been given explicitly in full
generality, we can follow the same approach as used for the general
$U(1)^k$ models to construct a class of potential Hodge numbers and
spectra that should be a superset of the set of allowed F-theory
possibilities.  Basically, for each possible $U(1)^k$ model, we
consider the Higgsings on charged matter that leave a residual
discrete gauge group.  In addition to the caveats discussed above for
the $U(1)^k$ models, there are also the issues that the $U(1)^k$ model
may in principle be singular even if the model with the discrete
symmetry is not, and that in principle there may be allowed models
with discrete symmetries that cannot be lifted to $U(1)^k$ models.
These are all good open questions for further research that would need
to be resolved to complete the classification process in this
direction.

\section{A tuning algorithm}
\label{sec:algorithm}

We now describe a general algorithm that, given any explicit choice of
base $B$, produces a finite list of possibilities for tuned
Weierstrass models.  In the most concrete case of toric bases and
tunings over toric divisors, this algorithm can be carried out in
an explicit way to enumerate and check all possibilities.  More
generally, the algorithm will produce a superset of possible tunings,
for which explicit realizations as Weierstrass models must be
confirmed.  We begin by describing the algorithm in a step-by-step
fashion.  We then summarize the outstanding issues related to this
algorithm.

\subsection{The algorithm}

\noindent
 {\sl i})  {\bf Choose a base}

We begin by picking a complex surface base that supports an
elliptically fibered Calabi-Yau threefold.  As summarized earlier,
from the work of Grassi \cite{Grassi} and the minimal model program,
this surface must be a blow-up of $\P^2$ or a Hirzebruch surface
$\F_m, m\leq 12$. The Enriques surface can also be used as a base, but
the canonical class is trivial up to torsion, so $f, g$ do not seem to
have interesting tunings.
The important data on the base that must be given includes the Mori
cone of effective divisors and the intersection form.
\vspace*{0.05in}

\noindent
{\sl ii})  {\bf   Tune nonabelian groups with generic matter on effective
  divisors of  self-intersection $\leq -1$}

The set of effective irreducible curves of negative self-intersection
forms a connected set.  This can be seen inductively:  The statement
certainly holds for all the Hirzebruch surfaces $\F_m$; for $m > 0$,
$\F_m$ contains a single curve of negative self-intersection, and for
$m = 0$ there are no such curves.  And any point
$p$ in a Hirzebruch surface, or any blow-up thereof, either lies on a
curve of negative self-intersection, or on a fiber of the original
Hirzebruch with self-intersection 0, so blowing up $p$ gives another
base with the desired property. 
(In the latter case, the fiber becomes a $-1$ curve after the
blow-up, which intersects the negative self-intersection curve on the
original $\F_m$.) 

Furthermore, at least one curve of
negative self-intersection  in any base containing such curves
will intersect an effective
curve of self-intersection 0.  This can be seen by taking, for example
the original $-m$ curve on any blow-up of $\F_m$, $m > 0$, and
noting that any base other than $\P^2$ and $F_0$ can be seen as a
blow-up of $\F_m, m > 0$.

Together, these statements and our analysis of \S\ref{sec:C-I},
\S\ref{sec:intersecting-2} are sufficient to prove that in principle
there are a finite number of possible tunings on all curves of
negative self-intersection as long as the Mori cone contains a finite
number of generators.  We can proceed by starting with a negative
self-intersection curve $\Sigma$
that intersects a 0-curve, construct the
finite set of possible tunings over  $\Sigma$, {\it etc.}, and then proceed by
constructing tunings over curves that intersect that curve, etc.,
checking consistency with previous curves at each stage.
This shows in principle that there is a finite algorithm for
constructing all tunings over curves of negative self-intersection
given a finitely generated Mori cone.

Note that
the Mori cone contains a finite number of curves of self-intersection
$-2$ or below.
In practice, we can proceed effectively by
using the results of Section \ref{sec:C-I}, \ref{sec:C-II}
to construct all possible tunings of nonabelian gauge groups on individual
effective divisors and non-Higgsable clusters represented by curves of
self-intersection $-2$ or below as units in the algorithm.  
The connection with $-1$ curves and at least one 0-curve are needed
in principle to bound the infinite families that otherwise
could be
tuned on chains of $-2$ curves or alternating $ -4, -1, -4, -1,
\ldots$ chains, and are also useful in practice to bound the
exponential complexity that would be encountered by independently
tuning the clusters without consideration of their connections.

Note also that in some unusual cases like $dP_9$, the Mori cone has an
infinite number of generators, associated with an infinite number of
$(-1)$-curves.  This algorithm appears inadequate in such cases,
however in all cases that we are aware of of this type, nothing can be
tuned on the infinite family of $-1$ curves due to a low number of
available moduli in $h^{2, 1}$.  This issue is discussed further below.
\vspace*{0.05in}

\noindent
{\sl iii})  {\bf Tune nonabelian groups on the remaining effective divisors}

We now consider tunings on remaining effective curves of non-negative
self-intersection.
We restrict attention to  cases
where the number of generators of the Mori cone is finite, and
there are a
finite number of effective curves with genus below any fixed
bound; we discuss below
situations where the number of Mori generators is infinite.  The
effective curves on which gauge groups can be tuned are generally
quite constrained.  From the analysis of
\S\ref{sec:intersecting-constraint}, no gauge group can be tuned on
any divisor that intersects a curve on which an algebra of $\gf_4$ or
above  is supported.  Thus, the non-negative curves on which we are
allowed to tune are perpendicular to all such curves, in particular
perpendicular to all curves of self-intersection $-5$ or below.  This
acts as a powerful constraint, particularly for bases with large
$h^{1, 1}(B)$, which can only arise in the presence of many curves
that have large non-Higgsable gauge factors.  An example is given in
\S\ref{sec:f12}.  Restricting attention to curves $C$ in the subspace
with $C \cdot D = 0$ for all $D$ of self-intersection $D \cdot D \leq
-5$, the genus grows as $g = 1 + (K + C) \cdot C/2$, which increases
rapidly with the self-intersection of $C$.  Practically, this rapidly
bounds the set of curves on which tunings are possible.  
While we do not give here an explicit algorithm for efficiently
enumerating these curves,
in general, the finiteness of the number of tunings
follows from an argument given in \cite{KMT-II}, which uses the
Hilbert Basis Theorem to show that the number of distinct strata of
tuning in the moduli space of Weierstrass models is finite.  We now
can in principle consider all possible tunings of the divisors that
admit tunings (using Tables~\ref{t:isolated} and~\ref{t:tunings-g}),
and constrain using the rules that govern connected divisors described
in Section \ref{sec:C-III}.  This gives a finite list of possible
nonabelian gauge factors tuned on divisors in the base, which by
construction satisfy the 6D anomaly cancellation conditions.
\vspace*{0.05in}

\noindent
{\sl iv})  {\bf Tune abelian gauge factors}

We can use the methods described in Section \ref{sec:abelian}
to identify the set of possible abelian models and spectra that could
in principle be realized from the Higgsing of additional nonabelian
gauge factors on effective divisors.  

\vspace*{0.05in}

\noindent
{\sl v})  {\bf  Tune exotic matter}

Finally, we can, in any specific case, identify a finite number of possible
tunings to anomaly-equivalent exotic matter content, as described in
Section \ref{sec:matter}.  Table~\ref{t:exotic-matter} gives the set of
possible such transitions that have been explicitly identified in
Weierstrass models.  This gives a finite set of possible matter
contents for a given nonabelian gauge content, though as discussed
earlier not all of these may have Weierstrass realizations.
Analogous transitions can be carried out for abelian factors, either
through the corresponding unHiggsed nonabelian theory, or in principle
directly through abelian matter transitions.
\vspace*{0.1in}

\subsection{Open questions related to the classification algorithm}

Here we summarize places where the algorithm encounters issues that
are not yet resolved.  Each of these is an interesting open research
problem.  Note that for tunings of nonabelian gauge groups with
generic matter over toric divisors in toric bases, there are no
outstanding issues, and the algorithm can in principle be carried out
for all bases and tunings.

\noindent
$\bullet$   {\bf Base issues}

The algorithm described here requires that the cone of effective
divisors on the base have a finite number of generators.  This is not
the case for some special cases of bases such as the 9th del Pezzo
surface $dP_9$.  The algorithm described here would not work for such
bases.  While the algorithm described here can be carried out for any
specific base with a finitely generated cone of effective divisors,
the full program of classifying all elliptic Calabi-Yau threefolds
also requires classifying the set of allowed bases.  In
\cite{wang-non-toric}, all non-toric bases that support elliptic
Calabi-Yau threefolds with $h^{2, 1}\geq 150$ were constructed (these
all have finitely generated Mori cones, so the algorithm here could be
applied without encountering this problem in classifying all tuned
elliptic Calabi-Yau threefolds with $h^{2, 1} \geq 150$).  To continue the algorithm
used there to arbitrarily low Hodge numbers would require resolving
several issues in addition to the finite cone issue.  In particular,
that algorithm used the intersection structure
of classes in the Mori cone.  In
some cases at lower Hodge numbers this does not uniquely fix the
intersection structure of effective divisors on the base; for example, one must distinguish
cases where 3 curves intersect one another in pairs from the case
where all three intersect at a point.
\vspace*{0.05in}

\noindent
$\bullet$ {\bf Apparent infinite families}

As mentioned above, in some bases such as $dP_9$ the Mori cone has an
infinite number of $-1$ curves.
Our algorithm would appear to break down in such situations.  Because
the number of tunings is proven to be finite, however, there cannot be
any tunings on an infinite family of distinct curves.  Thus, it seems
that the finite number of moduli in any given case must limit the
possibilities so that there are nonetheless a finite number of
tunings.  For example, for the base $dP_9$ we have $H =
273-29 \cdot 9 = 12$, giving insufficient moduli to tune even an SU(2)
on a  $-1$ curve, so in fact the number of tunings here is finite even
though the number of $-1$ curves is infinite.

We have also not given a completely rigorous proof and explicit
algorithm for enumerating the set of curves of self-intersection 0 or
above on a base with a finitely generated Mori cone.  While we believe
that this is in principle possible, and in explicit examples seems
straightforward, a more general analysis and explicit algorithm
relevant for non-toric bases would be desirable.

\vspace*{0.05in}

\noindent
$\bullet$ {\bf Explicit Weierstrass tunings}

In the work here we have carried out local analyses that ensure the
existence of Weierstrass models for any of the local tunings over
individual curves or clusters of curves of self-intersection -2 or
below, except some cases of large rank classical groups or complicated
-2 curve structures.  Beyond these cases, we have used anomaly
cancellation conditions to determine a superset of the set of allowed
models for tunings over general local configurations of arbitrary
curves, with Tate models used to produce most allowed constructions in
the case of
intersecting rational curves.  An explicit implementation of the
algorithm would need to confirm the existence of Weierstrass models to
determine which models in the superset admit explicit constructions.
We are not aware of any known exceptions to the existence of
Weierstrass models other than those discussed explicitly here, but we
cannot rule them out for example when considering multiple
intersecting curves supporting nonabelian gauge groups, or gauge
groups on higher genus curves.
\vspace*{0.05in}

\noindent
$\bullet$ {\bf Exotic matter representations}

We have listed in Table~\ref{t:exotic-matter} the set of non-generic
matter representations that have been found identified in the
literature through explicit Weierstrass models.  Even for these matter
representations, it is not clear whether all combinations of fields
that satisfy anomaly cancellation can be realized in F-theory
constructions.  It is also not known whether there are other exotic
matter representations that may admit realizations in F-theory.
We have also assumed that all exotic matter representations can be
realized through a transition from an anomaly-equivalent set of
generic representations; this statement is not proven.
\vspace*{0.05in}

\noindent
$\bullet$ {\bf Codimension two resolutions}

As mentioned in \S\ref{sec:Calabi-Yau}, while our algorithm in
principle could hope to classify the complete finite set of
Weierstrass models over a given base, there is a further challenge in
finding all resolutions of the Weierstrass model to a smooth
elliptic Calabi-Yau threefold.
Despite the recent work on
codimension two resolutions in the F-theory context \cite{mt-singularities, Esole-Yau,
  Lawrie-sn, Hayashi-ls, hlms, Esole-sy, Braun-sn}, there is as yet no
general understanding or systematic procedure for describing  such
resolutions, particularly in the context of the exotic matter
representations just mentioned where the curve in the base
supporting a nontrivial Kodaira singularity is itself singular.    
While the
number of distinct Weierstrass models must be finite by the argument
of \cite{KMT-II}, to the best of our knowledge there is no argument
known that the number of distinct resolutions of codimension two
singularities in a given Weierstrass model is finite \footnote{Thanks
  to D.\ Morrison for discussions on this point}, so a complete
classification of elliptic Calabi-Yau threefolds would require further
progress in this direction.
\vspace*{0.05in}

\noindent
$\bullet$ {\bf Abelian gauge groups}

We have outlined an approach to constructing a superset of the set of
possible abelian models over a given base.  For a single U(1) and
generic (charge 1, 2) matter, this can be done very explicitly using
the Morrison-Park form \cite{Morrison-Park} and Higgsing of SU(2)
models on higher genus curves.  For two U(1) factors and generic
matter this can in principle similarly be done following the analysis
of \cite{ckpt}, though we have not gone through the details of the
possibilities here.  For more U(1) factors, while the approach
described here and in \cite{ckpt} can in principle give a finite set
of abelian models through Higgsing nonabelian models, which should
represent a superset of the set of allowed possibilities, there is no
general construction known of the explicit multiple U(1) models.  For
non-generic U(1) charged matter, again while in principle a finite
list of possibilities compatible with anomaly cancellation can be
made, explicit Weierstrass constructions beyond those of charge 3
matter in \cite{Klevers-WT} are not known.  In principle, exotic
matter transitions could be classified directly in terms of the
abelian spectrum, though this has not yet been done.  For discrete
abelian groups, again in principle a finite set of possibilities can
be constructed by Higgsing the abelian models, but explicit
Weierstrass constructions are not known beyond the generic $\Z_2$
models and some $\Z_3$ models mentioned above.
\vspace*{0.1in}

Most of the complications and issues that arise in confirming the
existence of Weierstrass models for complicated gauge-matter
combinations arise only as the Hodge number $h^{2, 1}(X)$ becomes
small.  None of these issues were relevant in the classification of
Weierstrass models for elliptic Calabi-Yau threefolds with $h^{2,
  1}\geq 350$ in \cite{large-h21}, and we expect that one could go
quite a bit further down in $h^{2, 1}$ before encountering a problem
with the systematic classification that would require substantially
new insights into any of these problems.

We also emphasize that in principle, there is no obstruction to
carrying out this algorithm for arbitrary toric constructions, with
nonabelian gauge tunings only over toric divisors.

\section{Examples}
\label{sec:examples}

In this section we give some examples of applications of the methods
developed and described here.
In each case, the goal is not to be comprehensive, but to illustrate
the utility of the methodology developed in this paper and to suggest
directions for more comprehensive future work.

\subsection{Example: two classes of tuned elliptic fibrations in Kreuzer-Skarke}\label{sec:eg}

The rules that we have established so far must in particular be satisfied by any
Calabi-Yau elliptic fibration over toric surfaces. 
We have a complete set of rules that list
the allowed tunings on isolated toric curves; on multiple-curve NHCs; and on
clusters either neighboring or separated by a $-1$ curve. In each case
we have provided a formula for the shift in Hodge numbers of the
resulting threefold in comparison to a general elliptically fibered
Calabi-Yau over the same base. It is perhaps useful at this point to
see how these rules simplify practical computations.

To this end, we use our rules to explore two classes of tuned
fibrations within the Kreuzer-Skarke database
\cite{ks-data}, which contains all Calabi-Yau threefolds that can be
realized as hypersurfaces in toric varieties associated with reflexive
4D polytopes. In the future, these
rules may be useful to perform a more exhaustive study of all tuned
elliptic fibrations over toric (or more general) surfaces. For purposes
of illustration, we will consider the following simple classes of
fibrations over toric bases as classified in \cite{toric}:
\begin{itemize}
\item tunings of $\ge_6$, $\ge_7$ over $-4$ curves.\footnote{It is likely that a non-rank-enhancing tuning of $\gso(8)$ to $\gf_4$ does not yield a distinct Calabi-Yau but rather merely specializes to a subspace of the original's moduli space.} 
\item tunings of $\gsu(2)$ over $-2$ curves.
\end{itemize}

Proceeding to the first example, whenever a base contains a $-4$
curve, we enhance its generic $\gso(8)$ to $\ge_6$ and $\ge_7$ when
possible. By the $E_8$ rule, this will be possible whenever the
neighboring clusters (separated by $-1$ curves) support at most $\gsu(3)$ or
$\gsu(2)$ algebras, respectively. For instance, the $-4$ curve in the sequence
$(\cdots,-3,-2,-1,-4,-1-2,-3,\cdots)$ will admit enhancement to
$\ge_7$, whereas the $-4$ curve in the sequence
$(\cdots,-1,-3,-1,-4,-1,-3,-1,\cdots)$ will admit an enhancement only
to $\ge_6$, and no enhancement is possible on
\\ $(\cdots,-2,-3,-1,-4,-1,-3,-2,-1,\cdots)$, since the $-3$ curve of a
$(-3,-2)$ NHC supports the algebra $\gg_2$ and $\gg_2\oplus \ge_6
\nsubseteq \ge_8$.

Implementing this program yields the following results, as plotted in the diagram below. There are $1,906$ distinct Hodge numbers of generic fibrations over bases that support these tunings, and $1,562$ distinct Hodge numbers of tuned fibrations over these bases. The diagram is a scatterplot of both sets of Hodge numbers.

\begin{figure}[h]
\includegraphics[scale=0.5]{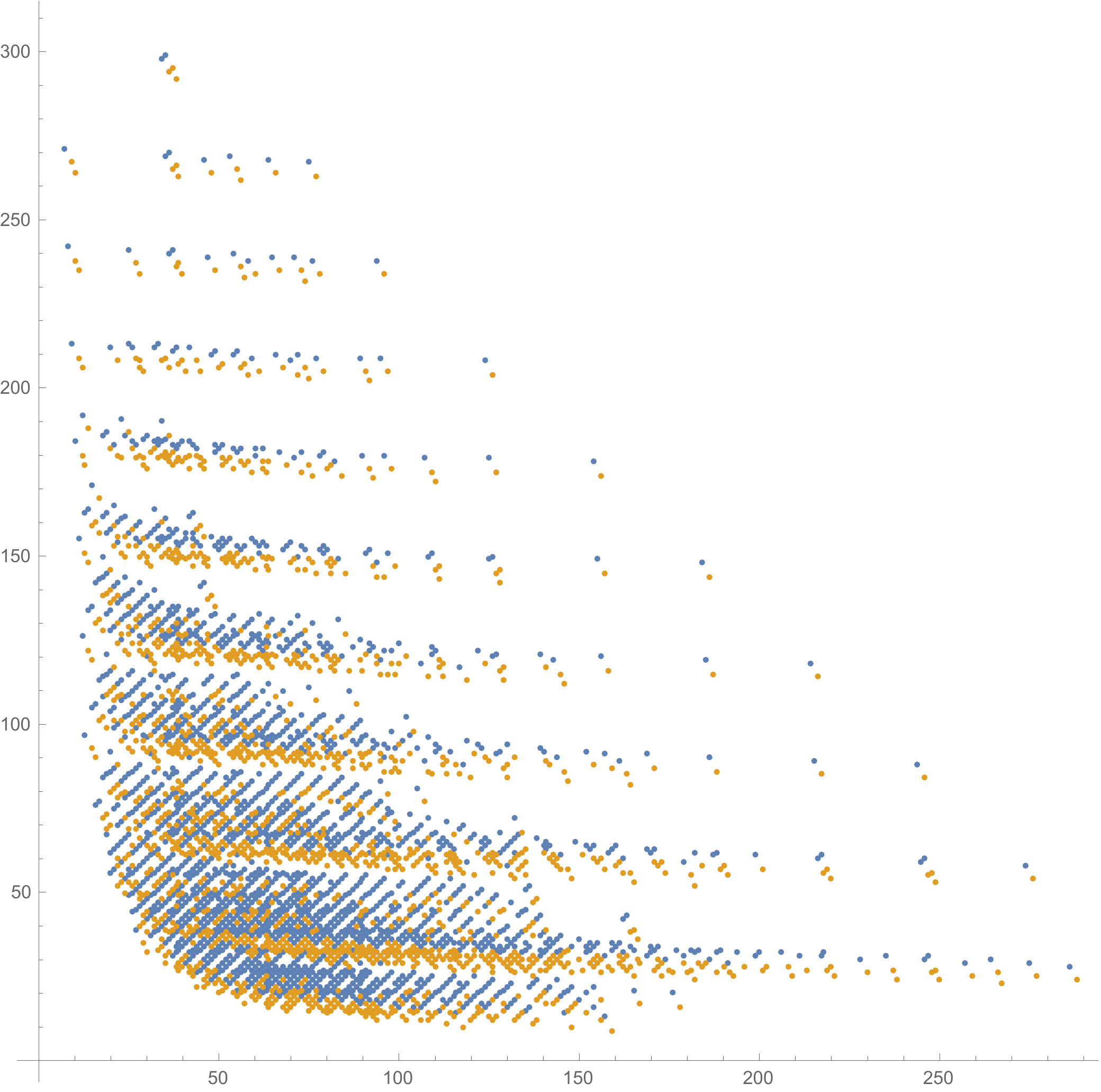}
\caption[x]{[Color online.] Tunings of $\ge_6$ and $\ge_7$ on $-4$
  curves. Blue dots mark Hodge numbers of untuned models over toric
  bases where the $\gso(8)$ gauge symmetry on a $-4$ curve can be
  enhanced to at least an $\ge_6$. Orange dots mark Hodge numbers of
  tuned models over the same bases (without distinguishing between
  $\ge_6$ and $\ge_7$. Here the $y$ axis $h^{2,1}$ is plotted versus the
   $x$ axis $h^{1,1}$.}
\end{figure}

Let us now consider our second example, namely tunings of $\gsu(2)$ on
$-2$ curves in toric bases. As we saw above, chains of $-2$ curves
have simple properties with respect to tunings. For $\gsu(2)$, the
allowed tunings are precisely controlled by the averaging rule: a
$\gsu(2)$ can be tuned on any divisor in a $-2$ chain, but once a
second $\gsu(2)$ is tuned on a different curve, all curves in between
are forced to carry $\gsu(2)$s as well (at least). Therefore, to find all these
tunings, we sweep all toric bases and identify $-2$ chains. For each
base, for each $-2$ chain, we choose a starting and ending point
(which could coincide) for the tuned $\gsu(2)$s. The total set of such
tunings on a given base is found by activating all independent
combinations of such tunings on the different $-2$ chains of the
base. Since we are here interested in a coarse classification of tuned
manifolds by their Hodge numbers, in this case there is a shift by
$\Delta(h^{1,1},h^{2,1}) = (+l,-(l+4))$ for each tuned group of
$\gsu(2)$s of length $l$.

Searching for tunable $-2$ curves, we find $8,517$ distinct Hodge
numbers of bases on which tunings are possible, resulting in $4,537$
distinct tuned Hodge numbers. In this example and in the above, all
Hodge numbers are in the Kreuzer-Skarke database, strongly suggesting
that these tuned elliptic fibrations represent different constructions
of the models in this database. This construction is a rather simple
and direct way to see that at least some
manifolds with these Hodge numbers are
elliptically fibered. The Hodge numbers of generic fibrations over
bases with $-2$ chains, together with the Hodge numbers of resulting
tunings, are graphed in the scatterplot.

\begin{figure}[h]
\includegraphics[scale=0.7]{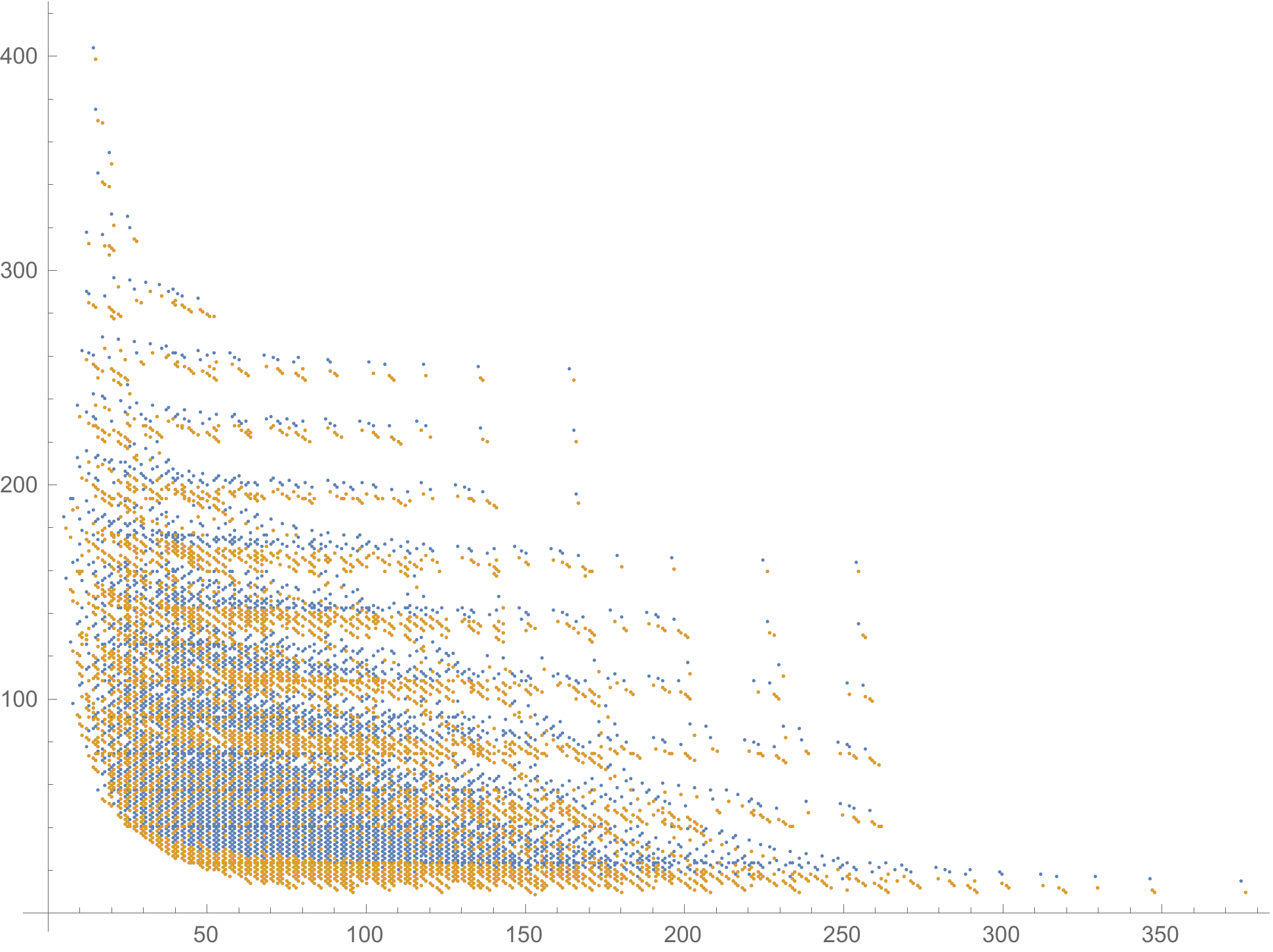}
\caption[x]{[Color online.] Tunings of $\gsu(2)$ on (chains of) $-2$
  curves. As for the previous example, blue dots mark Hodge numbers of
  untuned models over toric bases with $-2$ curves that can support a
  $\gsu(2)$ gauge symmetry. Orange dots mark Hodge numbers of tuned
  models over the same bases. Here the  $y$ axis $h^{2,1}$ is plotted
  versus the $x$ axis $h^{1,1}$.}
\end{figure}

\subsection{Example: tunings on $\F_{12}$}
\label{sec:f12}

As an example of the general tuning algorithm, we consider 
tuned Weierstrass models over the toric base $\F_{12}$, the twelfth
Hirzebruch surface.  This provides an illustration both of the general
principles of tunings and of some specific issues of interest.  We do
not strive for completeness in this listing, but give a broad class of
tunings to illustrate the ideas and methodology involved.

The Hirzebruch surface $\F_{12}$ has a cone of effective divisors
generated by the curves $S$ and $F$, where $S\cdot S= -12, S\cdot F =
0, F \cdot F = 0$.  The toric divisors are $S, F, \tilde{S}, F$ where
$\tilde{S}= S+ 12F$ has self-intersection $\tilde{S}\cdot \tilde{S}= +
12$.  The curve $S$ is a non-Higgsable cluster supporting a gauge
group $E_8$.   No curve intersecting $S$ can carry a gauge group,
since this would produce matter charged under the $E_8$ and a (4, 6)
singularity.  Thus, the only curves on which we can tune nonabelian
gauge factors are multiples of $\tilde{S}$, $k \tilde{S}$.
This simplifies the classification significantly.
The generic elliptic fibration over $\F_{12}$ has Hodge numbers (11,
491) \cite{WT-Hodge}, so this is our starting point.

\begin{table}
\begin{center}
\begin{tabular}{| l| c | c | c |}
\hline
$G$ & curve(s) & matter & $(h^{1, 1}(X),h^{2, 1} (X)) $\\
\hline
$\cdot$& $\cdot$ & $\cdot$ & (11, 491)\\
\hline
$\gsu(2)$ \hspace*{0.05in} (T) & $\tilde{S}$ & 88 $\times\ {\bf 2}$ & (12, 318)\\
$\gsu(3)$ \hspace*{0.05in} (T) & $\tilde{S}$  & 90 $\times\ {\bf 3}$  & (13, 229)\\
$\gsu(4)$ \hspace*{0.05in} (T) & $\tilde{S}$  & 64 $\times\ {\bf 4}$+ 14 $\times\ {\bf 6}$ & (14,
166)\\
$\gsu(5)$ \hspace*{0.05in} (T) & $\tilde{S}$  & 52 $\times\ {\bf 5}$+ 14 $\times\ {\bf  10}$ & (15,
115)\\
$\gsu(6)$ \hspace*{0.05in} (T, W$_{(r > 0)}$) & $\tilde{S}$  &   $(40 + r)$ $\times\ {\bf  6}$+ 
$(14-r)$ $\times
{\bf  15}$ & (16,
$76-r$)\\
$\gsu(7)$
\hspace*{0.05in} (T, W$_{(r > 0)}$) & $\tilde{S}$  &  $(28 + 5r)$ $\times\ {\bf 7}$+ 
$(14-3r)$ $\times\ {\bf  21}$ &
(17, $49-7r$)\\
$\gsu(8)$
\hspace*{0.05in} (H) & $\tilde{S}$  &   25 $\times\ {\bf  8}$+ 10 $\times\ {\bf
  28}$
+ 1 $\times\ {\bf 56}$ & (18, 18)
\\
$\gso(8)$ \hspace*{0.05in} (T) & $\tilde{S}$ &
16 $\times\ ({\bf 8}_v + {\bf 8}_s+ {\bf 8}_c)$ & (15, 135)\\
$\gso(10)$ \hspace*{0.05in} (T) & $\tilde{S}$ &
18 $\times\  {\bf 10}$
+ 16 $\times\ {\bf 16}$ & (16, 101)\\
$\gso(12)$ \hspace*{0.05in} (T) & $\tilde{S}$ &
20 $\times\  {\bf 12}$ + 8 $\times\ {\bf 32}$ & (17, 51)\\
$\ge_6$ \hspace*{0.05in} (T) & $\tilde{S}$ & 18 $\times\ {\bf 27}$ & (17, 83)\\
$\ge_7$ \hspace*{0.05in} (T) & $\tilde{S}$ & 10 $\times\ {\bf  56}$ & (18, 64)\\
\hline
$\gsu(2)$  $\times$ $\gsu(2)$ & 
$\tilde{S}$, $\tilde{S}$ & 64 $\times\ {\bf (1, 2)}$ 
+ 64 $\times\ {\bf (2,1 )}$ 
+ 12 $\times\ {\bf (2, 2)}$
& (13, 193)\\
$\cdots$ & $\cdots$ &$\cdots$ & $\cdots$ \\
$\gsu(4)$  $\times$ $\gsu(4)$  (?)& 
$\tilde{S}$, $\tilde{S}$ & 16 $\times\ {\bf (1, 4)}$ 
+ 16 $\times\ {\bf (4,1 )}$ 
+ 12 $\times\ {\bf (4, 4)}$
& (17, 3)\\
& & + 14 $\times\ {\bf (6, 1)}$ + 14 $\times\ {\bf (1, 6)}$ & \\
\hline
$\gsu(2)$  & $2 \tilde{S}$ & 128 $\times\ {\bf 2}$ +11 $\times\ {\bf 3}$ & (12, 205)\\
$\gsu(3)$  & $2 \tilde{S}$ & 118 $\times\ {\bf 3}$ +11 $\times\ {\bf 8}$ &
(13, 87)\\
$\gsu(4)$ (?)  & $2 \tilde{S}$ & 32 $\times\ {\bf 4}$ +
28 $\times\ {\bf 6}$ +11 $\times\ {\bf  15}$ & (14, 45)\\
$\gsu(2)$  & $3 \tilde{S}$ & 120 $\times\ {\bf 2}$ +34 $\times\ {\bf 3}$ & (12, 152)\\
$\gsu(3)$  (?) & $3 \tilde{S}$ &  54 $\times\ {\bf 3}$ + 34 $\times\ {\bf 8}$ &
(13, 65)\\
$\gsu(2)$ (?) & $4 \tilde{S}$ & 64 $\times\ {\bf 2}$ +69 $\times\ {\bf 3}$ &
(12,  159)\\
\hline
U(1)\hspace*{0.05in} (W) & [$2 \tilde{S}$] & 128 $\times$  $({\bf \pm
  1})$ +  10 $\times$   $({\bf \pm 2})$ &
(12, 216)\\
U(1)\hspace*{0.05in} (W) & [$3 \tilde{S}$] &  120 $\times$  $({\bf \pm
  1})$ +  33 $\times$   $({\bf \pm 2})$ &
(12,  186)\\
\hline
$\Z_2$\hspace*{0.1in} (W) & [$2 \tilde{S}$] & 256 $\times$ {\bf 1}  &
(11, 235)\\
$\Z_2$\hspace*{0.1in} (W) & [$3 \tilde{S}$] & 240 $\times$ {\bf 1}  &
(11, 251)\\
\hline
\end{tabular}
\end{center}
\caption[x]{\footnotesize Some tunings of elliptic Calabi-Yau
  threefolds and corresponding F-theory models
on the base $B =\F_{12}$.  Gauge group
  described is tuned on a multiple of $\tilde{S}$, additional
  non-Higgsable factor of $E_8$ from $S$ is dropped.  Curve for U(1)
  models is corresponding value of $[e_3]$ supporting associated
  unHiggsed SU(2).  Tunings with rank-equivalent descriptions ({\it
    e.g.} $\gg_2$) not included in list.  All tunings listed satisfy
  anomaly conditions and have no known inconsistency with F-theory.
  Those denoted by (T, W, H) have explicit tunings using Tate,
  Weierstrass, or heterotic descriptions.  Those denoted by (?) not
  only have no explicit F-theory description but are particularly
  suspicious as possible swampland residents.  Models without a
  denotation can all be understood as Higgsings of other models with
  known Tate or Weierstrass realizations.  }
\label{t:f12-tunings}
\end{table}

First, we consider tuning single nonabelian gauge factors on
$\tilde{S}$.  This curve has self-intersection $n = + 12$ and genus $g
= 0$.  Since $f, g$ can be described torically as functions in a local
coordinate $z$ of degrees $4, 7$, where $\tilde{S}=\{z = 0\}$, with at
least one monomial at each order, we can immediately tune the gauge
group types $\gsu (2), \ge_7$ that are enforced simply by the order of
vanishing of $f, g$.  The resulting tunings are tabulated in
Table~\ref{t:f12-tunings}.  It is also straightforward to confirm from
a direct toric monomial analysis that we can tune the orders of
vanishing of $f, g$ with a proper choice of monodromy conditions to
get any of the other gauge groups with algebras $\gsu (3),\gg_2, \gso
(7),\gso (8),\gf_4,\ge_6$.
This illustrates the general methods of \S\ref{sec:general-curves}
in a specific context.

The classical groups with algebras $\gsu(N), \gsp(N), \gso(N)$ must be
considered separately.  We focus attention particularly on the SU($N$)
groups.  For $\gsu(N), N = 6, 7$ there are exotic matter contents that
can be tuned using Weierstrass as described in
\S\ref{sec:matter} \cite{agrt}; these are
included in the Table.  For SU($N$) and Sp($k$), from the matter
spectrum in Table~\ref{t:isolated} it is clear that at $N = 2k = 10$
there is a problem as the number of fundamental representations
becomes negative.  In fact, a Tate analysis indicates that tuning
$\gsu(8)$ eliminates the single monomial of $a_6$ that has order $\leq
5$ on the $-12$ curve, forcing a non-minimal singularity there.  This
is an example of the constraint discussed in \S\ref{sec:e8-rule} that
a 0-curve with an $\ge_8$ on one side cannot have an $\gsu(N)$ with $N
\geq 8$ on the other side tuned using Tate.  This is an example of a
tuning in the swampland, which looks consistent from the low-energy
point of view but may not be allowed in F-theory. 
Note that the
groups that we can tune in this way  are subgroups of $E_8$,
matching with the expectation that any tuning on this base should have
a heterotic dual, with the resulting gauge group realized from an
$E_8$ bundle over K3 with instanton number 24.  An explicit
construction was given in \cite{agrt} for non-Tate Weierstrass tunings
of $\gsu(8)$ with $r$ matter fields in the triple-antisymmetric ({\bf 56})
representation; there it was argued that the only heterotic dual to a
tuning on the $+ 12$ curve of $\F_{12}$ has $r = 1$.  The other
Weierstrass tunings thus must have singularities\footnote{Thanks to
  N.\ Raghuram for confirming this in the case $r = 0$.} when restricted to
the compact base $\F_{12}$.  
From the low-energy point of view it seems quite obscure why an $E_8
\times$
SU($8$) theory constructed in this way would be inconsistent
with $r = 0$ but consistent with $r = 1$.
Understanding this kind of question
better and in more generality is an interesting area for further
research.

Since $\tilde{S}$ is a non-rigid divisor ($n\geq 0$), we can tune
multiple independent gauge factors on different curves $\Sigma_i$ in
this divisor class, which will then intersect one another with
$\Sigma_i \cdot \Sigma_j = 12$.  For example, tuning two SU(2) factors
on such curves gives a model with gauge group $(E_8 \times)\ SU(2)
\times SU(2)$, where there are 12 bifundamental fields in the ${\bf
  (2, 2)}$ representation.  From the spectrum and anomaly constraints
we would expect to be able to tune various product groups.  It is
fairly straightforward to see using Tate and the toric picture that it
should be possible to tune any product of groups SU($N_1$) $\times$
$\cdots \times $ SU($N_k$) with $\sum_{i} N_i < 8$, though we do not
  pursue the details here.  Cases with $\sum_{i} N_i = 8$ are less
  clear and would be interesting to explore in more detail.  Note that
  these product group configurations can be understood from Higgsing
  on $k$-index antisymmetric representations of the $\gsu(N)$ models
  on $\tilde{S}$ (see, {\it e.g.}, \cite{agrt} for an explicit
  discussion of several examples of such Higgsing in Weierstrass and
  dual heterotic descriptions). For example, Higgsing $\gsu(6)$ on a
  two-index antisymmetric ({\bf 15}) representation gives a model with
  gauge algebra $\gsu(2)\oplus \gsu(4)$, while Higgsing on a
  three-index antisymmetric ({\bf 20}) gives a model with gauge
  algebra $\gsu(3) \oplus \gsu(4)$.  The $\gsu(4) \oplus \gsu(4)$
  model that is allowed by anomalies would in principle come from an
  anomaly-allowed model with an $\gsu(8)$ gauge algebra and a 4-index
  antisymmetric matter field, but constructing a Weierstrass model for
  such configurations seems problematic \cite{agrt}.
Any model that can be realized in field theory from the Higgsing of a
model with a good F-theory description must also have a good F-theory
description, even if the explicit Weierstrass construction is
difficult to identify.  Thus, the product groups with $\sum_{i} N_i <
  8$ are also naturally understood as Higgsings of the models with a
  single $\gsu(N)$ algebra on $\tilde{S}$.

Now let us consider tunings on the curve $\Sigma = 2 \tilde{S}$.  This
curve has self-intersection $n =48$, and from $-K \cdot \Sigma = 2 (-K
\cdot \Sigma) = 28$, we have $g = 11$.  From Table~\ref{t:tunings-g}, we
see that anomalies suggest that
 we can tune such models for $SU(2), SU(3),$ and $SU(4)$.
Roughly, we are tuning on a curve described by a quadratic in the
coordinate $z$ that vanishes on $\tilde{S}$, so as for $\gsu(8)$ on
$\tilde{S}$, the $\gsu(4)$ algebra on $2 \tilde{S}$ may be problematic.
The $\gsu(2)$ and $\gsu(3)$ models can be understood in field theory
from the Higgsing of the $\gsu(2) \oplus \gsu(2)$ and $\gsu(3) \oplus
\gsu(3)$ models described above on a bifundamental field \cite{agrt},
and thus must exist in F-theory.  The $\gsu(4)$ model on $2 \tilde{S}$
is also unclear from this point of view.

Similarly, we can consider tuning SU(2) and SU(3) models on $3
\tilde{S}$ ($n = 108, g = 34$) and SU(2) on $4 \tilde{S}$ ($n = 192, g
= 69$).  We have not attempted to explicitly construct Weierstrass
models for these theories; the first of these should arise from
Higgsing of an $\gsu(2)\oplus \gsu(2) \oplus \gsu(2)$ model on
$\tilde{S}+ \tilde{S}+ \tilde{S}$, which should in turn come from a
Higgsing of $\gsu(6)$ on $\tilde{S}$, while the others are likely
problematic though acceptable from anomaly cancellation.

From these realizations of SU(2) and other nonabelian groups on higher
genus curves we can also proceed to implement the construction of U(1)
models as discussed in \S\ref{sec:abelian}.  As discussed in the main
body of the text, U(1) models and their spectra can be realized from
Higgsings of nonabelian models over the same base (which can be
allowed to have some singularities that are removed in the Higgsing).
In the simplest cases, we achieve the U(1) model by Higgsing an SU(2)
with generic matter; these can be explicitly realized using the
Morrison-Park form \cite{Morrison-Park}.  In this case, we can consider
the SU(2) realizations on $k \tilde{S}$ for $k = 2, 3$.  These give
U(1) theories with various spectra, as listed in the table.  Many
further U(1) models with explicit Weierstrass models could be
constructed with $q = 3$ charges through matter transitions in the
unHiggsed SU(2) theory to SU(2) models with {\bf 4} matter, as
discussed in \cite{Klevers-WT}.  and the superset of all such models
could be constructed by transferring an arbitrary number of groups of
3 adjoints into {\bf 4}'s, though Weierstrass models for all these are not
known.  Furthermore, many U(1) $\times$ U(1) models with generic matter
spectra could be constructed using the methods of \cite{ckpt}, and the
hypothetical superset of all $U(1)^k$ models may be constructable at
the level of spectra by considering all Higgsings of nonabelian models
including those constructed above, though we have neither a method for
explicitly constructing Weierstrass models in these cases, nor a proof
that this exhausts all possibilities for $k > 2$.  We do not explore
these considerations further here.

We conclude by constructing the model with discrete $\Z_2$ gauge group
 with what seems to be the largest value of $h^{2, 1}$.  This follows
 from taking the $k = 2$ generic U(1) model above and Higgsing on a
 field of charge 2, following \cite{mt-sections}.  Many more models
 with discrete gauge groups and various charges could be constructed,
 some explicitly in Weierstrass, and a larger superset by considering
 all Higgsings on non-unit charges of abelian $U(1)^k$ models.

\section{Conclusions and Outlook}\label{sec:conclusions}

In this paper we have reported on progress towards a complete
description of the set of Weierstrass tunings over a given complex
surface $B$ that supports elliptic Calabi-Yau threefolds.  These
Weierstrass tunings can be used to classify elliptic Calabi-Yau
threefolds and to study F-theory supergravity and SCFT models in six
dimensions.  In particular, for a given base $B$ the results
accumulated here give a set of constraints on the set of possible
tunings over $B$, which give a finite superset of the finite set of
consistent tunings.  While we have not completely solved the tuning
classification problem, we have framed the structure of the problem,
developed many of the components needed for a full solution, and
identified a few remaining components that need a more complete
analysis for a full understanding.

The tools developed in this work can be used in a
number of ways, including generating examples of elliptic Calabi-Yau
threefolds and F-theory models with particular features of interest,
the classification of elliptic Calabi-Yau threefolds and corresponding
6D supergravity theories, and exploration of the ``swampland'' of 6D
theories that seem consistent but cannot be realized explicitly in
F-theory.  In this concluding section we summarize the specific
results of this paper as well as a set of further issues to be
addressed, and we discuss the implications for the 6D swampland and
the potential extension of this kind of analysis to 4D F-theory
models.

\subsection{Summary of results}

This
progress extends previous work in the following ways:

\begin{itemize}
\item We have completely classified local tunings of arbitrary gauge
  groups with generic matter on a single rational
  curve, and shown in all cases except $\gsu(N)$ 
that
  the tunings allowed by anomaly cancellation can be realized in
  explicit local Weierstrass/Tate models;  for $\gsu(N)$ we
  have found Tate models for almost all cases, with a few exceptions
  at large odd $N$ and divisors of positive self-intersection
  for which Tate models are impossible and no Weierstrass models are known.
\item We have completely classified local tunings in the same way over
  all multiple-curve non-Higgsable clusters.
\item We have classified allowed local tunings on a pair of intersecting
  rational curves, and shown that a large fraction of anomaly-allowed tunings
  can be realized by Tate or Weierstrass models, but we have also
  identified quite a few exceptions.
\item We have identified some specific configurations, such as
  $\gsu(10) \oplus \gsu(3)$ and $\gg_2 \oplus \gsp(4)$ on a pair of
  intersecting curves of self-intersections $-2$ and $-1$, that are
  allowed by anomalies but cannot be realized through a Tate
  construction in F-theory.  These represent a component of the
  ``swampland'' both for supergravity theories and for SCFTs, and must
  be explained if the assertion of F-theory universality for 6D SCFT's
  is to be proven.  We have found a substantially larger number of
  configurations in the supergravity swampland that do not have
  low-energy field theory descriptions through an SCFT with gravity
  decoupled as they involve curves of nonnegative self-intersection.
\item We have identified extremal configurations of $-2$ curves,
  associated with degenerate elliptic curves satisfying $\Sigma \cdot
  \Sigma = -K \cdot \Sigma = 0$, as loci that in F-theory admit a finite
  number of $\gsu(N)$ tunings though low-energy consistency does not
  constrain $N$ in any known way.
\item
We have investigated the validity of the ``$E_8$'' rule \cite{SCFT-I},
which constrains the gauge factors that can be tuned on divisors that
intersect a $-1$ curve carrying no gauge group,
in the context of tuning enhanced gauge symmetries.  We found that the
$E_8$ rule is a necessary  condition for tunings to
be possible, but may not be a sufficient condition as some
combinations of gauge groups that live within $E_8$ cannot be realized
using Tate.  We also identified an analogous rule for curves of
self-intersection 0.
\item Combining the preceding results gives a complete set of tools
  that can in principle produce the finite set of all possible
 tunings over toric curves in toric bases. 
Work in this direction is in progress \cite{Huang-WT}.
In the toric case, each
 prospective tuning can be checked for global consistency in a global
 Weierstrass model using toric methods.
\item These tools, in the context of 6D SCFT's, give
a systematic description of tunings of an SCFT in terms of a
Weierstrass or Tate model on the set of
  contracted curves, complementing the analysis of
  \cite{SCFT-II}. In particular, this work goes beyond that reported
  in
\cite{SCFT-II}  in that we systematically construct explicit Weierstrass models for the
  configurations allowed by anomaly constraints, and identify some new
  configurations that do
  not admit Tate tunings and do not have known or straightforward
  Weierstrass models, yet which satisfy low-energy consistency
  conditions.
These tunings can also be applied in the closely related
context of F-theory
realizations of little string theories \cite{little}.
\item We have used anomaly cancellation to classify the set
  of possible tunings over curves of arbitrary genus that are
  acceptable from low-energy considerations.
\item We have computed explicitly the Hodge number shifts for the
  elliptic Calabi-Yau threefold for
all the preceding tunings.
\item We have provided geometric proofs of strong constraints on local
  combinations of allowed tunings, matching constraints from anomaly
  considerations.  In particular, we have shown that the only possible
  pairs of gauge group factors that can arise on intersecting
  divisors, {\it and hence the only combinations of gauge groups that
    can share matter in any low-energy theory arising from F-theory,}
  have one of the five combinations of algebras listed in
  Table~\ref{t:intersecting-possibilities} (or arise as a product
  subgroup of one of the allowed realized individual
or product groups after an
  appropriate breaking).
\item We have given a general procedure for classifying allowed
  tunings of non-generic matter and abelian gauge fields, which will
  give a finite set of tunings allowed over any given base, and which should
  be a superset of the complete set that can be explicitly realized in
  F-theory.
\end{itemize}

\subsection{The 6D ${\cal N} = 1$ ``tuning'' swampland}
\label{sec:tuning-swamp}

In general, one of the goals of this work is to narrow down the
``swampland'' of models that seem consistent from low-energy
considerations but that lack UV descriptions in string/F-theory
\cite{universality}.  For 6D supergravity models, this problem can be
broken into two parts: first, the matching of completely Higgsed 6D
supergravity models to F-theory constructions, and second the matching
of all possible gauge enhancements through tuning/unHiggsing in the
F-theory and supergravity models.  There are still substantial
outstanding questions related to the first part; in particular, we do
not have a proof that a low-energy model with a BPS dyonic string of
Dirac self-charge -3 or below implies the presence of a non-Higgsable
gauge field, while F-theory implies this condition.  In this paper we
address the second part of the question: given a completely Higgsed 6D
supergravity theory with an F-theory realization we ask whether all
possible unHiggsings of the 6D SUGRA theory that are consistent with
anomaly cancellation can be realized as tunings of the corresponding
F-theory model.  By comparing field theory and F-theory geometric
analysis of various local combinations of gauge groups over different
curve types, we have shown that in almost all cases, F-theory
reproduces precisely the set of gauge groups and matter through
tunings that are allowed by anomaly cancellation conditions and other
low-energy consistency constraints.  We have also, however, identified
some situations where field theory and F-theory are not in agreement,
or seem from our current understanding not to be in agreement.  We
list these here.
\vspace*{0.05in}

\noindent {\bf Tunings on a single divisor} For local tunings of
generic matter types over a single rational curve, we found that
virtually everything that is allowed by anomaly cancellation has an
explicit Tate or Weierstrass realization.  The only class of
exceptions were the tunings of large-rank $\gsu(N)$ algebras listed in
Table~\ref{t:su}.  For those cases, Tate models are not possible.  In
some examples such as $\gsu(21)$ and $\gsu(23)$ on a $+ 1$ curve, a
straightforward approach to Weierstrass models also fails; although we
have not proven rigorously that a Weierstrass realization is
impossible this seems likely to be true as other known non-Tate
Weierstrass models realize exotic matter.  These swampland examples
may have a low-energy inconsistency, may be realizable through exotic
Weierstrass models or may be stuck in the swampland. The large
$\gsu(N)$ tunings constitute the complete set of single-divisor
swampland examples encountered in this work.
\vspace*{0.05in}

\noindent {\bf Tunings on a pair of divisors}
For local tunings of generic matter over a pair of rational curves
that intersect at a single point, we found a larger class of instances
of models that are allowed through anomalies but not through Tate
constructions.  In addition to a couple of known examples such as
$\gso(8) \oplus \gsu(2)$ when the $\gsu(2)$
is on a $-2$ curves (which is
known to have field theory inconsistencies \cite{Tachikawa}), we found
other examples of algebras $\gsu(N) \oplus \gsp(k)$, $\gsu(N) \oplus
\gsu(M)$, $\gg_2 \oplus \gsp(k)$, and $\gso(N) \oplus \gsp(k)$ that
are acceptable according to anomaly cancellation but do not have Tate
realizations.  A simple example is 
$\gsu(2j + 1) \oplus \gsu(2j + 8)$ on a pair of curves of
self-intersection $-1, -2$.
A further list of  examples of
tunings on two intersecting curves
without Tate forms is
given in \S\ref{sec:intersecting-2}.  Like the examples on a single rational
curve, we do not have a proof that Weierstrass models cannot be found
for any of these cases, though we do not expect a non-Tate Weierstrass
model in any of these cases with generic matter content.
\vspace*{0.1in}

\noindent {\bf Tunings on degenerate elliptic curves} 
As discussed in \S\ref{sec:nonlinear-2},  \S\ref{sec:generalized-intersection},
there are some local combinations of
divisors, for example a sequence of two or more $-2$ curves mutually
intersecting in a loop, which naively admit an infinite number of
gauge group tunings with a finite number of moduli needed for the
tuning.  These correspond to low-energy 6D supergravity theories with
$T \geq 9$ with no apparent inconsistency.  
We have identified these configurations in F-theory
as degenerate elliptic curves satisfying $\Sigma \cdot \Sigma = -K
\cdot \Sigma = 0$.  From the F-theory point of view $\gsu(N)$ gauge
groups can be tuned on such curves with $N$ taking values only up to a
specific bound associated with the constraint $\Delta =-12K$.  As
discussed in \cite{KMT-II}, however, there is no low-energy
understanding at this time of this ``Kodaira constraint,'' so that for
effective cones containing such $-2$ curve configurations there is
effectively an infinite swampland.  This is an example of the more
general issue that adding a gauge group with only adjoint matter
(essentially an ${\cal N} = (1, 1)$ multiplet)
 does not affect anomaly conditions, and can be limited in F-theory
 but not in the low-energy theory.
\vspace*{0.1in}

\noindent {\bf Exotic matter} We have listed in \S\ref{sec:matter}
some exotic matter
content for which there are known constructions.  Other types of
matter appear to be allowed by the anomaly constraints, but are at
this point lacking Weierstrass constructions.
The resolution of this part of the tuning swampland will be addressed
further
elsewhere.
\vspace*{0.1in}

\noindent {\bf Constraints from divisors without gauge factors}

Constraints such as the $E_8$ rule that involve divisors that do not
carry gauge factors are not currently understood from 6D supergravity,
though they can be partly explained in SCFT limits.  We have
identified some possible
exceptions to the sufficiency of $E_8$ rule for
determining allowed tunings; for example, $\gsu(5) \oplus \gsu(5)$
and $\gsu(9) \oplus \cdot$
cannot be tuned using the Tate procedure
on a pair of divisors that intersect a $-1$ curve
without forcing a gauge factor on the $-1$ curve.  Such examples
suggest some new low-energy consistency condition in both supergravity
and SCFT or a novel construction of non-Tate Weierstrass models with
generic matter for certain groups.  We have also sketched
out the analogue of the $E_8$ rule for
divisors of self-intersection 0.  These kinds of constraints are
somewhat similar to the constraint that, for example, a low-energy
theory containing a BPS dyonic string of Dirac self-charge $-3$, which
would correspond in F-theory to a $-3$ curve in the base, needs to
carry a gauge group $\gsu(3)$ in this case.  A related set of issues
is the distinction between type $III, IV$, and $I_2, I_3$ realizations
of $\gsu(2), \gsu(3)$, which have slightly different rules for
intersections but are not easily distinguished in the low-energy
theory.  Understanding the $E_8$ rule and these other related
conditions from low-energy considerations is an important part of the
outstanding problem of clearing the 6D swampland.
\vspace*{0.1in}

\noindent {\bf Abelian gauge factors} We have outlined an algorithm,
following \cite{ckpt}, which in principle gives a superset of the set
of possible F-theory models with abelian gauge field content.  This
algorithm is based on Higgsing of nonabelian gauge factors with
adjoint matter.  Proving that all abelian F-theory models can be
constructed in this fashion, and matching precisely with low-energy
anomaly constraints, particularly for higher-rank abelian groups,
remains an outstanding research problem.

\subsection{Tate vs. Weierstrass}

One interesting question that arises in attempting to do generic
tunings is the extent to which Tate models can produce the full set of
possible tunings.  It is known that there are Weierstrass tunings of
$\gsu(N)$ with $N = 6, 7, 8$ that cannot be realized through Tate
\cite{kmss-Tate, mt-singularities, agrt}, though these are associated with
exotic matter ({\it e.g.}, in the three-index antisymmetric
representation).  We have also identified cases of $\gso(N)$ with $N =
13$ on curves of self-intersection $n$ with $n$ even but not $n = -4$,
where Weierstrass models can be realized but Tate cannot.  It is known
that Weierstrass and Tate tunings of $\gsu(N)$ are equivalent for $N <
  6$, and are believed to be
more 
generally equivalent when exotic matter representations
  are not included.  It therefore
seems possible that many of the tunings found
  here on a single curve or a pair of curves that do not have Tate
  realizations also do not have Weierstrass realizations with generic
  matter content
and represent
  elements of the swampland.  
\footnote{(Added in v2:)
David Morrison has  pointed out that some models with groups that
  cannot be realized using Tate can have Weierstrass realizations when
there is a discrete Mordell-Weil group associated with a 
quotient of the usual
simply-connected simple group associated with a given algebra.}
But identifying precisely the set of
  cases where Tate and Weierstrass forms are not equally valid, is a
  remaining task that needs to be completed to clear out this part of
  the F-theory swamp.

\subsection{4D F-theory models}

The focus of this paper has been on 6D supergravity theories described
by F-theory compactifications on complex surfaces.  An analogous
set of constructions give 4D ${\cal N} = 1$ supergravity theories from
F-theory compactifications on complex threefolds.
While the 4D case is much less well understood, and the connection between
the underlying F-theory geometry and low-energy physics is made more
complicated by fluxes, D-brane world-volume degrees of freedom, 
a nontrivial superpotential, and a weaker set of anomaly constraints,
the basic principles of tuning that we have developed here are
essentially the same in 4D.  For threefold bases, divisors that have a
local toric description can again be analyzed torically, and we can
write Weierstrass and Tate models for the kinds of gauge groups and
matter that can be tuned.  

Perhaps the clearest result of this paper that
immediately generalizes to 4D F-theory constructions is the constraint
derived in \S\ref{sec:intersecting-constraint} that limits the
possible products of gauge groups on intersecting divisors to only
the five (families of) algebra pairings  listed in
Table~\ref{t:intersecting-possibilities}. This constraint is also
valid for tunings in 4D F-theory models, with the same microscopic
derivation from the Weierstrass analysis.
A consequence of this result is that we have shown in general that
any ${\cal N} = 1$ low-energy theory of supergravity coming from
F-theory can only have matter charged under multiple gauge factors
when the factors are either among those in
Table~\ref{t:intersecting-possibilities}, or both factors come from
the breaking of a larger single group
(or product group) such as $\ge_8$ that is realized
in F-theory.

More generally,
the  methods summarized in \S\ref{sec:AG}
in association with (\ref{eq:sheaves}), which were developed
in \cite{Anderson-Taylor, 4D-NHC}, can be used to identify the
non-Higgsable gauge group and local tunings on any local combination
of divisors an a base threefold, with explicit Weierstrass and Tate constructions for
local toric geometries.
For single divisors, the set of possible
tunings will follow a similar pattern to that found here in
\S\ref{sec:C-I}.  
The non-Higgsable clusters over single toric divisors were analyzed
(in the context of $\P^1$ bundles) in \cite{Anderson-Taylor,
  Halverson-WT}, and the  finite set of possible tunings over such
combinations of
divisors can be constructed using the same methods as those used
here and is again finite in most cases.  A similar analysis is also
possible for divisors with a positive or less negative normal bundle.
For example, in analogy with a $+ 1$ curve,
it is straightforward to confirm that
any of the exceptional gauge algebras can be tuned on a divisor with
the geometry of $D = \P^2$ with a normal bundle of $+ H$, and that
a Tate form for $Sp(k)$ and SU($2k$) can be realized in a toric model,
for $k \leq 16$, in analogy with the bound of 12 for the same tunings
on the $+ 1$ curve \cite{mt-singularities}.

Also for multiple intersecting divisors, a similar analysis can be
carried out.  

The difficult part of generalizing the analysis of this paper to 4D is
the absence of strong low-energy constraints for 4D ${\cal N} = 1$
supergravity models.  While in 6D, as we have shown here, the set of
constraints imposed by low-energy anomaly cancellation conditions is
almost precisely equivalent to the constraints imposed by Weierstrass
tuning, in 4D the known low-energy constraints are much weaker, so the
apparent swampland is much larger.  Whether this is an indication that
F-theory describes a much smaller part of the space of consistent 4D
supergravity theories, or we are simply lacking insight into 4D
low-energy consistency conditions, is an important open question for
further research (see \cite{Grimm-Taylor} for some initial investigations
in this direction).

\subsection{Outstanding questions}

In this paper we have made progress towards a complete classification
of allowed tunings for Weierstrass models over a given base.  A
desirable final goal of this program would be a complete set of local
constraints (in terms only of gauge algebras, matter representations,
and the self-intersection matrix of the base) such that a fibration
exists that produces a Weierstrass model for a Calabi-Yau threefold
and corresponding 6D supergravity model {\it if and only if} that
fibration satisfies all of the local constraints.  Here we summarize
some questions that still need to be addressed to complete the
classification and to match Weierstrass tunings in 6D F-theory models
to low-energy supergravity theories.

\begin{itemize}
\item The remaining local configurations in the ``tuning swampland'' summarized
  in \S\ref{sec:tuning-swamp} should hopefully be able to be
  identified either as allowed by as-yet-unknown Weierstrass tunings,
  or as inconsistent in UV-complete quantum 6D supergravity theories.
\item We have addressed in this paper local constraints associated
  with the tunings of gauge groups and matter over a single divisor
  corresponding to a curve in the base
  and the set of other curves intersecting that divisor.  We do not
  know that every model that satisfies all local constraints of this
  type is globally consistent, though we do not know of any
  counterexamples.  It would be desirable to either prove that local
  constraints are sufficient or identify conflicts that can arise
  nonlocally.
\item We have only checked Weierstrass/Tate realizations for rational
  curves with local toric descriptions.  It would be desirable to
  expand the methodology to higher-genus curves without a local toric
  description.
\item 
  Tuned gauge factors such as $\gg_2$ that can be broken without decreasing rank
  give additional contributions to $h^{2, 1} (X)$ from the
  associated charged matter fields that are uncharged under the
  smaller group.  
This should be understood better geometrically, and
 may also be related to the issue of
  tuning $\gsu(N)$ factors on degenerate elliptic curves; in both
  cases the essential issue is the addition of a $(1, 1)$ multiplet
  with gauge bosons and hypermultiplets that precisely cancel anomaly
  constraints.  The connection between this phenomenon and the 
  topology of the elliptic Calabi-Yau threefold should be better
  understood, and will be described further elsewhere
 \cite{Berglund-note}.
\item The classification of exotic matter representations,
  particularly those realized by gauge groups supported on singular
  divisors in the base, must be completed.
In particular, the method of analysis in \S\ref{sec:matter} assumed
that any exotic matter configuration can be realized as a tuning of a
generic matter configuration --- that is, that there are no
non-Higgsable exotic matter configurations possible.  While we believe
that this is true we do not have a rigorous proof of this statement.
\item It needs to be shown whether the approach used here of constructing
  abelian gauge factors from Higgsing of nonabelian tuned gauge
  factors is able to produce all abelian gauge structures of arbitrary
  rank; even if this is possible, a systematic understanding of how
  this can be implemented for higher-rank gauge groups and what
  singularities are possible in the nonabelian enhanced model for a
  consistent Higgsed abelian theory must be better understood.
\end{itemize}

A further set of questions, which fit into this general framework but which
go beyond the goal of classifying
all tunings over a given base surface, include the following:

\begin{itemize}
\item We have not addressed the question of different resolutions of
  the Kodaira singularities, which are not relevant for the low-energy
  6D physics, but would need to be addressed in general for a complete
  classification of smooth elliptic Calabi-Yau threefolds, as
  discussed in \S\ref{sec:Calabi-Yau}.
\item While substantial progress has been made towards the complete
  classification of non-toric bases that support elliptic Calabi-Yau
  threefolds \cite{wang-non-toric}, technical issues remain to be
  solved for a complete classification of all allowed bases for $h^{2,
    1}(X) < 150$.  
\item The complete elimination of the 6D swampland would require
  progress on relating apparently consistent  completely Higgsed low-energy models (such
  as those with -3 dyonic strings but no gauge group) to F-theory
  constructions and/or developing new constraints on low-energy 6D
  supergravity theories.
\item  As discussed in the previous subsection, much work remains to
  be done to generalize this story to 4D F-theory constructions.
\end{itemize}

The partial progress presented in this paper, then, should be viewed
as both a set of tools for Weierstrass constructions and as a framing
of the remaining challenges and an invitation to meet them. It is
becoming increasingly clear that the sets of elliptically fibered
Calabi-Yau threefolds, associated Weierstrass models, and 6D
supergravity theories are tightly controlled, richly structured, and
closely related.
Moreover, as discussed in \S\ref{sec:Calabi-Yau}, elliptically fibered
Calabi-Yau manifolds may represent a very large fraction of the total
number of Calabi-Yau varieties in any dimension, so that the analysis
of elliptic Calabi-Yau spaces may give insight into the more general
properties of Calabi-Yau manifolds.  Following this general line of
inquiry will doubtless reveal many other physical and geometric
insights yet undiscovered.

\acknowledgments{During the course of this work, we benefited from
  many conversations with Yu-Chien Huang, who had many helpful
  comments on an early draft of this work.  
We would like to particularly thank Nikhil Raghuram and Yinan Wang for
numerous technical discussions and helpful input 
on various particular aspects of this
work and for comments on the draft.
We would also like to thank
  Lara Anderson, Per Berglund, Clay Cordova, James Gray, Jim Halverson, Denis Klevers, David
  Morrison, Tom Rudelius, Yuji Tachikawa
and Andrew Turner for helpful discussions. The authors are supported by DOE grant
  DE-SC00012567.  }

\appendix

\section{Tabulated results}\label{sec:results}

Ultimately, one of the principal utilities of our results is to enable
easy calculations. In practical calculations, it is convenient to have
explicit lists of all {\em a priori} allowed tunings. Therefore, in
this appendix, we unpack the formulas for Hodge shifts in terms of
self-intersection number $n$ (and possibly a group parameter $N$),
re-packaging them in tables that explicitly list all allowed tunings
on a given self-intersection number curve or given cluster. We give a
table for each isolated curve or multi-curve cluster with negative
self-intersection number. These tables for isolated curves are
presented first, followed by the tables concerning multi-curve
clusters. For each curve or cluster, the data listed include: algebra,
matter representations, Hodge shifts from the generic fibration, and
finally the global symmetry group as determined in
\cite{global-symmetries}. This last piece of information is reproduced
here because it provides a field theory constraint on which algebras
can be tuned on intersecting divisors as discussed in section
\S\ref{sec:intersecting-2}.

One final note for using these tables: instead of displaying changes
in $h^{2,1}$, we display changes in $\hon$, the number of neutral
hypermultiplets of the theory. 
For
rank-preserving tunings (in which $h^{1,1}$ does not change),
the actual Hodge numbers of the Calabi-Yau threefold do not change,
and the Calabi-Yau threefolds may be related \cite{Berglund-note}.
Thus, from the point of view of Hodge numbers, the Hodge numbers of
the threefold for
the theory with the larger group are equal to those with that reached
after rank-preserving breaking.
Nonetheless, it seems that the resulting physical
theories are distinct, so from the standpoint of studying 6D SUGRA and
SCFT, these tuned models must be included. Brackets $[.]$ are placed
around shifts $\Delta \hon$ that cannot be equated to $\Delta
h^{2,1}$; such models do not contribute distinct Hodge number combinations.

\begin{table}[h]
\begin{center}
\begin{tabular}{|c|c|c|c|}
\hline
$\gg$ & matter & $(\Delta h^{1,1},\Delta \hon)$ & global symmetry algebra(s) \\
\hline
$\gsu(2)$ & $10\times {\bf 2}$ & $(1,-17)$ & $\gso(20)$ \dots \\
$\gsu(3)$ & $12\times {\bf 3}$ & $(2,-28)$ & $\gsu(11)$ \dots \\
$\gsu(N)$ & $(8+N)\times{\bf N}+\frac{{\bf N(N-1)}}{\bf 2}$ & $(N-1,-\frac{15N+N^2}{2}-1)$ & $\gsu(8+N)$\\
$\gsp(N/2)$ & $(8+N)\times \bf{N}$ & $(\frac{N}{2},-8N-N^2-1)$  & $\gso(16+2N)$ \dots \\
$\gso(7)$ & $2\times{\bf 7}+6\times S$ & $(3,[-41])$ & $\gsp(6)\oplus \gsp(2)$ \\
$\gso(8)$ & $3\times (\bf{8}_f+\bf{8}_s+\bf{8}_s)$ & $(4,-44)$ & $\gsp(3)\oplus \gsp(3)\oplus \gsp(1)^{\oplus 3}$ \\
$\gso(9)$ & $4\times {\bf9}+3\times S$ & $(4,[-48])$ & $\gsp(4)$ \\
$\gso(10)$ & $5\times {\bf 10} + 3\times S$ & $(5,-53)$ & $\gsp(5) $\\
$\gso(11)$ & $6\times {\bf 11} + \frac{3}{2}\times S$ & $(5,[-59])$ & $\gsp(6)$\\
$\gso(12)$ & $7\times {\bf 12} + \frac{3}{2}\times S$ & $(6,-66)$ & $\gsp(7)$\\
$\gg_2$ & $7\times \bf{7}$ & $(2,[-35])$ & $\gsp(7)$ \\
$\gf_4$ & $4\times {\bf 26}$ & $(4,[-52])$ & - \\
$\ge_6$ & $5\times{\bf 27}$ & $(6,-57)$ & - \\
$\ge_7$ & $\frac{7}{2}\times \bf{56}$ & $(7,-63)$ & - \\
\hline
\end{tabular}
\end{center}
\caption[x]{\footnotesize Table of all tunings on an isolated -1 curve. The ellipses ``$\dots$'' indicate that $\gsu(2)$, $\gsu(3)$, and the $\gsp(N)$ series may have different global symmetry algebras other than those listed here depending on the details of their tuning; see \cite{global-symmetries}. We have simply listed global symmetry for the most generic tuning.}
\label{t:-1}
\end{table}

\begin{table}[t!]
\begin{center}
\begin{tabular}{|c|c|c|c|}
\hline
$\gg$ & matter & $(\Delta h^{1,1},\Delta \hon)$ & global symmetry algebra(s) \\
\hline
$\gsu(2)$ & $4\times {\bf 2}$ & $(1,-5)$ & $\gsu(4)$ ($I_2$); $\gso(7)_s$ ($III/IV$)  \\
$\gsu(3)$ & $6 \times {\bf 3}$ & $(2,-10)$ & $\gsu(6)$ \dots \\
$\gsu(N)$ & $2N\times {\bf N}$ & $(N-1,-N^2-1)$ & $\gsu(2N)$ \\
$\gso(7)$ & $\bf{7}+4\times S$ &  $(3, [-18])$ & $\gsp(4)\oplus \gsp(1)$ \\
$\gso(8)$ & $2\times (\bf{8}_f+\bf{8}_f+\bf{8}_s)$ & $(4,-20)$ & $\gsp(2)\oplus \gsp(2) \oplus \gsp(1)^{\oplus 2}$ \\
$\gso(9)$ & $3\times {\bf 9}+2\times S$ & $(4,[-23])$ & $\gsp(3)$\\
$\gso(10)$ & $4\times {\bf 10} + 2 \times S$ & $(5,-27)$ & $\gsp(4)$ \\
$\gso(11)$ & $5 \times {\bf 11} + S$ & $(5,[-32])$ & $\gsp(5)$ \\
$\gso(12)$ & $6 \times {\bf 12} + S$ & $(6,-38)$ & $\gsp(6)$ \\
$\gso(13)$ & $7\times {\bf 13}+\frac{1}{2}\times S$ & $(6,[-46])$ & $\gsp(7)$ \\
$\gg_2$ & $4\times \bf{7}$ & $(2,[-14])$ & $\gsp(4)$ \\
$\gf_4$ & $3\times{\bf 26}$ & $(4,[-26])$ & -\\
$\ge_6$ & $4\times {\bf 27}$ & $(6,-30)$ & -\\
$\ge_7$ & $3\times {\bf 56}$ & $(7,-35)$ & -\\
\hline
\end{tabular}
\end{center}
\caption[x]{\footnotesize Table of all tunings on an isolated -2 curve. We have explicitly included two global symmetry algebras for $\gsu(2)$, depdening on whether it is tuned as an $I_2$ or $III/IV$ singularity type. Again, an ellipsis ``\dots'' denotes that there are other symmetry algebras for $\gsu(3)$ when it is tuned in a non-generic way.\cite{global-symmetries}}
\label{t:-2}
\end{table}

\begin{table}[t!]
\begin{center}
\begin{tabular}{|c|c|c|c|}
\hline
$\gg$ & matter & $(\Delta h^{1,1},\Delta \hon)$ & global symmetry algebra \\
\hline
$\gsu(3)$ & $\emptyset$ & $(0,0)$ & - \\
$\gg_2$ & ${\bf 7}$ & $(0,-1)$ & $\gsp(1)$ \\
$\gso(7)$ & $2\times S$ & $(1,-3)$ & $\gsp(2)$ \\
$\gso(8)$ & $\bf{8}_f+\bf{8}_s+\bf{8}_c$ & $(2,-4)$ & $\gsp(1)\oplus \gsp(1) \oplus \gsp(1)$ \\
$\gso(9)$ & $2\times {\bf 9} + S$ & $(2,[-6])$ & $\gsp(2)$\\
$\gso(10)$ & $3\times {\bf 10} + S$ & $(3,-9)$ & $\gsp(3)$ \\
$\gso(11)$ & $4 \times {\bf 11} +\frac{1}{2}\times S$ & $(3,[-13])$ & $\gsp(4)$\\
$\gso(12)$ & $5 \times {\bf 12} + \frac{1}{2}\times S$ & $(4,-18)$ & $\gsp(5)$ \\
$\gf_4$ & $2\times {\bf 26}$ & $(2,[-8])$ & -\\
$\ge_6$ & $3 \times {\bf 27}$ & $(4,-11)$ & -\\
$\ge_7$ & $\frac{5}{2}\times {\bf 52}$ & $(5,-15)$ & - \\
\hline
\end{tabular}
\end{center}
\caption[x]{\footnotesize Table of all tunings on an isolated -3 curve.}
\label{t:-3}
\end{table}

\begin{table}[t!]
\begin{center}
\begin{tabular}{|c|c|c|c|}
\hline
$\gg$ & matter & $(\Delta h^{1,1},\Delta \hon)$ & global symmetry algebra \\
\hline
$\gso(8)$ & $\emptyset$ & $(0,0)$ & - \\
$\gso(N>8)$ & $(N-8)\times {\bf N}$ & $(\lfloor (N-8)/2, -N\frac{N-15}{2}+28)$ & $\gsp(N-8)$ \\
$\gf_4$ & ${\bf 26}$ & $(0,[-2])$ &  -\\
$\ge_6$ & $2 \times {\bf 27}$ & $(2,-4)$ & - \\
$\ge_7$ & $2\times {\bf 52}$ & $(3,-7)$ & -\\
\hline
\end{tabular}
\end{center}
\caption[x]{\footnotesize Table of all tunings on an isolated -4 curve.}
\label{t:-4}
\end{table}

\begin{table}[t!]
\begin{center}
\begin{tabular}{|c|c|c|}
\hline
$\gg$ & matter & $(\Delta h^{1,1},\Delta \hon)$ \\
\hline
$\gf_4$ & ${\bf 26}$ & $(0,0)$ \\
$\ge_6$ & ${\bf 27}$ & $(2,-1)$ \\
$\ge_7$ & $\frac{3}{2}\times {\bf 52}$ & $(3,-3)$ \\
\hline
\end{tabular}
\end{center}
\caption[x]{\footnotesize Table of all tunings on an isolated -5 curve. All matter has trivial global symmetry algebra.}
\label{t:-5}
\end{table}

\begin{table}[t!]
\begin{center}
\begin{tabular}{|c|c|c|}
\hline
$\gg$ & matter & $(\Delta h^{1,1},\Delta \hon)$ \\
\hline
$\ge_6$ & $\emptyset$ & $(0,0)$ \\
$\ge_7$ & ${\bf 52}$ & $(1,-1)$ \\
\hline
\end{tabular}
\end{center}
\caption[x]{\footnotesize Table of all tunings on an isolated -6 curve. All matter has trivial global symmetry algebra.}
\label{t:-6}
\end{table}

\begin{table}
\begin{center}
\begin{tabular}{|c|c|c|}
\hline
cluster & $\gg$ & $(\Delta h^{1,1},\Delta \hon)$ \\
\hline
22 & $\gg_2\oplus \gsu(2)$ & $(3,[-12])$ \\
& $\gso(7)\oplus \gsu(2)$ & $(4,[-15])$ \\
\hline
222 & $\gsu(2)\oplus \gg_2\oplus \gsu(2)$ & $(4,[-8])$ \\
& $\gsu(2)\oplus\gso(7)\oplus\gsu(2)$ & $(5,[-12])$ \\
& $\gg_2\oplus\gsu(2)\oplus \cdot $ & $(3,[-11])$ \\
\hline
2222 & $\gsu(2)\oplus\gg_2\oplus\gsu(2)\oplus\cdot$ & $(4,[-8])$ \\ 
\hline
22222 & $\cdot \oplus \gsu(2)\oplus \gg_2\oplus \gsu(2)\oplus \cdot$ & $(4,[-8])$\\
\hline
$2_1\cdots 2_k$ & (Only $\gsu(n)$'s; see \S\ref{sec:-2}) & see (\ref{eq:2-shift}) \\
\hline
\end{tabular}
\end{center}
\caption[x]{\footnotesize Table of possible tunings on $-2$ chains. (Chains are listed with self-intersections sign-reversed.) Because matter is very similar between these cases, we do not list it explicitly, preferring to display the shift in Hodge numbers resulting from that matter. For convenience, we summarize the relevant matter content here: $4{\bf 2}$ for $\gsu(2)$, $4{\bf 7}$ for $\gg_2$, $4{\bf 8_s}+{\bf 7}$ for $\gso(7)$ and $4{\bf 8_s}+2{\bf 8_f}$ for $\gso(8)$. $\gsu(2)$ shares a half-hypermultiplet with all groups but itself, where it shares a whole hyper; with $\gso$'s, it is the spinor representation which is shared.}
\label{t:2chains-reproduced}
\end{table}

\begin{table}
\begin{center}
\begin{tabular}{|c|c|c|c|}
\hline
cluster & $\gg$ &| $(\Delta h^{1,1},\Delta \hon)$ & matter \\
\hline
(-3,-2) & $\gg_2\oplus\gsu(2)$ & (0,0) & $({\bf 7},\frac{1}{2}{\bf 2})+\frac{1}{2}{\bf 2}$ \\
& $\gso(7)\oplus\gsu(2)$ & (1,-1) & $({\bf 8}_s, \frac{1}{2}{\bf 2})+{\bf 8}_s $ \\
\hline
(-3,-2,-2) & $\gg_2\oplus\gsu(2)$ & (0,0) & $({\bf 7},\frac{1}{2}{\bf 2})+\frac{1}{2}{\bf 2}$ \\
\hline
(-2,-3,-2) & $\gsu(2)\oplus \gso(7)\oplus \gsu(2)$ & (0,0) &
$(\frac{1}{2}{\bf 2}, {\bf 8_s}, \cdot)+(\cdot, {\bf 8_s} ,\frac{1}{2}{\bf 2})$ \\
\hline
\end{tabular}
\end{center}
\caption[x]{\footnotesize Table of possible tuned gauge algebras,
  together with matter and Hodge shifts, on the NHCs with multiple
  divisors.  }
\label{t:multiple-reproduced}
\end{table}

We should emphasize: all of the information in this section is in
principle contained in the body of the paper, {\it e.g.} Table
\ref{t:isolated}. At first glance, it may appear that the tables
presented here {\em do} have more information, namely the information
about when certain series can no longer be consistently
tuned. However, it is possible to read this off from the original
tables as well: given a general formula for the matter multiplicies of
an algebra $\gg$ in terms of self-intersection number $n$ (and
possibly a group parameter $N$), a group will be impossible to tune
whenever one of the following occurs: the formula predicts either
negative multiplicity representations or fractional representations
(This last requires some care, since $\frac{1}{2}$-multiplicity
representations may occur when the representation is self-conjugate, in
which case this denotes a half-hypermultiplet.). This discussion also
makes it clear that while the information in this appendix is already
contained throughout the body of the paper, unpacking it requires some
work. Hence, the motivation to collect these expressions more
explicitly here.

Also note that the two final tables in this section, which pertain to to multiple-curve clusters, are reproduced here for convenience; the identical tables also appear in the body of the paper.

\section{Tabulations of group theory coefficients}
\label{sec:group-coefficients}

In this appendix, we present tables of the coefficients A, B, and C,
which appear in anomaly cancellation conditions in 6D. All these
coefficients have been calculated elsewhere, but as the existing
calculations and results are somewhat scattered throughout the
literature, we collect these results here for ease of
reference. Many of these coefficients were originally derived in
\cite{Erler}; additional
classical group coefficients are reproduced from
\cite{finite}, and normalization coefficients $\lambda$ are defined as in
 \cite{KMT-II}.  

\begin{table}
\centering
\begin{tabular}{|c|c|c|c|c|c|c|}
\hline
Group & representation $R$ & dimension & $A_R$ & $B_R$ & $C_R$ \\
\hline
\multirow{7}{*}{$SU(N)$} & ${\tiny\yng(1)}$ & $N$ & 1 & 1 & 0 \\
 & Adjoint & $N^2-1$ & $2N$ & $2N$ & 6 \\
 & ${\tiny\yng(1,1)}$ & $ \frac{N(N-1)}{2} $ & $ N-2 $ & $ N-8 $ & 3 \\
 & ${\tiny\yng(2)}$ & $ \frac{N(N+1)}{2} $ & $ N+2 $ & $ N+8 $ & 3 \\
 & ${\tiny\yng(1,1,1)}$ & $ \frac{N(N-1)(N-2)}{6} $& $ \frac{N^2-5N+6}{2} $&$ \frac{N^2-17N+54}{2} $ & $ 3N-12 $ \\
 & ${\tiny\yng(3)}$ & $ \frac{N(N+1)(N+2)}{6} $ &$ \frac{N^2+5N+6}{2} $&$ \frac{N^2+17N+54}{2} $ & $ 3N+12 $ \\
\hline
\multirow{4}{*}{$SO(N)$} & ${\tiny\yng(1)}$ & $N$ & 1 & 1 & 0 \\
 & ${\tiny\yng(1,1)} = $  Adjoint & $\frac{N(N-1)}{2}$ & $N-2$ & $N-8$ & 3 \\
 & ${\tiny\yng(2)}$ & $\frac{N(N+1)}{2}$ & $N+2$ & $N+8$ & 3\\
 & $S$ & $2^{\lfloor (N-1)/2\rfloor}$ & $2^{\lfloor (N+1)/2 \rfloor-4}$ & $-2^{\lfloor (N+1)/2\rfloor-5}$ & $3\cdot 2^{\lfloor (N+1)/2\rfloor-7}$ \\
\hline
\multirow{3}{*}{$Sp(\frac{N}{2})$} & ${\tiny\yng(1)}$ & $N$ & 1 & 1 & 0 \\
 & ${\tiny\yng(1,1)}$ & $\frac{N(N-1)}{2}-1$ & $N-2$ & $N-8$ & 3 \\
& & & & &\\
 & ${\tiny\yng(1,1,1)}$ & $ \frac{N(N-4)(N + 1)}{6} $& $ \frac{N^2-5N+6}{2} $&$ \frac{N^2-17N+54}{2} $ & $ 3N-12 $ \\
 & ${\tiny\yng(2)} = $ Adjoint & $\frac{N(N+1)}{2}$ & $N+2$ & $N+8$ & 3\\
 \hline
\end{tabular}
\caption{Values of the group-theoretic coefficients $A_R, B_R, C_R$
for some representations of $SU(N)$, $SO(N)$ and $Sp(N/2)$.  
Also note that we do not distinguish between $S_{\pm}$ spin
representations of $SO(N)$ for even $N$, as these representations have
identical group theory coefficients.
For SU(2) and SU(3), there is no quartic Casimir;
$B_R = 0$ for all representations, and $C_R^{(SU(2, 3))} = C_R +
B_R/2$ in terms of the values given in the table.} 
\label{table:infinite-series-coefficients}
\end{table}

\begin{table}
\begin{center}
\begin{tabular}{|c|c|c|c|c|}
\hline
Group & representation $R$ & dimension & $A_R$ & $C_R$ \\
\hline
\multirow{2}{*}{$G_2$} & ${\tiny\yng(1)}$ & $7$ & $1$ & $\frac{1}{4}$ \\
& Adjoint & $14$ & $4$ & $\frac{5}{2}$ \\
\hline
\multirow{2}{*}{$F_4$} & ${\tiny\yng(1)}$ & $26$ & $1$ & $\frac{1}{12}$\\
& Adjoint & $52$ & $3$ & $\frac{5}{12}$ \\
\hline
\multirow{2}{*}{$E_6$} & ${\tiny\yng(1)}$ & $27$ & $1$ & $\frac{1}{12}$  \\
& Adjoint & $78$ & $4$ & $\frac{1}{2}$\\
\hline
\multirow{2}{*}{$E_7$} & ${\tiny\yng(1)}$ & $56$ & $1$ & $\frac{1}{24}$ \\
& Adjoint & $133$ & $3$ & $\frac{1}{6}$\\
\hline
$E_8$ & ${\tiny\yng(1)} = $
Adjoint & $248$ & $1$ & $\frac{1}{100}$ \\
\hline
\end{tabular}
\end{center}
\caption{Group theoretic coefficients $A_R$ and $C_R$ for the exceptional groups. Note that $B_R$ is not included as it vanishes for all exceptional groups.}
\label{t:exceptional-coefficients}
\end{table}

\begin{table}
\begin{center}
\begin{tabular}{|c|c|c|c|c|c|c|c|c|}
\hline
Group & SU($N$) & Sp($N$) & SO($N$) & $G_2$ & $F_4$ & $E_6$ & $E_7$ & $E_8$ \\
\hline
$\lambda$ & 1 & 1 & 2 & 2 & 6  & 6 & 12  & 60 \\
\hline
\end{tabular}
\end{center}
\caption{Group theoretic normalization constants $\lambda$ for all simple Lie groups.}
\label{t:lambdas}
\end{table}

\section{Complete HC Calculations}\label{sec:appendix}

Here we construct a local model of each tuning possible on an NHC. We perform both anomaly calculations and calculations in local geometry to show which tunings are allowed and which cannot be realized. For each tuning, we find the matter representations using anomaly cancellation arguments, and calculate $\Delta \hon$ using anomaly cancellation as well as local geometry. The results of this section, as well as other results of this paper, are summarized in appendix ref{sec:results}.

\subsection{The Cluster (-3,-2)}

An appropriate local toric model has the fan $\{v_i\}_{i=0}^4 = \{(3,1),(1,0),(0,1),(-1,2)\}$. Let us now consider the base (untuned) case and check
that it corresponds to $\gg_2\oplus\gsu(2)$ on $\{-3,-2\}$, as has
already been derived as part of the NHC classification.  Although we
have already given several anomaly calculations, it is useful to
follow the anomaly calculations in these cases, because the ``A''
condition allows one to determine the shared matter, and highlights an
interesting feature of $\gsu(2)$ shared matter. This is a feature of
these calculations that cannot be seen in clusters besides those
containing multiple divisors.

The ``C'' calculations straightforwardly yield $N_f=1$ for $\gg_2$ on a $(-3)$-curve, as we already saw when discussing the $(-3)$-cluster, above. Similarly, for an $\gsu(2)$ on a $(-2)$-curve, we obtain
\begin{eqnarray}
\Sigma \cdot \Sigma & = & \frac{\lambda^2}{3}\left(\sum_RC_R-C_{Adj}\right) \nonumber \\
-2 & = &\frac{1}{3} \left(\frac{N}{2}-8\right) \nonumber \\
N & = & 4
\end{eqnarray}
fundamentals. The ``A'' condition of bifundamental matter is more interesting in this case, because it dictates
\begin{eqnarray}
\xi_i\cdot \xi_j=\lambda_i\lambda_j\sum_RA^i_RA^j_Rx^{ij}_R
\end{eqnarray}
Recall,we are considering $\xi_i\cdot \xi_j=1$ for neighbors and is
zero otherwise. The sum on the right hand side is over all shared
representations between the two gauge groups, which (consistent with
the string theory description) can occur only between divisors that
intersect. The $A_i$ are all integer-valued (and must be positive),
the $x_{ij}$ denote the multiplicities of the representations, and the
$\lambda$ are group theory coefficients introduced earlier. The key
fact in this case is that, whereas $\lambda=1$ for $\gsu(2)$,
$\lambda=2$ for $\gg_2$, $\gso(7)$, and $\gso(8)$ ($\lambda$ is always
the same for differnet gauge algrebras distinguished only by
monodromy). Since $A=1$ by definition for fundamental representations,
the equation now reads $1=2x$, so it is crucial that $x$, the
multiplicity of the shared hypermultiplet, can be
$x=\frac{1}{2}$. This is the case because the fundamental
representation of $\gsu(2)$ is self-conjugate, and in six dimensions,
a half-hypermultiplet can be shared. Indeed, organizing these
representations as explicitly as possible, we have
\begin{equation}
\begin{array}{cccc}
& \gg_2 &\oplus & \gsu(2) \\
\hline
 & (7, && \frac{1}{2}{\bf 2}) \\
& && \frac{1}{2}{\bf 2}
\end{array}
\end{equation} 
and it is clear that if $\gsu(2)$ could not share
half-hypermultiplets, then the bifundamental would imply 7 (as opposed
to $3\frac{1}{2}$ shared fundamentals of $\gsu(2)$, which would be a
contradiction because $\gsu(2)$ only has $4$ fundamentals total. This
same feature in principle, from anomaly cancellations alone, would
allow $\gsu(2)$ to remain adjacent to $\gso(7)$ and $\gso(8)$, with 7
and 8 dimensional fundamental representations, as well as 8
dimensional spinor representations. Returning to the main calculation
for $\gg_2 \oplus \gsu(2)$,
we have a total contribution to $\hon$ of $14+3-\frac{1}{2}7\times
2-\frac{1}{2}\times2=+9$. Thus, to calculate shifts from this generic
case, we must subtract $9$.
This is the base, untuned value against which the further tunings are compared.

Similar anomaly calculations for $\gso(7)\oplus \gsu(2)$ yield matter
$N_s=2$ spinors and $N_f=0$ fundamentals, giving a contribution to
$\hon$ of $+21+3-\frac{1}{2}8\times2-8=+8$; in other words, 
a shift of $\delta \hon = 8-9=-1$. 

Likewise, for $\gso(8) \oplus \gsu(2)$, we obtain $N_s=2$, $N_f=1$,
for a contribution of $+28+3-\frac{1}{2}8\times 2 - 2\times 8 = +7$;
in other words, this would be
a further shift of $-1$ from $\gso(7)$, or a total
shift of $-2$ from the untuned case.
From anomaly cancellations alone, this configuration appears to be allowed.

Let us now match these results with those of the local geometric
model. By inspecting the model, we see that $(f,g)$ vanish to orders
$(2,3)$ on the $-3$ curve $\Sigma_1$ corresponding to $v_1$, and to
orders $(1,2)$ on the curve $C_2$ corresponding to $v_2$. Moreover,
expanding in a local coordinate $w$ that defines $\{w = 0\}=\Sigma_2$
and $z$ such that $\{z=0\}=\Sigma_1$, let us define $f_i$ and $g_i$ by
the expansion $f=\sum_i f_iz^i$, $g_i=\sum_i g_iz^i$. Then $f_2 =
f_{1,1}w + f_{1,2}w^2$ and $g_3 = g_{3,2}w^2+g_{3,3}w^3$. Hence we
immediately see that on $\Sigma_2$ we are in Kodaira case $III$, an
$\gsu(2)$, whereas on $\Sigma_1$ we are in case $I_0^*$ with generic
$g_3\neq *^2$; hence this curve carries $\gg_2$. So far, this just
confirms that our local model reproduces the known gauge algebras of
this NHC.

Proceeding to the tunings, we implement $\gso(7)$ by imposing the appropriate condition 
\begin{eqnarray}
x^3+f_2x+g_3 & = &(x-A)(x^2+Ax+B) \nonumber \\
x^3+(f_{1,1}w + f_{1,2}w^2)x+(g_{3,2}w^2+g_{3,3}w^3) & = & x^3+(B-A^2)x-AB 
\end{eqnarray}
This immediately implies that $A$ must be proportional to $w$; with this restriction, $B$ can be of the form $B_1w +B_2w^2$, so we lose only one degree of freedom. This is in accorance with the anomaly calculation.

Proceeding to $\gso(8)$, we find that this tuning is impossible. It
requires the factorization
\begin{eqnarray}
x^3+(f_{1,1}w + f_{1,2}w^2)x+(g_{3,2}w^2+g_{3,3}w^3) & = & (x-A)(x-B)(x+(A+B)) \nonumber \\
& = & x^3+(AB-(A+B)^2)x+AB(A+B)
\end{eqnarray}
which requires now that $B \propto w$ have no quadratic term; hence we
would lose one further degree of freedom, in perfect agreement with
anomaly cancellation. However, let us ask what form $f$ and $g$ now
take. Indeed, $f\propto w^2z^2+z^3(w^2+\cdots)$ and $g\propto
w^3z^3+z^4(w^2+\cdots)+\cdots$, which implies that at the intersection
point $\Sigma_1 \cdot \Sigma_2 = \{ z = w = 0\}$, $(f,g)$ vanish to
orders $(4,6)$. Hence the $\gso(8)$ tuning is not allowed.
This is another example of the result discussed in the main text that
global symmetries prevent such a gauge group from intersecting with an
$\gsu(2)$ on a $-2$ curve realized through a Kodaira type III or IV singularity.

\subsection{The Cluster $(-3,-2,-2)$}

One might naively expect that the previous analysis would extend without modification to this cluster. However we will see that even enhancement to $\gso(7)$ is impossible, {\it i.e.} no monodromy at all is allowed for the type $I^*_0$ singularity that generically gives rise to $\gg_2$. To see this, again exlicitly construct the local model, which can be simply obtained from the previous model by adding the vector $v_5 = (-2,3)$ to the fan. This modifies the monomials in such a way that the generic orders of $(f,g)$ have orders $(2,2)$ on $\Sigma_2$ and $(2,3)$ on $\Sigma_1$. Moreover, making the same expansion of $f$ and $g$ in powers of $z$ as above, we have
\begin{eqnarray}
f_2 & = & f_{2,2}w^2 \nonumber \\
g_3 & = & g_{3,2}w^2+g_{3,3}w^3
\end{eqnarray}
Notice that the only change insofar as we are concerned from the
cluster $(-3,-2)$ is that $f_2$ now has no linear term. In fact, this
prevents any tuning at all. To see this, we will attempt to implement
the monodromy condition for $\gso(7)$, the most modest enhancment. The
failure of this tuning will imply the failure of all potential higher
tunings. 

Indeed, $\gso(7)$ requires $f_2=B-A^2$ and $g_3=-AB$, whence $B$ must be linear and therefore $g_2$ must be purely cubic in $w$. But this implies that the total $f$ and $g$ have the lowest order terms $f \propto w^2z^2+\cdots$ and $g \propto w^3z^3+w^2z^4+\cdots$ where in each case ellipses indicate higher order terms. In other words, we again have a $(4,6)$ singularity at the intersection $\Sigma_1\cdot \Sigma_2 = \{z = w = 0\}$.

To summarize, no tunings are allowed on this cluster.

\subsection{The Cluster (-2,-3,-2)}

Our local model will be a portion of $\F_3$ blown up four times, with the overal divisor structure $(+1,-2,-1,-2,-3,-2,-1,-2)$, which is to be cyclically identified, as always. By omitting all but 5 rays in the fan (the middle three of which correspond to the sequence $(-2,-3,-2)$, we obtain a local model of this geometry.

The anomaly calculation proceeds as above, for the cluster $(-3,-2)$. The only difference is that now the original (untuned) gauge algebra on the $-3$ curve is $\gsu(7)$, which must as above have matter $2\times {\bf 8}_s$. Therefore {\it all} matter is shared, in the form
\begin{equation}
\begin{array}{ccc}
\gsu(2) \oplus & \gso(7) \oplus & \gsu(2) \\
\hline
(\frac{1}{2}{\bf 2} & {\bf 8}_s) & \\
& ({\bf 8}_s & \frac{1}{2}{\bf 2})
\end{array}
\end{equation}
This yields a contribution to $\hon$ of $+21+3+3-2\times8=+11$. This configuration is enhanced to $\gsu(2)\oplus\gso(8)\oplus \gsu(2)$, which is identical in matter except that the $\gso(8)$ now carries one additional multiplet in the fundamental ${\bf 8}_f$, for a contribution to $\hon$ of $+28+3+3-3\times 8=+10$; in other words, $\Delta \hon = -1$ upon performing this tuning. We will find that these tunings are not possible in the following monomial analysis. Although consistent with anomaly cancellation, these tunings suffer from a field theory inconsistency identified in \cite{Tachikawa}.

The monomial calculations first confirm the untuned gauge / matter content: consulting the figures, it is clear that there are no monomials for $f_0$ or for $g_{i \leq 1}$ on the divisor $v_1=(1,0)$. Similarly for $v_3$. These two divisors, adjacent to the $(-3)$-curve ($v_4$) are the $(-2)$-curves on which $\gsu(2)$'s are forced. Similarly, on the middle curve $v_4$, one directly sees that the degrees of vanishing on $v_4$ are $(f,g)=(2,4)$. This falls into the $I_0^*$ case. In order to distinguish monodromies, we first read off the available monomials for $f_2$ ($\{w^1\}$) and for $g_3$ ($\{ \}$). The polynomial to investigate then takes the form
\begin{eqnarray}
x^3+f_{2,1}wx+0 & = & x^3+(B-A^2)+AB) \nonumber \\
\text{\ \ \ or \ \ \ } & = & x^2+(AB-(A+B)^2)x+AB(A+B) 
\end{eqnarray}
corresponding to $\gso(7)$ and $\gso(8)$ respectively. In order for the first equation to hold, the constant term requires that either $A$ or $B$ be equal to zero. Investigating the $x^1$ term, we must choose $A=0$, while $B$ can be proportional to $w^1$. Hence we begin with one degree of freedom (in $f_2$), and end with one. This implies that the $\gso(7)$ is indeed the minimal gauge group. In the second case of tuning an $\gso(8)$, it is clear that this is only possible when $A=B=0$, resulting in the loss of one degree of freedom. This is in accordance with the anomaly calculation. However, we must pause to examine the reality check allowed by the local toric model. This factorization can only be satisfied with $(f,g)=(3,4)$, hence at the point of intersection of $-3$ with either $-2$, we have a total vanishing of order $(f,g)=(4,6)$, so this tuning cannot be achieved. In other words, the Non-Higgsable Cluster $(-2,-3,-2)$ is completely rigid: it admits no tunings.

\subsection{The Cluster (-4)}

Our model is $\F_4$ (with $+4$ curve removed), and the analysis
proceeds along nearly identical lines to that of $\F_3$. For instance,
the conditions defining the monomials for $f$ and $g$ are identical
save for the one modification: the slope of the line bounding the top
of the triangle is now $-\frac{1}{4}$. (It intercepts the $b=0$ axis
at $n=4,6$ still for $f$ and  $g$, respectively.) 

From the anomaly point of view, the initial
(forced) gauge algebra is $\gso(8)$; one way to see this is to note
that no group of lesser rank can satisfy all anomaly cancellation
conditions on a curve of self-intersection $\leq -3$. (For example a
$\gsu(N\geq 4)$ has adjoint $C_{Adj}=6$, which means that on a $-n$
curve, $-3n= \sum_RN_RC_R-6$. Since $C_R \geq 0$ for all
representations $R$, it is clearly impossible to satisfy this equation
on a $-n$ curve for $n\geq 3$.) Investigating the anomaly conditions
for $\gso(8)$ reveals that they are satisfied with no matter, leading
to a contribution to $\hon$ of $+28$ vectors (in its
adjoint). Enhancement to $\gf_4$ is accompanied by the appearance of 1
fundamental hypermultiplet, for a contribution of $+52-26=+26$ to
$\hon$, or a change of $\Delta \hon = - 2$. Finally, enhancement to $\ge_6$ is
accompanied by 2 fundamentals, for a contribution to $\hon$ of
$+78-2\times 27=+24$, {\it i.e.} $\Delta \hon = -4$ from the generic fibration.

In the monomial counting picture, we have the following explanation:
the untuned version is $\gso(8)$ because the slope of the triangle's
upper boundary ($-\frac{1}{4}$) implies that from $(6,0)$ the boundary
rises to a maximum of height at $(-6,3)$; this is the unique monomial
in $g_3$, and its first component is even, which implies that
$g_3=w^0$ is a perfect square. It is clear that to increase the order
of $f$ and $g$ from $(2,3)$ to $(3,4)$, only one monomial from each of
$-4K$ and $-6K$ need be removed; we lose 2 degrees of
freedom. Enhancing to $\ge_6$ requires imposing the monodromy
condition that $g_4$ be a perfect square. The available monomials are
$\{w^0,w^1,w^2,w^3,w^4\}$, so we may impose the condition that this be
a perfect square by setting it equal to the square of a generic
quadradic. This restricts to a three dimensional subspace of the
original 5 parameter space; in other words, we lose 2 more degrees of
freedom beyond the $\gf_4$ tuning. To tune to $\ge_7$ requires that we
enhance the order of $(f,g)$ from $(2,3)$ to $(3,5)$ in other words
eliminating the 1 monomial of $f_2$ as well as all $1+5$ monomials of
$g_3$ and $g_4$; so we shift by $-7$ in $\hon$ from the untuned
$\gso(8)$, or by $-3$ subsequent to a tuning to $\ge_6$. This is all
in accordance with the anomaly results.

\subsection{The Cluster (-5)}

We begin with the base (untuned) case: $\gf_4$, which can be enhanced
exactly twice, to $\ge_6$ and further to $\ge_7$. Anomaly calculations
yield no matter for the $\gf_4$ (so it contributes the dimension of
its adjoint, $+52$, to $\hon$), whereas for $\ge_6$ we find 1
fundamental hypermultiplet, yielding a contribution of $+78-27=+51$ to
$\hon$; {\it i.e.} $\Delta \hon = 51-52 = -1$. A final enhancement to $\ge_7$ reveals $1\frac{1}{2}$
hypermultiplets (the fundamental also enjoys the self-conjugate
property as for $\gsu(2)$'s), which yields a contribution of
$+133-\frac{3}{2}56=+49$ to $\hon$,  {\it i.e.} $\Delta \hon = -3$ from the generic fibration.

A monomial analysis confirms this. Examining the local model, we find $f_{i\leq 2}$ and $g_{i\leq 3}$ have no monomials, hence $f$ and $g$ are forced to vanish to degree at least $(3,4)$ on $\Sigma$. We also see that $g_4$ is the span of $\{ w^0, w^1, w^2\}$, so that generically there is no factorization. To tune to $\ge_6$, we need only set this quadratic to be the square of a general linear function, thereby losing one degree of freedom. This is in accordance with anomaly results. In order to enhance to $\ge_7$, we need only increase the order of $g$ from $4$ to $5$, {\it i.e.} to eliminate all 3 monomials in $g_4$. This represents a shift of $-3$ from the original (untuned) $\gf_4$, or a shift of $-2$ subsequent to tuning an $\ge_6$, also in agreement with our anomaly calculations.

\subsection{The Cluster (-6)}

The local model is $\F_6$ (with the $+6$ curve removed). The initial gauge algebra is $\ge_6$, which has no matter, and hence contributes $V=78$ adjoint vectors to the count of $\hon$; this can be confirmed by investigating the ``C'' condition on a $-6$ curve. The analagous calculation for $\ge_7$ reveals 1 fundamental hypermultiplet, which leads to a contribution of $133-56=77$ to $\hon$, which leads to a shift $\Delta \hon=-1$.

A monomial analysis confirms these results: the upper boundary of the triangle of monomials now has slope $-\frac{1}{6}$, which implies there is only one monomial in $g_4$. This must be removed in order to obtain a degree of vanishing of $(f,g)=(3,5)$ so that the resulting algebra will be $\ge_7$; hence we indeed lose just one degree of freedom.

\subsection{The Clusters (-7), (-8), and (-12)}

The algebras of these clusters cannot be enhanced. At this point, it
bears mentioning that we have never enhanced a cluster to
$\ge_8$. Indeed, we are interested in tunings which do not change the
base geometry, {\it i.e.} require no blowups of the base alone. However, an
$\ge_8$ on any curve $\Sigma$ other than a $-12$ curve will
necessitate blowups in the base. The reason is straightforward: to
tune $\ge_8$, $f$ and $g$ must be of order $4$ and $5$. Yet $f$ and
$g$ restricted to $\Sigma$ are polynomials, and will generically have
isolated zeroes. Such points will lead to $(4,6)$ (non-minimal)
singularities, which require blowups of the base. In fact, the number
of blowups required on a curve $\Sigma$ with a tuned $\ge_8$ is always
equal to that required to bring the self-intersection of $\Sigma$ to
$-12$. (It is not difficult to confirm this. In fact, it is zeroes of
$g_5$ that lead to these singularities. Using equation
\ref{eq:sheaves}, we see that $\text{deg}(g_5)=12+n$, where $n$ is the
self-intersection of the curve on which $\ge_8$ appears. Hence $g_5$
has the correct number of zeroes to bring the self-intersection to
$-12$ after blowing up.) Without loss of generality then, we can
simply restrict to tunings of $\ge_8$ only on existing $-12$ curves.


\begin{thebibliography}{99}


\bibitem{Vafa-F-theory}
  C.~Vafa,
  ``Evidence for F-Theory,''
  Nucl.\ Phys.\  B {\bf 469}, 403 (1996)
  {\tt arXiv:hep-th/9602022}.

\bibitem{Morrison-Vafa}
  D.~R.~Morrison and C.~Vafa,
  ``Compactifications of F-Theory on Calabi--Yau Threefolds -- I,''
  Nucl.\ Phys.\  B {\bf 473}, 74 (1996)
  {\tt arXiv:hep-th/9602114};
  D.~R.~Morrison and C.~Vafa,
  ``Compactifications of F-Theory on Calabi--Yau Threefolds -- II,''
  Nucl.\ Phys.\  B {\bf 476}, 437 (1996)
  {\tt arXiv:hep-th/9603161}.

\bibitem{phenom-1}
J.~J.~Heckman,
``Particle Physics Implications of F-theory,''
Ann.\ Rev.\ Nucl.\ Part.\ Sci.\  {\bf 60}, 237 (2010)
{\tt arXiv:1001.0577}

\bibitem{phenom-2}
T.~Weigand,
``Lectures on F-theory Compactifications and Model Building,''
Class.\ Quant.\ Grav.\  {\bf 27}, 214004 (2010)
{\tt arXiv:1009.3497}


\bibitem{gs-west}
  M.~B.~Green, J.~H.~Schwarz and P.~C.~West,
 ``Anomaly Free Chiral Theories In Six-Dimensions,''
  Nucl.\ Phys.\  B {\bf 254}, 327 (1985).

\bibitem{Sagnotti}
  A.~Sagnotti,
  ``A Note on the Green-Schwarz mechanism in open string theories,''
  Phys.\ Lett.\  B {\bf 294}, 196 (1992)
  {\tt arXiv:hep-th/9210127}.

\bibitem {Grassi-Morrison}
A.~Grassi, D.~R.~Morrison, ``Group representations and the Euler characteristic of elliptically fibered Calabi-Yau threefolds'',
 J.  Algebraic Geom.  12 (2003), 321-356
{\tt 	arXiv:math/0005196}.


\bibitem{KMT}
  V.~Kumar, D.~R.~Morrison and W.~Taylor,
  ``Mapping 6D ${\cal N} = 1$ supergravities to F-theory,''
  JHEP {\bf 1002}, 099 (2010)
  {\tt arXiv:0911.3393 [hep-th]}.

\bibitem{KMT-II}
  V.~Kumar, D.~R.~Morrison and W.~Taylor,
  ``Global aspects of the space of 6D ${\cal N} = 1$ supergravities,''
  JHEP {\bf 1011}, 118 (2010)
  {\tt arXiv:1008.1062 [hep-th]}.

\bibitem {Grassi-Morrison-2}
  A.~Grassi and D.~R.~Morrison,
  ``Anomalies and the Euler characteristic of elliptic Calabi-Yau threefolds,''
Commun.\ Num.\ Theor.\ Phys.\  {\bf 6}, 51 (2012)
  {\tt arXiv:1109.0042 [hep-th]}.

\bibitem{Grimm-Kapfer} 
  T.~W.~Grimm and A.~Kapfer,
  ``Anomaly Cancelation in Field Theory and F-theory on a Circle,''
{\tt arXiv:1502.05398 [hep-th]}.

\bibitem{Seiberg-Witten}
  N.~Seiberg and E.~Witten,
  ``Comments on String Dynamics in Six Dimensions,''
  Nucl.\ Phys.\  B {\bf 471}, 121 (1996)
  {\tt arXiv:hep-th/9603003}.

\bibitem{clusters} 
  D.~R.~Morrison and W.~Taylor,
  ``Classifying bases for 6D F-theory models,''
Central Eur.\ J.\ Phys.\  {\bf 10}, 1072 (2012),
{\tt arXiv:1201.1943 [hep-th]}.

\bibitem{toric}
  D.~R.~Morrison and W.~Taylor,
  ``Toric bases for 6D F-theory models,''
Fortsch.\ Phys.\  {\bf 60}, 1187 (2012)
{\tt arXiv:1204.0283}

\bibitem{martini}
G.~Martini, W.~Taylor,
``6D F-theory models and elliptically fibered Calabi-Yau threefolds
over semi-toric base surfaces,''
JHEP {\bf 1506}, 061 (2015)
{\tt arXiv:1404.6300}

\bibitem{wang-non-toric}
W.~Taylor, Y.~Wang,
``Non-toric bases for elliptic Calabi-Yau Threefolds and 6D F-theory Vacua,''
{\tt arXiv:1504.07689} 

\bibitem{large-h21}
S.~Johnson, W.~Taylor,
``Calabi-Yau Threefolds with Large $h^{2,1}$,''
JHEP {\bf 1410}, 23 (2014),
{\tt arXiv:1406.0514}

\bibitem{SCFT-I}
J.~J.~Heckman, D.~R.~Morrison, and C.~Vafa,
``On the Classification of 6D SCFTs and Generalized ADE Orbifolds,''
JHEP {\bf 1405}, 028 (2014)
{\tt arXiv:1312.5746 [hep-th]}.

\bibitem{SCFT-II}
J.~J.~Heckman, D.~R.~Morrison, T.~Rudelius, and C.~Vafa,
``Atomic Classification of 6D SCFTs,''
Fortsch.\ Phys.\  {\bf 63}, 468 (2015)
{\tt arXiv:1502.05405 [hep-th]}.

\bibitem{Nakayama}
N.~Nakayama, ``On {W}eierstrass models,'' in {\em Algebraic geometry and
  commutative algebra, Vol.\ II}, pp.~405--431.
\newblock Kinokuniya, 1988.

\bibitem{Grassi}
A.~Grassi, ``On minimal models of elliptic threefolds,'' Math. Ann. {\bf 290}
  (1991) 287--301.

\bibitem{Gross}
M.\ 
Gross, 
``A finiteness theorem for elliptic Calabi--Yau threefolds,''
Duke Math.\ Jour.\  {\bf  74}, 271 (1994). 

\bibitem{universality}
  V.~Kumar and W.~Taylor,
  ``String Universality in Six Dimensions,''
Adv.\ Theor.\ Math.\ Phys.\  {\bf 15}, no. 2, 325 (2011),
  {\tt arXiv:0906.0987 [hep-th]}.

\bibitem{Vafa-swampland}
  C.~Vafa,
  ``The string landscape and the swampland,''
  {\tt arXiv:hep-th/0509212}.

\bibitem{Morrison-Rudelius}
  D.~R.~Morrison and T.~Rudelius,
  ``F-theory and Unpaired Tensors in 6D SCFTs and LSTs,''
{\tt arXiv:1605.08045 [hep-th]}.
%

\bibitem{Morrison-TASI} 
  D.~R.~Morrison,
  ``TASI lectures on compactification and duality,''
{\tt hep-th/0411120}.

\bibitem{Denef-F-theory}
  F.~Denef,
  ``Les Houches Lectures on Constructing String Vacua,''
  {\tt arXiv:0803.1194 [hep-th]}.

\bibitem{WT-TASI}
  W.~Taylor,
  ``TASI Lectures on Supergravity and String Vacua in Various Dimensions,''
  {\tt arXiv:1104.2051 [hep-th]}.

\bibitem{bhpv}
W.\ P.\ Barth, K.\ Hulek, C.\ A.\ M.\ Peters, A.\ Van de Ven,
``Compact complex surfaces,'' Springer, 2004.  

\bibitem{Tate}
J. Tate, ``Algorithm for determining the type of a singular fiber in an elliptic pencil, Modular
functions of one variable, IV'' (Proc. Internat. Summer School, Univ. Antwerp, Antwerp,
1972), Lecture Notes in Math., vol. 476, Springer, Berlin, 1975, pp. 33-52.

\bibitem{Bershadsky-all}
  M.~Bershadsky, K.~A.~Intriligator, S.~Kachru, D.~R.~Morrison, V.~Sadov and C.~Vafa,
  ``Geometric singularities and enhanced gauge symmetries,''
  Nucl.\ Phys.\  B {\bf 481}, 215 (1996)
  {\tt arXiv:hep-th/9605200}.

\bibitem{kmss-Tate}
S.~Katz, D.~R.~Morrison, S.~Sch{\"a}fer-Nameki and J.~Sully, ``Tate's Algorithm and F-theory,''
JHEP {\bf 1108}, 094 (2012)
{\tt arXiv:1106.3854 [hep-th]}.

\bibitem{mt-singularities} 
D.\ R.\  Morrison and W.\ Taylor, 
``Matter and singularities,''
JHEP {\bf 1201}, 022 (2012)
{\tt arXiv:1106.3563 [hep-th]}


\bibitem{ckpt} 
  M.~Cvetic, D.~Klevers, H.~Piragua and W.~Taylor,
  ``General U(1)xU(1) F-theory Compactifications and Beyond: Geometry
  of unHiggsings and novel Matter Structure,''
JHEP {\bf 1511}, 204 (2015)
{\tt arXiv:1507.05954 [hep-th]}.

\bibitem{agrt} 
  L.~B.~Anderson, J.~Gray, N.~Raghuram and W.~Taylor,
  ``Matter in transition,'' 
JHEP {\bf 1604}, 080 (2016)
{\tt  arXiv:1512.05791 [hep-th]}.  
  
  

\bibitem{Klevers-WT} 
  D.~Klevers and W.~Taylor,
  ``Three-Index Symmetric Matter Representations of SU(2) in F-Theory
  from Non-Tate Form Weierstrass Models,''  
{\tt arXiv:1604.01030 [hep-th]}.  

\bibitem{Zariski}
O. Zariski, 
``The theorem of {R}iemann-{R}och for high multiples of an effective divisor on an algebraic surface,''
Ann. of Math. (2) {\bf 76} (1962)
560--615.

\bibitem{4D-NHC}
  D.~R.~Morrison and W.~Taylor,
  ``Non-Higgsable clusters for 4D F-theory models,''
JHEP {\bf 1505}, 080 (2015)
{\tt arXiv:1412.6112 [hep-th]}.

\bibitem{Fulton} 
William Fulton, ``Introduction to Toric Varieties,'' Annals of
Mathematics Study 131, Princeton University Press, Princeton, 1993.

\bibitem{Erler}
  J.~Erler,
  ``Anomaly Cancellation In Six-Dimensions,''
  J.\ Math.\ Phys.\  {\bf 35}, 1819 (1994)
  {\tt arXiv:hep-th/9304104}.

\bibitem{finite}
  V.~Kumar and W.~Taylor,
  ``A bound on 6D ${\cal N} = 1$ supergravities,''
    JHEP {\bf 0912}, 050 (2009)
  {\tt arXiv:0910.1586 [hep-th]}.

\bibitem{b-Grimm}
  F.~Bonetti and T.~W.~Grimm,
  ``Six-dimensional (1,0) effective action of F-theory via M-theory on Calabi-Yau threefolds,''
JHEP {\bf 1205}, 019 (2012)
{\tt arXiv:1112.1082 [hep-th]}.

\bibitem{WT-Hodge} 
  W.~Taylor,
  ``On the Hodge structure of elliptically fibered Calabi-Yau threefolds,''
JHEP {\bf 1208}, 032 (2012)
{\tt arXiv:1205.0952 [hep-th]}.

\bibitem{stw}
R.\ Wazir, ``Arithmetic on elliptic threefolds,''
{\sl Compos.\ Math.,} {\bf 140},
pp.\ 567-580, (2004).

\bibitem{Berglund-note}
P.\ Berglund,
Y.\ Huang, H.\ Smith, W.\ Taylor, Y.\ Wang, 
{\it work in progress}.

\bibitem{Danielsson-Sundborg} 
  U.~H.~Danielsson and B.~Sundborg,
  ``Exceptional equivalences in N=2 supersymmetric Yang-Mills theory,''
Phys.\ Lett.\ B {\bf 370}, 83 (1996)
{\tt hep-th/9511180}

\bibitem{akms} 
  A.~C.~Avram, M.~Kreuzer, M.~Mandelberg and H.~Skarke,
  ``Searching for K3 fibrations,''
Nucl.\ Phys.\ B {\bf 494}, 567 (1997)
{\tt hep-th/9610154}.

\bibitem{Candelas-cs} 
  P.~Candelas, A.~Constantin and H.~Skarke,
  ``An Abundance of K3 Fibrations from Polyhedra with Interchangeable
  Parts,''  Commun.\  Math.\  Phys.\  {\bf 324}, 937 (2013)
{\tt arXiv:1207.4792 [hep-th]}.  


\bibitem{Gray-hl} 
  J.~Gray, A.~S.~Haupt and A.~Lukas,
  ``Topological Invariants and Fibration Structure of Complete
  Intersection Calabi-Yau Four-Folds,''
JHEP {\bf 1409}, 093 (2014)
{\tt arXiv:1405.2073 [hep-th]}.

\bibitem{Anderson-aggl} 
  L.~B.~Anderson, F.~Apruzzi, X.~Gao, J.~Gray and S.~J.~Lee,
  ``A New Construction of Calabi-Yau Manifolds: Generalized CICYs,''
Nucl.\ Phys.\ B {\bf 906}, 441 (2016)
{\tt arXiv:1507.03235 [hep-th]}.

\bibitem{Esole-Yau} 
  M.~Esole and S.~-T.~Yau,
  ``Small resolutions of SU(5)-models in F-theory,''
Adv.\ Theor.\ Math.\ Phys.\  {\bf 17}, no. 6, 1195 (2013)
{\tt arXiv:1107.0733 [hep-th]}.

\bibitem{Lawrie-sn} 
  C.~Lawrie and S.~Sch\"afer-Nameki,
  ``The Tate Form on Steroids: Resolution and Higher Codimension Fibers,''
JHEP {\bf 1304}, 061 (2013)
{\tt arXiv:1212.2949 [hep-th]}.

\bibitem{Hayashi-ls} 
  H.~Hayashi, C.~Lawrie and S.~Schafer-Nameki,
  ``Phases, Flops and F-theory: SU(5) Gauge Theories,''
JHEP {\bf 1310}, 046 (2013)
{\tt arXiv:1304.1678 [hep-th]}.

\bibitem{hlms} 
  H.~Hayashi, C.~Lawrie, D.~R.~Morrison and S.~Schafer-Nameki,
  ``Box Graphs and Singular Fibers,''
JHEP {\bf 1405}, 048 (2014)
{\tt arXiv:1402.2653 [hep-th]}.

\bibitem{Esole-sy} 
  M.~Esole, S.~H.~Shao and S.~T.~Yau,
  ``Singularities and Gauge Theory Phases,''
{\tt arXiv:1402.6331 [hep-th]};
  ``Singularities and Gauge Theory Phases II,''
{\tt arXiv:1407.1867 [hep-th]}.

\bibitem{Braun-sn} 
  A.~P.~Braun and S.~Schafer-Nameki,
  ``Box Graphs and Resolutions I,''
Nucl.\ Phys.\ B {\bf 905}, 447 (2016)
{\tt arXiv:1407.3520 [hep-th]}.


\bibitem{Kreuzer-Skarke} 
  M.~Kreuzer and H.~Skarke,
  ``Complete classification of reflexive polyhedra in four-dimensions,''
Adv.\ Theor.\ Math.\ Phys.\  {\bf 4}, 1209 (2002)
{\tt hep-th/0002240}.

\bibitem{ks-data}
M.~Kreuzer and H.~Skarke, data available online at
{\tt http://hep.itp.tuwien.ac.at/~kreuzer/CY/}

\bibitem{Katz-Vafa}
  S.~H.~Katz and C.~Vafa,
  ``Matter from geometry,''
  Nucl.\ Phys.\  B {\bf 497}, 146 (1997)
  {\tt arXiv:hep-th/9606086}.

\bibitem{Tachikawa}
  K.~Ohmori, H.~Shimizu, Y.~Tachikawa and K.~Yonekura,
  ``6d $\mathcal{N}=\left(1,\;0\right) $ theories on S$^{1}$ /T$^{2}$ and class S theories: part II,''
JHEP {\bf 1512}, 131 (2015)
{\tt arXiv:1508.00915}

\bibitem{Persson}
U.\ Persson, ``Configurations of Kodaira fibers on rational elliptic
surfaces,'' Math.\ Z. {\bf 205} (1990), 1-47

\bibitem{Miranda}
R.\ Miranda, ``Persson's list of singular fibers for a rational
elliptic surface,'' Math.\ Z. {\bf 205} (1990), 191-211

\bibitem{mpt}
D.\ Morrison, D.\ Park and W.\ Taylor, {\it to appear}.

\bibitem{global-symmetries}
M.~Bertolini, P.~.R~Merkx, D.~R.~Morrison,
``On the global symmetries of 6D conformal theories,''
{\tt arXiv:1510.08056}

\bibitem{phase-transitions}
E.~Witten,
``Phase Transitions in M-theory and F-theory,''
Nucl.\ Phys.\ B {\bf 471}, 195 (1996)
{\tt arXiv:hep-th/9603150}

\bibitem{Raghuram-17}
N.\ Raghuram, {\it private communication}.

\bibitem{Schwarz-infinite}
  J.~H.~Schwarz,
  ``Anomaly-Free Supersymmetric Models in Six Dimensions,''
  Phys.\ Lett.\  B {\bf 371}, 223 (1996)
  {\tt arXiv:hep-th/9512053}.

\bibitem{little} 
  L.~Bhardwaj, M.~Del Zotto, J.~J.~Heckman, D.~R.~Morrison, T.~Rudelius and C.~Vafa,
 ``F-theory and the Classification of Little Strings,''
Phys.\ Rev.\ D {\bf 93}, no. 8, 086002 (2016)
{\tt arXiv:1511.05565 [hep-th]}.

\bibitem{Sadov}
  V.~Sadov,
  ``Generalized Green-Schwarz mechanism in F theory,''
  Phys.\ Lett.\  B {\bf 388}, 45 (1996)
  {\tt arXiv:hep-th/9606008}.

\bibitem{kpt} 
  V.~Kumar, D.~S.~Park and W.~Taylor,
  ``6D supergravity without tensor multiplets,''
JHEP {\bf 1104}, 080 (2011)
{\tt arXiv:1011.0726 [hep-th]}.


\bibitem{Morrison-Park} 
  D.~R.~Morrison and D.~S.~Park,
  ``F-Theory and the Mordell-Weil Group of Elliptically-Fibered Calabi-Yau Threefolds,''
JHEP {\bf 1210}, 128 (2012)
{\tt arXiv:1208.2695} [hep-th].

\bibitem{mt-sections} 
  D.~R.~Morrison and W.~Taylor,
  ``Sections, multisections, and U(1) fields in F-theory,''
  {\tt arXiv:1404.1527 [hep-th]}.

\bibitem{Klevers-3} 
  D.~Klevers, D.~K.~Mayorga Pena, P.~K.~Oehlmann, H.~Piragua and J.~Reuter,
  ``F-Theory on all Toric Hypersurface Fibrations and its Higgs Branches,''
JHEP {\bf 1501}, 142 (2015)
{\tt arXiv:1408.4808 [hep-th]}

\bibitem{Grimm-kk} 
  T.~W.~Grimm, A.~Kapfer and D.~Klevers,
``The Arithmetic of Elliptic Fibrations in Gauge Theories on a Circle,''
{\tt arXiv:1510.04281 [hep-th]}.


\bibitem{Lawrie-sw} 
  C.~Lawrie, S.~Schafer-Nameki and J.~M.~Wong,
  ``F-theory and All Things Rational: Surveying U(1) Symmetries with Rational Sections,''
JHEP {\bf 1509}, 144 (2015)
{\tt arXiv:1504.05593 [hep-th]}


\bibitem{Park-Taylor}
  D.~S.~Park and W.~Taylor,
  ``Constraints on 6D Supergravity Theories with Abelian Gauge Symmetry,''
JHEP {\bf 1201}, 141 (2012)
{\tt arXiv:1110.5916 [hep-th]}.

\bibitem{Park-abelian} 
  D.~S.~Park,
  ``Anomaly Equations and Intersection Theory,''
JHEP {\bf 1201}, 093 (2012)
{\tt arXiv:1111.2351 [hep-th]}.

\bibitem{Mayrhofer:2012zy} 
  C.~Mayrhofer, E.~Palti and T.~Weigand,
  ``U(1) symmetries in F-theory GUTs with multiple sections,''
JHEP {\bf 1303}, 098 (2013)
{\tt arXiv:1211.6742} [hep-th].

\bibitem{Braun:2013yti} 
  V.~Braun, T.~W.~Grimm and J.~Keitel,
  ``New Global F-theory GUTs with U(1) symmetries,''
JHEP {\bf 1309}, 154 (2013)
{\tt arXiv:1302.1854} [hep-th].


\bibitem{Borchmann:2013jwa} 
  J.~Borchmann, C.~Mayrhofer, E.~Palti and T.~Weigand,
  ``Elliptic fibrations for $SU(5)\times U(1)\times U(1)$ F-theory vacua,''
Phys.\ Rev.\ D {\bf 88}, no. 4, 046005 (2013)
{\tt arXiv:1303.5054} [hep-th].
 

\bibitem{Cvetic-Klevers-1} 
  M.~Cvetic, D.~Klevers and H.~Piragua,
  ``F-Theory Compactifications with Multiple U(1)-Factors: Constructing Elliptic Fibrations with Rational Sections,''
JHEP {\bf 1306}, 067 (2013)
{\tt arXiv:1303.6970} [hep-th].

\bibitem{Cvetic:2013uta} 
  M.~Cvetic, A.~Grassi, D.~Klevers and H.~Piragua,
  ``Chiral Four-Dimensional F-Theory Compactifications With SU(5) and Multiple U(1)-Factors,''
JHEP {\bf 1404}, 010 (2014)
{\tt arXiv:1306.3987 [hep-th]}


\bibitem{Borchmann:2013hta} 
  J.~Borchmann, C.~Mayrhofer, E.~Palti and T.~Weigand,
  ``SU(5) Tops with Multiple U(1)s in F-theory,''
Nucl.\ Phys.\ B {\bf 882}, 1 (2014)
{\tt arXiv:1307.2902} [hep-th].

\bibitem{Cvetic-Klevers-2} 
  M.~Cvetic, D.~Klevers, H.~Piragua and P.~Song,
  ``Elliptic fibrations with rank three Mordell-Weil group: F-theory with U(1) x U(1) x U(1) gauge symmetry,''
JHEP {\bf 1403}, 021 (2014)
{\tt arXiv:1310.0463} [hep-th].

\bibitem{Klemm-lry}
  A.~Klemm, B.~Lian, S.~S.~Roan, S.~-T.~Yau,
  ``Calabi-Yau fourfolds for M theory and F theory compactifications,''
  Nucl.\ Phys.\  {\bf B518}, 515-574 (1998).
  {\tt hep-th/9701023}.


\bibitem{Mayrhofer:2014opa} 
  C.~Mayrhofer, D.~R.~Morrison, O.~Till and T.~Weigand,
  ``Mordell-Weil Torsion and the Global Structure of Gauge Groups in F-theory,''
JHEP {\bf 1410}, 16 (2014)
{\tt arXiv:1405.3656 [hep-th]}.


\bibitem{Anderson:2014yva} 
  L.~B.~Anderson, I.~Garcia-Etxebarria, T.~W.~Grimm and J.~Keitel,
  ``Physics of F-theory compactifications without section,''
JHEP {\bf 1412}, 156 (2014)
{\tt arXiv:1406.5180 [hep-th]}.


\bibitem{Garcia-Etxebarria:2014qua} 
  I.~Garcia-Etxebarria, T.~W.~Grimm and J.~Keitel,
  ``Yukawas and discrete symmetries in F-theory compactifications without section,''
JHEP {\bf 1411}, 125 (2014)
{\tt arXiv:1408.6448 [hep-th]}.


\bibitem{Mayrhofer:2014haa} 
  C.~Mayrhofer, E.~Palti, O.~Till and T.~Weigand,
  ``Discrete Gauge Symmetries by Higgsing in four-dimensional F-Theory Compactifications,''
JHEP {\bf 1412}, 068 (2014)
{\tt arXiv:1408.6831 [hep-th]}.


\bibitem{Mayrhofer:2014laa} 
  C.~Mayrhofer, E.~Palti, O.~Till and T.~Weigand,
  ``On Discrete Symmetries and Torsion Homology in F-Theory,''
JHEP {\bf 1506}, 029 (2015)
{\tt arXiv:1410.7814 [hep-th]}.


\bibitem{Cvetic:2015moa} 
  M.~Cvetic, R.~Donagi, D.~Klevers, H.~Piragua and M.~Poretschkin,
  ``F-theory vacua with $\mathbb Z_3$ gauge symmetry,''
Nucl.\ Phys.\ B {\bf 898}, 736 (2015)
{\tt arXiv:1502.06953 [hep-th]}.


\bibitem{Grimm:2015ona} 
  T.~W.~Grimm, T.~G.~Pugh and D.~Regalado,
  ``Non-Abelian discrete gauge symmetries in F-theory,''
JHEP {\bf 1602}, 066 (2016)
{\tt arXiv:1504.06272 [hep-th]}.

\bibitem{Oehlmann:2016wsb} 
  P.~K.~Oehlmann, J.~Reuter and T.~Schimannek,
  ``Mordell-Weil Torsion in the Mirror of Multi-Sections,''
{\tt arXiv:1604.00011 [hep-th]}.
%

\bibitem{Huang-WT}
Y.\ Huang and W.\ Taylor,
{\it work in progress}.

\bibitem{Anderson-Taylor} 
  L.~B.~Anderson and W.~Taylor,
  ``Geometric constraints in dual F-theory and heterotic string compactifications,''
  JHEP {\bf 1408}, 025 (2014),
{\tt arXiv:1405.2074 [hep-th]}.

\bibitem{Halverson-WT} 
  J.~Halverson and W.~Taylor,
  ``$\mathbb{P}^1$-bundle bases and the prevalence of non-Higgsable
  structure in 4d F-theory models,''  
JHEP {\bf 1509}, 086 (2015)
{\tt arXiv:1506.03204 [hep-th]}.

\bibitem{Grimm-Taylor} 
  T.~W.~Grimm and W.~Taylor,
  ``Structure in 6D and 4d N=1 supergravity theories from F-theory,''
  JHEP {\bf 1210}, 105 (2012)  {\tt arXiv:1204.3092 [hep-th]}.



\end{thebibliography}
\end{document}